\documentclass [11pt]{article}
\usepackage{amsfonts}
\usepackage{graphicx}
\usepackage{amsmath,amssymb, dsfont}
\usepackage{latexsym}
\usepackage{mathrsfs}
\usepackage{bbold}
\usepackage{color}
\usepackage{slashed}
\usepackage{cancel}
\usepackage{soul} %
\usepackage{setspace} %
\usepackage{comment}
\usepackage[dvipsnames]{xcolor}
\usepackage{booktabs}
\usepackage{multirow}
\usepackage{ifthen,multicol,pifont,epsfig,latexsym,amsmath,amssymb,tikz-feynhand, tikz-feynman, feynmf,bbm}
\usepackage{cite}
\usepackage[small]{caption}
\usepackage{epsfig}
\usepackage{slashed}
\usepackage{tabu}
\usepackage{xcolor}
\usepackage{epstopdf}
\usepackage{latexsym}
\usepackage{graphicx}
\usepackage{rotating}
\usepackage{booktabs}
\usepackage{bbm}
\usepackage{bbold}
\usepackage{enumitem}
\usepackage[numbers,compress]{natbib}

\definecolor{red}{rgb}{1,0,0}

\makeatletter
\def\section{\@startsection {section}{1}{\z@}{-3.5ex plus -1ex minus
 -.2ex}{2.3ex plus .2ex}{\large\bf}}
\def\subsection{\@startsection{subsection}{2}{\z@}{-3.25ex plus -1ex
minus -.2ex}{1.5ex plus .2ex}{\normalsize\bf}}
\makeatother
\makeatletter

\@addtoreset{equation}{section}

\makeatother

\newcommand{\bea}{\begin{equation} \begin{aligned}} \newcommand{\eea}{\end{aligned} \end{equation}}
\def\be{\begin{equation}} \def\ee{\end{equation}} 
\def\nn{\nonumber}

\usepackage[colorlinks,linkcolor=black,citecolor=blue,urlcolor=blue,linktocpage,pagebackref]{hyperref}

\allowdisplaybreaks

\begin{document}

\thispagestyle{empty}

\begin{center}

	\vspace*{-.6cm}

	\begin{center}

		\vspace*{1.1cm}

		{\centering \Large\textbf{Free Energy on the Sphere for Non-Abelian Gauge Theories}}

	\end{center}

	\vspace{0.8cm}
	{\bf Fabiana De Cesare$^{b,c}$, Lorenzo Di Pietro$^{a,b}$ and Marco Serone$^{b,c}$}

	\vspace{1.cm}
	
	${}^a\!\!$
	{\em  Dipartimento di Fisica, Universit\`a di Trieste, \\ Strada Costiera 11, I-34151 Trieste, Italy}
		
	\vspace{.3cm}

	${}^b\!\!$
	{\em INFN, Sezione di Trieste, Via Valerio 2, I-34127 Trieste, Italy}

	\vspace{.3cm}

	${}^c\!\!$
	{\em SISSA, Via Bonomea 265, I-34136 Trieste, Italy}

	\vspace{.3cm}

\end{center}

\vspace{1cm}

\centerline{\bf Abstract}
\vspace{2 mm}
\begin{quote}

We compute the $S^d$ partition function of the fixed point of non-abelian gauge theories in continuous $d$, using the $\epsilon$-expansion around $d=4$. We illustrate in detail the technical aspects of the calculation, including all the factors arising from the gauge-fixing procedure, and the method to deal with the zero-modes of the ghosts. We obtain the result up to NLO, i.e. including two-loop vacuum diagrams. Depending on the sign of the one-loop beta function, there is a fixed point with real gauge coupling in $d>4$ or $d<4$. In the first case we extrapolate to $d=5$ to test a recently proposed construction of the UV fixed point of $5d$ $SU(2)$ Yang-Mills via a susy-breaking deformation of the $E_1$ SCFT. We find that the $F$ theorem allows the proposed RG flow. In the second case we extrapolate to $d=3$ to test whether QCD$_3$ with gauge group $SU(n_c)$ and $n_f$ fundamental matter fields flows to a CFT or to a symmetry-breaking phase. We find that within the regime with a real gauge coupling near $d=4$ the CFT phase is always favored. For lower values of $n_f$ we compare the average of $F$ between the two complex fixed points with its value at the symmetry-breaking phase to give an upper bound of the critical value $n_f^*$ below which the symmetry-breaking phase takes over. 

\end{quote}

\newpage

\tableofcontents
	%%%%%%%%%%%%%%%%%%%%%%%%%%%%%%%%%%%%%%%
\section{Introduction}

The intuitive idea that the number of degrees of freedom should decrease along a Renormalization Group (RG) flow can be made precise in Quantum Field Theory (QFT). This is done by assigning a quantity to Conformal Field Theories (CFT) --which typically are the endpoints of RG flows-- with the property of monotonicity along the flow, i.e. its value is always larger for the UV CFT than it is for the IR CFT.  Such quantities were first found in even spacetime dimensions $d$ \cite{Zamolodchikov:1986gt,Cardy:1988cwa,Komargodski:2011vj,Komargodski:2011xv} and identified with the $a$ coefficient in the Weyl anomaly
\be
\left\langle T_{\mu}^{\mu}\right\rangle \sim(-1)^{d / 2} a E_{d}+\sum_{i} c_{i} I_{i}~,
\ee
where $E_{d}$ is the Euler density term, and $c_{i}$ are the coefficients of other Weyl invariant curvature terms $I_i$. The inequality $a_{\mathrm{UV}}>a_{\mathrm{IR}}$ was proved in $d=2$ \cite{Zamolodchikov:1986gt}, in which case it reduces to the celebrated $c$-theorem because $a=c/3$, and in $d=4$ \cite{Komargodski:2011vj,Komargodski:2011xv}. Attempts to generalize the proof to $d=6$ \cite{Elvang:2012st} so far have only succeeded in the supersymmetric case \cite{Cordova:2015fha}. 

In odd dimensions, even though there are no Weyl anomalies, a monotonic quantity can still be defined using the (renormalized) free energy on the sphere
\be
F = -\log  Z_{S^d}~.
\ee
The inequality $F_{\mathrm{UV}}>F_{\mathrm{IR}}$, known as the $F$-theorem, was proposed and checked in \cite{Jafferis:2011zi, Klebanov:2011gs} and then proved in $d=3$ using the relation to entanglement entropy across a spherical entangling surface \cite{Casini:2011kv, Casini:2012ei}, which can also be used to prove monotonicity of the $a$ coefficient in $d=2,4$ \cite{Casini:2006es,  Casini:2017vbe}. For other odd dimensions it is conjectured that the decreasing quantity becomes $(-1)^{\frac{d+1}{2}}F$ \cite{Klebanov:2011gs}. This has not been proved yet but is motivated by several examples.

It is sometimes possible to continue RG flows to non-integer dimensions, at least formally. When a flow to an interacting fixed point can be continued to the vicinity of its upper or lower critical dimension, it becomes short and controllable in perturbation theory. This strategy, known as $\epsilon$-expansion \cite{Wilson:1971dc, Wilson:1973jj}, can lead to a useful approximation of strongly coupled fixed points. Motivated by this method, ref. \cite{Giombi:2014xxa} proposed to unify all the previously mentioned inequalities in a single relation valid in continuous dimensions. The authors defined 
\be
\widetilde{F}=-\sin\left(\frac{\pi d}{2}\right)F,
\label{eq:FtildeDef}
\ee
which in odd $d$ exactly reproduces the $(-1)^{(d+1)/2}F$ term, while in even $d$ provides a smooth limit proportional to the $a$ anomaly: the factor  $\sin(\frac{\pi d}{2})$ cancels  the pole in the free energy leading to the finite limit $\widetilde{F}=\pi a/2$. Therefore, the inequality
\be
\widetilde{F}_\text{UV}>\widetilde{F}_\text{IR}
\label{eq:Ftheo}
\ee
automatically encodes all the previous relations, and extends them to non-integer values of $d$. 

In this paper we compute the quantity $\widetilde{F}$ for the fixed point in non-abelian gauge theories, in an expansion around $d=4$.  The existence of such fixed points can be inferred from the leading terms in the $\beta$ function for the gauge coupling, which in $d=4+2\epsilon$ read
\begin{equation}
\beta_g = \epsilon g + \beta_0 g^3 + \mathcal{O}(g^5)~.
\end{equation}
For a small number of matter fields the one-loop coefficient $\beta_0$ is negative and  leads to a one-loop fixed point $g^{* 2}_{\text{1-loop}}>0$ for $d>4$, while for $d<3$  a minimal number of matter fields is required to have $\beta_0>0$, so that $g^{* 2}_{\text{1-loop}}>0$. The quantity $\widetilde{F}$ was computed in \cite{Giombi:2015haa} for $d$-dimensional QED with $n_f$ four-component fermionic matter fields, for which $\beta_0$ is always positive. It was then extrapolated to $d=3$ to study the existence of an interacting IR CFT for QED in 3 spacetime dimensions, by comparing with the quantity $F$ for the spontaneously broken phase of $2 n_f^2 +1$ massless Goldstone bosons. 

The calculation in non-abelian gauge theories presents several new challenges compared to the abelian case. Firstly, the gauge fixing requires a more careful analysis, because it becomes unavoidable to include the interaction with the ghost fields. On the sphere massless scalar fields like the ghosts have zero modes. Due to the fermionic nature of ghosts, this naively leads to a zero in the partition function, which manifests as an IR divergence in $\widetilde{F}$. This divergence needs to be cured by an appropriate regulator (or alternatively by appropriately modifying the gauge-fixing procedure, as we describe in an appendix). Note that, in order to obtain $\widetilde{F}$, it is crucial to carefully keep track of the normalization of the path integral on $S^d$ when implementing the gauge-fixing through the Faddeev-Popov procedure \cite{Faddeev:1967fc}. Secondly, the derivative self-interaction of the gluon leads to diagrams with two derivatives acting on the propagator, and it is important to include also the contact-term contributions in order to evaluate correctly the integrals over the positions of the vertices. Thirdly, unlike QED the renormalization in the gauge sector is not simply encoded in the definition of a renormalized gauge coupling, instead one needs to consider also wave-function counterterms for the gluons and the ghosts. We perform the calculation, taking care of all these issues, up to the next-to-leading (NLO) order, i.e. including up to two-loop vacuum diagrams. The result is in eq.~\eqref{eq:fFinal}. Note that, while we compute the two-loop diagrams in generic $\xi$-gauge, which allows us to compare with heat-kernel results for generic background \cite{Jack:1982sn}, we keep track of the normalization of the path integral only in the special case of the Landau gauge, i.e. $\xi = 0$.

We then apply this result to the fixed points of $SU(n_c)$ non-abelian gauge theories in $d=3$ and in $d=5$. In $d=3$, just like in the QED case mentioned above, the theory is known to flow to a CFT in the IR for a sufficiently large number of matter flavors $n_f$ \cite{Appelquist:1989tc}, and it is conjectured to change its behavior for $n_f$ smaller than an unknown critical value $n_f^*$, flowing instead to a phase with spontaneous breaking of the global symmetry  \cite{Vafa:1983tf,Vafa:1984xh,Appelquist:1989tc,Komargodski:2017keh}. We adopt the same logic as in \cite{Giombi:2015haa}, and compare $F$ of the fixed point to that of the putative Goldstone bosons phase. We find that when $\beta_0>0$, so that $g^{* 2}_{\text{1-loop}}>0$, the conformal phase is always favored compared to the symmetry-breaking phase. For $\beta_0<0$ the fixed point is complex in the $\epsilon$-expansion, but a unitary fixed point in $d=3$ can still exist.\footnote{The opposite situation can also occur, a fixed point for $\epsilon\ll 1$ which disappears in physical integer dimensions.} We propose a more speculative approach to estimate $F$ of the $3d$ CFT in this case, by taking an average value of $\widetilde F$ among the two complex fixed points. With this method we find that the Goldstone boson phase becomes favored for small $n_f$, allowing us to put an upper bound on $n_f^*$. The values found for $2\leq n_c \leq 5$ are reported in eq.~\eqref{eq:nfbounds}. The result for $n_c=2$ favorably compares with previous bounds found using again the $F$-theorem combined with supersymmetry \cite{Sharon:2018apk}, or lattice methods \cite{Karthik:2018nzf}.  We also give an estimate for the upper bound on $x^*$ in the Veneziano limit in eq.~\eqref{eq:Veneziano}, where $x=n_f/n_c$.

In $d=5$ we use our calculation to investigate the existence of interacting CFTs that UV complete $5d$ non-abelian gauge theories. If such CFTs exist they would be an example of a non-supersymmetric interacting CFT in $d>4$. An interesting construction in the case of $SU(2)$ Yang-Mills theory was recently proposed in \cite{BenettiGenolini:2019zth}, and further refined in \cite{Bertolini:2021cew}, using the $E_1$ superconformal field theory that UV completes $SU(2)$ Super Yang-Mills. The putative non-supersymmetric CFT is obtained as the IR endpoint of the RG flow triggered by a certain non-supersymmetric deformation of $E_1$, and by construction it is endowed with a relevant deformation that flows to ordinary $SU(2)$ Yang-Mills theory. Using our extrapolation to $5d$ we can compare the quantity $F$ of the non-supersymmetric CFT with that of the $E_1$ SCFT, known from supersymmetric localization \cite{Chang:2017cdx}, and test if the RG flow is allowed. 
We can also easily repeat this check in the case with fundamental flavors $n_f$ and compare with the $F$ quantity of the $E_{n_f + 1}$ SCFT that UV completes the supersymmetric gauge theory with flavors. In all cases in which we have evidence for a fixed point in $d=5$, namely $n_f \leq 4$ \cite{DeCesare:2021pfb}, we obtain that the $F$-theorem allows the proposed RG flow.

The rest of the paper is organized as follows: in section 2 we explain some generalities about the calculation of the sphere partition function, we perform the gauge-fixing and compute the one-loop determinants for non-abelian gauge theories. In section 3 we derive the Feynman rules on the sphere, including the gauge field propagator in an arbitrary $\xi$-gauge. In section 4 we compute the two-loop vacuum-vacuum diagrams and obtain our main result. In section 5 we apply the result to the $d=3$ and $d=5$ models described above. In section 6 we draw our conclusions and outline some possible future directions. Most of the technical points of the calculation are relegated to the first three appendices. In appendix D 
 we show a sanity check of our results, by comparing in detail the UV divergences obtained for pure Yang-Mills theory in ref.~\cite{Jack:1982sn} in the Feynman gauge $\xi=1$ with our results.
In appendix E we explain a possible alternative gauge-fixing procedure (used already in \cite{Pestun:2007rz})  where ghost zero modes are treated more carefully by introducing ghosts for ghosts,
which we also use to partially check the results in the main body. 

Finally, a comment on notation: in this paper $n_f$ always refers to the number of $4d$ Dirac fermions. Given the way we analytically continue fermions, $n_f$ $4d$ Dirac fermions
give rise to $2n_f$ Dirac fermions in $3d$ and $n_f$ Dirac fermions in $5d$.

%%%%%%%%%%%%%%%%%%%%%%%%%%%%%%%%%%%%%%

\section{Free energy of gauge theories on the sphere: leading order}

Let us consider a non-abelian gauge theory with $n_f$ massless Dirac fermions in the fundamental representation.
We want to compute the sphere free energy in $d=4+2\epsilon$, defined as
\be
F  = -\log  Z_{S^d}~, \qquad 
Z_{S^d}  = \frac{1}{\mathrm{vol}(\mathcal{G})} \int \mathcal{D}A\mathcal{D}\psi\mathcal{D}\bar{\psi} \, \exp\left(-S[A,\psi,\bar{\psi}, h]\right)~.
\label{eq:a}
\ee
Here $h$ denotes the round metric $h_{\mu\nu}$ on $S^d$ with radius $R$ and coordinate $x$, while $\mathrm{vol}(\mathcal{G})$ is the volume of the space of all gauge transformations, which in our choice of normalization does not depend on the gauge coupling $g$. We can split the action on the sphere in
\begin{align}
\begin{split}
S=S_\text{YM}+S_\text{Ferm}+S_\text{curv}\ ,
\label{eq:b}
\end{split}
\end{align}
with
\begin{align}
S_\text{YM} & =\int d^{d} x \sqrt{h}\left(\frac{1}{4 g_{0}^{2}}  \mathrm{Tr}[F_{\mu\nu}(x)F^{\mu\nu}(x)]\right)\ \label{eq:c1} \,, \\
S_\text{Ferm} &=\int d^{d} x \sqrt{h}\left(-\sum_{i=1}^{n_f} \bar{\psi}_{i} \gamma^{\mu}\left(\nabla_{\mu}+i A_{\mu}\right) \psi^{i}\right)\ ,\label{eq:c2} \\
S_\text{curv} & =\int d^{d} x \sqrt{h}\left(b_{0} E +c_{0} \mathcal{R}^{2} /(d-1)^{2}\right),
\label{eq:c3}
\end{align}
where $h={\rm det}\, h_{\mu \nu}$,\footnote{We use the same symbol $h$ for the metric determinant ${\rm det}\, h_{\mu \nu}$ and for the metric tensor $h_{\mu\nu}$ whenever indices are omitted. The difference should be clear from the context.} $g_0$ is the bare gauge coupling constant, $\psi^{i}$ are $n_f$ four-component Dirac fermions and $\nabla_{\mu}$ is the curved space covariant derivative which includes the spin connection term when acting on fermions. As the action should contain all operators that are marginal in $d=4$, we have added the curvature terms together with their bare coupling parameters $b_{0}$ and $c_{0}$.\footnote{In a generic Euclidean manifold we should also include a term with the square of the Weyl tensor, omitted here as it vanishes on the sphere.} For future purposes, we recall the expression for the Ricci scalar $\mathcal{R}$ and the Euler density $E$ on $S^d$:
\begin{equation}
\begin{aligned}
&\mathcal{R}=\frac{d(d-1)}{R^2}~,~~E=\mathcal{R}_{\mu \nu \lambda \rho} \mathcal{R}^{\mu \nu \lambda \rho}-4 \mathcal{R}_{\mu \nu} \mathcal{R}^{\mu \nu}+\mathcal{R}^{2}=\frac{d(d-1)(d-2)(d-3)}{R^4}~.
\label{eq:curvScala}
\end{aligned}
\end{equation}

\subsection{One-loop determinants}
\label{sec:oneloop}

At leading order in a loopwise expansion the free energy is determined by one loop determinants. 
As a consequence of the splitting in eq.~\eqref{eq:b}, we can divide the leading term of the sphere free 
energy $F_\text{Free}$ in three parts:
\begin{equation}
F_\text{Free}=F_\text{free-YM}+F_\text{free-ferm}+F_\text{curv}\ ,
\end{equation}
with
\begin{align}
F_\text{free-YM} & =-\log\bigg(\frac{1}{\mathrm{vol}(\mathcal{G})} \int \mathcal{D}Ae^{-S_{\text{free-YM}}[A, h]}\bigg)
\label{eq:d1}\,, \\
F_\text{free-ferm} & =-\log \Big(\int\mathcal{D}\psi\mathcal{D}\bar{\psi} \, e^{-S_{\text{free-ferm}}[\psi, h]}\Big)\,,
 \label{eq:d2} \\
F_\text{curv} & =\Omega_d R^{d-4}(d(d-1)(d-2)(d-3))b_0+ d^2 c_0),
\label{eq:d3}
\end{align}
where $S_\text{free-YM}$ is the quadratic part of the Yang-Mills action, $S_{\text{free-ferm}}$ the free fermion action and 
$\Omega_d= 2\pi^{\frac{d+1}{2}}/\Gamma(\frac{d+1}{2})$ is the volume of the $d$-dimensional sphere with unit radius. 

The expression for $F_{\text{free-ferm}}$ was found in ref.~\cite{Giombi:2014xxa}. The result for a single four-component Dirac spinor is 
\be
F_\text{free-ferm}(d)=-\frac{4}{\sin(\frac{\pi d }{2})\Gamma(1+d)}\int_0^1 du\ \cos\left(\frac{\pi u}{2}\right)\Gamma\left(\frac{1+d+u}{2}\right) \Gamma\left(\frac{1+d-u}{2}\right)\ .
\label{eq:r}
\ee

Let us now focus on the computation of $F_\text{free-YM}$.
The gauge field $A^\mu$ on the sphere can be written as the sum of a longitudinal part $A^\mu_{(0)}$ and a transverse part $A^\mu_{(1)}$, which can be separately decomposed in orthonormal eigenvectors of the sphere Laplacian $-\nabla^2$:
\begin{equation}
\begin{split}
A^\mu&=A_{(0)}^\mu+A_{(1)}^\mu~,\quad\text{such that} \quad  \nabla_\mu A_{(1)}^\mu=0~,\\
A_{(0)}^\mu&=\sum_{l>0}a_{(0)}^\ell A_{(0)}^{\mu\ \ell}~,  \quad A_{(1)}^\mu=\sum_{l>0}a_{(1)}^\ell A_{(1)}^{\mu\ \ell}~,
\label{eq:h}
\end{split}
\end{equation}
with corresponding eigenvalues $\lambda_{\ell}^{(1)}$, $\lambda_{\ell}^{(0)}$ and degeneracies  $g_{\ell}^{(1)}$, $g_{\ell}^{(0)}$ given by \cite{Rubin:1984tc}
\bea
\lambda_{\ell}^{(1)} & =\frac{(\ell(\ell+d-1)-1)}{R^2}~, \quad \quad \;\;\; g_{\ell}^{(1)} =\frac{\ell(\ell+d-1)(2 \ell+d-1) \Gamma(\ell+d-2)}{\Gamma(\ell+2) \Gamma(d-1)}~,~~ \ell > 0~, \\
\lambda_{\ell}^{(0)} & = \frac{\ell(\ell+d-1)-(d-1)}{R^2}~, \quad g_{\ell}^{(0)} =\frac{(2 \ell+d-1) \Gamma(\ell+d-1)}{\Gamma(\ell+1) \Gamma(d)}~, \,~~\quad \qquad \qquad \ell > 0~.
\label{eq:f}
\eea
Note that the eigenfunctions of the longitudinal part can be rewritten in terms of the covariant derivative of the spherical harmonics $Y_\ell(x)$
\begin{equation}\label{eq:p2}
A_{(0)}^{\mu\ \ell}=\frac{1}{\sqrt{\lambda_{\ell}^{(S)}}}\nabla^\mu Y_\ell(x)~,~~\text{for } \ell\ge1~.
\end{equation}
We take the spherical harmonics to be normalized as
\be
\int  d^dx\sqrt{h}\ Y_\ell(x)Y_{\ell'}(x)=\delta_{\ell\ell'}~.
\label{eq:u2}
\ee
In order to make the basis $A_{(0)}^{\mu\ \ell}$ orthonormal, we have fixed the normalization factor in terms of the eigenvalue of the laplacian operator associated to $Y_\ell(x)$
\begin{equation}
\begin{split}
\lambda_{\ell}^{(S)}=\frac{\ell(\ell+d-1)}{R^2}~,
\label{eq:f2}
\end{split}
\end{equation}
which has degeneracy $g_\ell^{(0)}$. 
Note a crucial difference between the spectrum for a scalar and for the longitudinal modes of a vector:  the former includes a constant mode with eigenvalue $\lambda_{0}^{(S)}=0$ and degeneracy $g_0^{(0)}=1$, while for the latter the modes are restricted to $\ell>0$ and as a result the constant is excluded.

In dimensional regularization the following identities are valid, which will be useful later in the computation:
\begin{equation}
\sum_{\ell=1}^{\infty} g_{\ell}^{(1)}=1 \quad \text{and} \quad  \sum_{\ell=1}^{\infty} g_{\ell}^{(0)}=-1~.
\label{eq:o}
\end{equation}
With this decomposition in longitudinal and transverse mode the path integral measure can be rewritten as
\be
\int\mathcal{D}A=\int\prod_{\ell=1}^\infty d a_{(0)}^\ell\ \int\prod_{\ell=1}^\infty d a_{(1)}^\ell\ \ .
\ee

\subsection{Computation in Landau gauge}
\label{sec:Landau}

We want to compute 
\be
F_\text{free-YM}=-\log \bigg(\frac{1}{\mathrm{vol}(\mathcal{G})} \int \mathcal{D}A\,e^{-S_{\text{free-YM}}[A, h]} \bigg)\,,
\ee
with
\begin{equation}
   S_\text{free-YM}= \int d^{d} x \sqrt{h}\,\frac{1}{2 g_{0}^{2}}\mathrm{Tr}\left[A_{\nu}(-\delta^\nu_\mu\nabla^2+R^\nu_\mu+\nabla^\nu\nabla_\mu) A^\mu\right]
\end{equation}
and $R^\nu_\mu=\frac{d-1}{R^2}\delta^\nu_\mu$ on $S^d$.
In order to perform the explicit computation it is convenient to add a gauge-fixing term to the action. We work in Landau gauge and set to zero the longitudinal component of the gauge field. In order to do that we insert in the  path integral of eq.~\eqref{eq:a} the following functional identity, valid for any fixed $A_\mu(x)$:
\begin{equation}
1 = \int_\mathcal{G'} \mathcal{D}\mu_g(U) \delta(\nabla^\mu A_\mu^{U} )\left | \mathrm{det}\frac{\delta \nabla^\mu A_\mu^{U}}{\delta \epsilon}\right | ~,
\label{eq:e}
\end{equation}
where $A^U_\mu(x)$  is the gauge-transformed field under $U(x)$
\begin{equation}
A_\mu(x) \to A^U_\mu(x) = U(x)(\nabla_\mu + i A_\mu(x) )U^{-1}(x) \equiv U(x)D^A_\mu U^{-1}(x)~.
\end{equation}
Taking the components in the Lie Algebra, denoted with indices $a,b,c,\dots$, and also writing $U=\exp(i \epsilon^a T^a)$ in terms of the parameter $\epsilon^a$ and the generators $T^a$, we get the infinitesimal transformation
\begin{equation}
\delta A^a_\mu(x) = (D^A_\mu\epsilon)^a(x) = \nabla_\mu \epsilon^a(x) + i f^{abc}A^b_\mu(x)\epsilon^c(x)~.
\end{equation}
The integration in eq.~\eqref{eq:e} is performed over the functional Haar measure $\mu_g$ and is restricted to the set of gauge transformations $\mathcal{G}'$ that act non-trivially on $A_\mu(x)$, i.e. those that give a non-zero functional determinant. In the functional derivative the variation $\delta \epsilon$ is an infinitesimal variation away from $U$ (the integration variable) and tangential to $\mathcal{G}'$, hence $\delta \epsilon$ is any fluctuation not annihilated by the covariant derivative with connection $A_\mu^U$. So we have
\begin{equation}
\left|\mathrm{det}\frac{\delta \nabla^\mu A_\mu^U}{\delta \epsilon}\right | = \mathrm{det}' \left(-\nabla^\mu D^{A^U}_\mu\right)~,
\end{equation}
where the prime denotes that we need to exclude the zero eigenvalue  and the minus sign is taken to ensure positivity of the determinant, at least perturbatively. At this point in order to proceed we restrict ourselves to the case of Landau gauge, and use that in Landau gauge the operator is self-adjoint as  $\nabla^\mu$ and $D^{A^U}_\mu$ commute. Therefore, we can implement the prime by excluding constant modes instead of covariantly constant ones. We will always assume this meaning of the prime from now on, as this will lead to a great simplification in the following manipulations.

Inserting the identity in the path integral and exchanging the order of the integrals we obtain
\begin{equation}
F =-\log \bigg( \frac{1}{\mathrm{vol}(\mathcal{G})} \int_\mathcal{G'} \!\mathcal{D}\mu_g(U) \! \int\! \mathcal{D}A \, \exp\left(-S[A,\psi,\bar{\psi}, h]\right) \,\delta(\nabla^\mu A_\mu^{U}) \mathrm{det}' \left(-\nabla^\mu D^{A^U}_\mu\right)\! \bigg)~.
\end{equation}
Using gauge invariance of the integration measure and of the action the  integral in $A$ can be rewritten in terms of the variable $A^U$, renamed $A$. As a result the integral over $\mu_g$ yields just the volume of $\mathcal{G}'$ and we get
\begin{equation}
F  =-\log\Bigg( \frac{\mathrm{vol}(\mathcal{G'})}{\mathrm{vol}(\mathcal{G})}   \int \mathcal{D}A \, \exp\left(-S[A,\psi,\bar{\psi}, h]\right) \,\delta(-\nabla^\mu A_\mu) \mathrm{det}' \left(-\nabla^\mu D^A_\mu\right) \Bigg)~.
\end{equation}
The ratio of the two infinite-dimensional volumes gives the volume of the constant gauge transformations, i.e. the volume of the group $G$, multiplied by an
 additional factor that arises by requiring an orthonormal mode decomposition in the path integral.\footnote{The normalization of the path integral is chosen following ref.~\cite{Pestun:2007rz}. There is however a difference in the computation of the volume of the gauge group as in our notation the coupling does not appear in the volume expression. }
 In order to explain this factor, consider separating a generic gauge transformation $f:S^d\rightarrow G$ in a constant and a non-constant part $f(x)=f_0+f'(x)$. This can be done via the decomposition in spherical harmonics: $f(x)=\sum_{\ell=0}^\infty F_\ell Y_\ell(x)$. In terms of this decomposition the measure of the path integral is 
\be\int\mathcal{D}f=\int\prod_{\ell=0}^\infty d F_\ell~.
\label{eq:g}
\ee
Because of the normalization in \eqref{eq:u2} we have $Y_0=1/\sqrt{ \mathrm{vol}(S^d)}$, which implies ${f}_0=F_0/\sqrt{ \mathrm{vol}(S^d)}$ and
\begin{equation}
\mathrm{vol}(\mathcal{G}) =\int dF_0 \,\mathrm{vol}(\mathcal{G'})= \mathrm{vol}(S^d)^{\frac{\text{dim}(G)}{2}}\,\mathrm{vol}(G) \,\mathrm{vol}(\mathcal{G'})~.
\end{equation}
This leads to
\begin{equation}
F  =-\log \Bigg(  \frac{ \mathrm{vol}(S^d)^{\frac{-\text{dim}(G)}{2}}}{ \mathrm{vol}(G)}
 \int \mathcal{D}A \, \exp\left(-S[A,\psi,\bar{\psi}, h]\right) \,\delta(\nabla^\mu A_\mu ) \mathrm{det}' \left(-\nabla^\mu D^A_\mu\right) \Bigg)~.
\label{eq:a6}
\end{equation}
We then introduce non-constant $c'$ and $\bar{c}'$ ghost modes to rewrite the $\text{det}'$ as
\begin{equation}
\mathrm{det}' \left(-\nabla^\mu D^A_\mu\right)=\int \mathcal{D}c'\,\mathcal{D}\bar{c}' \exp{\left(-\int d^d x \sqrt{h(x)} \,\mathrm{Tr}[\bar{c}'(x) \nabla^\mu D^A_\mu c'(x)] \right)} ~.
\label{eq:Sghost}
\end{equation} 
The final step is to use the decomposition \eqref{eq:h} to rewrite the $\delta$-functional in eq.~\eqref{eq:a6} in terms of the coefficients of the decomposition
\begin{equation}
\delta(\nabla^\mu A_\mu )=\delta\left(\sum_{\ell=1}^\infty\frac{a^\ell_{(0)}}{\sqrt{\lambda_\ell^{(S)}}}\nabla^2 Y_\ell(x) \right)=\prod_{\ell=1}^\infty{\left(\frac{\ell(\ell+d-1)}{R^2}\right)^{-\frac{g_\ell^{(0)}}{2}\mathrm{dim}(G)}}\delta\left(a^\ell_{(0)}\right) \ .
\label{eq:nablaExp}
\end{equation}
This sets to zero the longitudinal modes and provides a crucial factor in the path integral. 
Plugging eq.~\eqref{eq:nablaExp} in eq.~\eqref{eq:a6} and focusing on the Yang-Mills leading contribution gives
\begin{equation}
\begin{split}
F_{\text{free-YM}}  &= -\log \Bigg(\frac{1}{\mathrm{vol}(G)\sqrt{ \mathrm{vol}(S^d)^{\text{dim}(G)}}} \prod_{\ell=1}^\infty{\left(\frac{\ell(\ell+d-1)}{R^2}\right)^{-\frac{g_\ell^{(0)}}{2}\mathrm{dim}(G)}}\\ &\int \mathcal{D}A_{(1)}\, \mathcal{D}c'\,\mathcal{D}\bar{c}' \, \exp\left(-S_{\text{YM-Free}}[A_{(1)}, h]-\int d^d x \sqrt{h(x)} \,\mathrm{Tr}[\bar{c}'(x) \nabla^2 c'(x)]\right)\Bigg) ~.
\end{split}
\label{eq:freeym}
\end{equation}

We are finally ready to compute the integral. We start from the integration over $A_{(1)}$. Using the decomposition in eqs.(\ref{eq:h}-\ref{eq:f}) and the normalization in eq.~\eqref{eq:g} we get
\begin{equation}
\begin{split}
\int \mathcal{D}A_{(1)} \, &\exp\left({-\int d^{d} x \sqrt{h}\frac{1}{2 g_{0}^{2}}\left(A^a_{(1)\mu}(-\nabla^2+(d-1)) A_{(1)a}^\mu\right)}\right)\\&=\prod_{\ell=1}^\infty\left(\frac{2\pi g_0^2R^2}{(\ell+1)(\ell+d-2)}\right)^{\frac{g_\ell^{(1)}}{2}\mathrm{dim}(G)}.
\label{eq:v}
\end{split}
\end{equation}
For the computation of the ghost path integral we again decompose in spherical harmonics: 
\be c'(x)=\sum_{\ell=1}^\infty C_\ell Y_\ell(x), \quad\int\mathcal{D}c'=\int\prod_{\ell=1}^{\infty}dC_\ell\,.\ee As we are dealing with Grassmann variables, we have \be\int\mathcal{D}{C_\ell}\mathcal{D}\bar{C}_\ell\exp\left(\bar{C}_\ell C_\ell\right) = 1~,\ee
implying
\begin{equation}
     \int\mathcal{D}c'\,\mathcal{D}\bar{c}' \, \exp\left(-\int d^d x \sqrt{h(x)} \,\mathrm{Tr}[\bar{c}'(x) \nabla^2 c'(x)]\right)=\prod_{\ell=1}^\infty\left(\frac{\ell(\ell+d-1)}{R^2}\right)^{{g_\ell^{(0)}}\mathrm{dim}(G)}\,.
\end{equation}
Replacing in eq.~\eqref{eq:freeym}, we get
\begin{align}
\begin{split}
F_{\text{free-YM}} &=\log\mathrm{vol}(G)+ \frac{\text{dim}(G)}{2}\left(\log \mathrm{vol}(S^d)+ \sum_{\ell=1}^{\infty} g_{\ell}^{(1)} \log \left(\frac{(\ell+1)(\ell+d-2)}{2\pi g_0^{2}R^2}\right)\right.\\&\left.-\sum_{\ell=1}^{\infty} g_{\ell}^{(0)} \log \left({\frac{\ell(\ell+d-1)}{R^2}}\right) \right)~.\label{eq:z7}
\end{split}
\end{align}
In order to find an explicit expression for these series one can follow \cite{Giombi:2015haa}, who performed the same computation in the abelian case. Their procedure is based on the rewriting of the logarithms appearing in eq.~\eqref{eq:z7} with the identities
\be
\log (y)=\int_{0}^{\infty} \frac{d t}{t}\left(e^{-t}-e^{-y t}\right)\,, \quad \frac{1}{t}=\frac{1}{1-e^{-t}} \int_{0}^{1} d u \, e^{-u t} \ .
\label{eq:u}
\ee
Then, using gamma function identities, eq.~\eqref{eq:o}, and performing the $t$-integrals, one can find an analytical expression for $F_{\text {free-YM}}$. The only subtle point regards the ghost determinant. It is necessary to add and remove the zero mode, regulating with a mass parameter $\delta$ which is set to zero in the end. This provides
\be -\sum_{\ell=1}^{\infty} g_{\ell}^{(0)} \log \left({\ell(\ell+d-1)}\right)=\lim _{\delta \rightarrow 0}\left[-\sum_{\ell=0}^{\infty} g_{\ell}^{(0)} \log \left({(\ell+\delta)(\ell+d-1)}\right)+ \log \left(\delta(d-1)\right)\right]\ ,\ee
For the sum over $\ell$ we use again eq.~\eqref{eq:u}, while for the $ \log \left(\delta(d-1)\right)$ we use \cite{Giombi:2015haa} 
\begin{equation}
\log(\delta)=-\int_0^1\frac{1}{u+\delta}+\log(1+\delta)\,.
\end{equation} 
Putting everything together we find a smooth limit $\delta\rightarrow0$, which reads
\be
F_{\text {free-YM}}(d)=\mathrm{dim}(G)F_{\mathrm{Max}}(d)-\frac{\mathrm{dim}(G)}{2} \log \left({ g_0^2 R^{4-d}}\right)+\log\frac{\mathrm{vol}(G)}{(2\pi)^{\text{dim}(G)}}\,.
\label{eq:s}
\ee
where $F_{\mathrm{Max}}(d)$ reads
\be
\begin{aligned}
F_{\mathrm{Max}}(d)=&\frac{1}{2}\log(2\pi(d-1)^{2}\Omega_d)-\frac{1}{\sin \left(\frac{\pi d}{2}\right)} \int_{0}^{1} du\biggl((2 u-d) \sin \left(\frac{\pi}{2}(d-2 u)\right) \frac{\Gamma(d-u) \Gamma(u)}{\Gamma(d+1)}\\&+\left(d^{2}+\right.1-3 d(1+u)+2 u(u+2))  \frac{\sin \left(\frac{\pi}{2}(2 u-d)\right)\Gamma(d-2-u) \Gamma(1+u)}{2 \Gamma(d)}\\&+\frac{\sin \left(\frac{\pi d}{2}\right)(d-2)}{(d-2)^{2}-u^{2}} +\frac{\sin \left(\frac{\pi d}{2}\right)}{u}\biggr)\,.
\label{eq:t}
\end{aligned}
\ee
\section{Feynman rules on the sphere}
\label{rules}

In this section we discuss the Feynman rules on $S^d$ for non-abelian gauge theories. We start by reviewing some preliminary notion on  maximally symmetric spaces. We then generalize the computation of the vector propagator presented in \cite{Allen_1986} in the Feynman gauge to an arbitrary $\xi$-gauge. The ghost propagator requires some care in order to remove the zero mode, while the propagator of the Dirac fermion is computed by a Weyl transformation from flat space. We then derive the Feynman rules for the vertices. 

\subsection{Bitensors in maximally symmetric spaces}
\label{prel}

The two-point function of a spinning operator in a curved space $M$ defines a bitensor, namely a bilocal function that is a tensor with respect to both of its arguments. In maximally symmetric spaces bitensors can be expressed as sums and products of a few building blocks.
Let us briefly review these building blocks following ref.~\cite{Allen_1986}. Starting with the geodesic distance $\mu(x,x')$, which is a biscalar, other basic geometric objects are the parallel propagator $h^\nu_{\ b'}(x,x')$ transporting vectors along geodesics from $x$ to $x'$, and the unit vectors $n_\nu(x,x')$ and $n_{\nu'}(x,x')$  tangent to the geodesic at $x$ and $x'$ respectively:
\begin{equation}
n_{\nu}\left(x, x^{\prime}\right)=\nabla_{\nu} \mu(x, x) \quad \text { and } \quad n_{\nu^{\prime}}\left(x, x^{\prime}\right)=\nabla_{\nu^{\prime}} \mu\left(x, x^{\prime}\right).
\end{equation} 
$h^\nu_{\ b'}(x,x'), n_\nu(x,x')$ and $n_{\nu'}(x,x')$ are examples of bitensors. We use the following notation: a bitensor $(n,m)$ is a rank $n$ tensor at $x$ and a rank $m$ tensor at $x'$. So for instance $h^\nu_{\ b'}(x,x'), n_\nu(x,x')$ and $n_{\nu'}(x,x')$ are respectively $(1,1)$, $(1,0)$ and $(0,1)$ bitensors.
In general objects written as the contraction of two bitensors depend on both $x$ and $x'$, even if they contain only primed or unprimed indices. An exception is the following identity relating the metric $h_{ \nu \lambda}$ to the parallel propagator
\begin{equation}
h_{\nu \lambda}(x)=h_{\nu}^{\ \rho^{\prime}}(x, x^{\prime}) h_{\rho^{\prime} \lambda}\left(x^{\prime}, x\right)\,,
\end{equation}
and similarly for $h_{\nu' \lambda'}(x')$. Covariant derivatives of bitensors can be taken with respect to either $x$ or $x'$ and are denoted by $\nabla_\nu$ and $\nabla_{\nu'}$ respectively. 

It is possible to prove that any bitensor in a maximally symmetric space can be expressed as sums and products of the building blocks $h_{\nu \lambda}$, $h_{\nu '\lambda'}$, $n_\nu$, $n_{\nu'}$ and $h_{\nu \lambda'}$, with coefficients that are only functions of $\mu$. This provides a remarkable simplification in finding the structure of propagators and their explicit expressions. 

Let us list some properties, useful for the derivation of propagators:
\begin{equation}
\begin{aligned}
\nabla_{\nu} n_{\lambda} & =A\left(h_{\nu \lambda}-n_{\nu} n_{\lambda}\right)\,, \\
\nabla_{\nu} n_{\lambda^{\prime}} & =C\left(h_{\nu \lambda^{\prime}}+n_{\nu} n_{\lambda}\right)\,, \\
\nabla_{\nu} h_{\lambda c^{\prime}} & =-(A+C)\left(h_{\nu \lambda} n_{\rho^{\prime}}+h_{\nu \rho} n_{\lambda}\right),
\end{aligned}
\label{prop}
\end{equation}
where
\begin{equation}
\begin{aligned}
&  A(\mu)= \frac{1}{R} \cot (\mu / R) \,, \\&
  C(\mu)= -\frac{1}{R} \frac{1}{\sin (\mu / R)}\,,
  \label{eq:A&C}
\end{aligned}
\end{equation}
where $R$ is the radius, defined in terms the constant value of the Ricci curvature scalar in eq.~\eqref{eq:curvScala}. For future convenience it is useful to introduce the variable 
\begin{equation}
z(x,x')\equiv \cos^2\left(\frac{\mu(x,x')}{2R}\right)\,.
\label{eq:zed}
\end{equation}
which is the chordal distance between the points.

Let us now specify to a sphere $S_R^d$ of radius $R$. Using stereographic coordinates $x^\mu$ we write the metric as
\begin{equation}
ds^2= h_{\mu \nu} dx^\mu dx^\nu \,, \qquad h_{\mu \nu}=\frac{4 R^{4}}{\left(R^{2}+|x|^{2}\right)^{2}} \delta_{\mu \nu}~.
\label{eq:stereo}
\end{equation}
The geodesic distance is given by the following identity
\begin{equation}\label{eq:zdef}
\cos \left(\frac{\mu(x, x')}{R}\right) =
1-\frac{2 R^{2}|x-x'|^{2}}{\left(R^{2}+|x|^{2}\right)\left(R^{2}+|x'|^{2}\right)} = 2z(x,x')-1\,.
\end{equation}
When $x'=0$, we denote for simplicity
\be
z\equiv z(x,0) = \frac{R^2}{R^2+x^2}\,.
\label{eq:zDef}
\ee
The variable $z$ will be useful to write propagator expressions and, in particular, their expansion around coincident points. 

\subsection{Vector propagator on $S^d$}
\label{sec:gaugeprop}

Vector propagators for maximally $d$-dimensional symmetric spaces have been computed in \cite{Allen_1986}. For our purpose we need the expression of the massless vector field on the sphere. It follows from the quadratic part of the gauge action
\begin{equation}
   S_\text{free-YM}= \int d^{d} x \sqrt{h}\frac{1}{2 g_{0}^{2}}\mathrm{Tr}\left[A_{\nu}(-\delta^\nu_\mu\nabla^2+R^\nu_\mu+\left(1-\frac{1}{\xi}\right)\nabla^\nu\nabla_\mu) A^\mu\right],
   \label{eqn:1}
\end{equation}
that the vector propagator 
$
Q_{\nu \lambda'}^{ab}(x,x')= \langle A^a_\nu (x)  A^b_{\lambda'} (x') \rangle = g_0^2\delta^{ab}Q_{\nu \lambda'}(x,x')
$
satisfies the equation
\begin{equation}
\left(-h^{\mu \nu}\nabla^2 -R^{\mu\nu}+\left(1-\frac{1}{\xi}\right)\nabla^\mu\nabla^\nu\right)Q_{\nu \lambda'}(x,x')=\delta(x-x')h^\mu _{ \lambda'}~.
\label{eqn:44}
\end{equation}
The propagator $Q_{\nu \lambda'}(x,x')$ is a maximally symmetric (1,1) bitensor and can be decomposed as
\begin{equation}
Q_{\nu \lambda'}(x,x')=\alpha(\mu)h_{\nu \lambda'}+\beta(\mu) n_{\nu}n_{\lambda'},
\label{eq:gaugeprop}\end{equation}
where $\alpha$ and $\beta$ are generic functions of the geodesic distance. Their expression is found in eqs.~\eqref{eq:alphasol},~\eqref{eq:alphatilda} and \eqref{eq:betaExp} in  appendix \ref{app:prop}, where the interested reader can also find their detailed derivation.

\subsection{Ghost propagator on  $S^d$}
\label{sec:sphere}

The ghost propagator $G^{ab}(x,x') =\langle c^{\prime\,a}(x) \bar{c}^{\prime\,b}(x') \rangle$ satisfies
\begin{equation}
\nabla^2G^{ab}(x,x')=\delta(x-x')\delta^{ab}~.
\end{equation}
As explained in section \ref{sec:Landau}, $c'$ has the zero mode removed, so we need to subtract the constant part from this propagator. This is also clear from the expansion of the propagator in terms of the spherical harmonics~\eqref{eq:f2}:
\begin{equation}
G^{ab}(x,x')=\sum_{\ell> 0}\frac{R^2 }{-\ell(\ell+d-1)}Y_{\ell}(x)Y_{\ell}(x')\delta^{ab}~,
\label{eqn:3}
\end{equation}
where the constant mode $\ell=0$ is excluded from the sum, otherwise giving a divergence. In order to resum this expression we need to introduce a small regulator, as we did for the one-loop computation of the free energy:
\begin{equation}
G^{ab}(x,x')=\lim_{\delta \rightarrow 0}\Biggl [\sum_{\ell\ge 0}\frac{R^2 Y_{\ell}(x)Y_{\ell}(x')}{-\ell(\ell+d-1)+\delta(d-1+\delta)}-\frac{R^2 Y_{0}^2}{\delta(d-1+\delta)}\Biggr]\delta^{ab} \,.
\label{eqn:24}
\end{equation}  
The first term corresponds to the propagator $G_{\mathrm{reg}}(x, x')$ associated to a scalar field with mass $m^2=\delta(d-1+\delta)/R^2$, whose expression as a function of $z$ is
\begin{equation}
G_{\mathrm{reg}}(z;\delta)=-\frac{\Gamma(d-1+\delta)}{4(4\pi)^{\frac{d}{2}-1}R^{d-2}\Gamma(1+\delta)\sin(\pi \delta)\Gamma(\frac{d}{2})} {}_2F_1\Big(-\delta,-1+d-\delta,\frac{d}{2},z\Big).
\end{equation}
Plugging in eq.~\eqref{eqn:24} and taking the limit $\delta\rightarrow 0$, we find a well-defined expression for the ghost propagator:
\be\begin{split}
G^{ab}(z) = \delta^{ab}G(z)= & \frac{\delta^{ab}}{4(4\pi)^{\frac{d}{2}-1}R^{d}\sin(\pi d)\Gamma(2-d)\Gamma(\frac{d}{2})}\Big(H(d-2) \\ 
& -\frac{2(d-1)z}{d} {}_3F_2\big(1,1,d;2,1+\tfrac{d}{2};z\big)\Big) \,,
\end{split}
\label{eqn:13b}
\ee
where $H$ denotes the harmonic number, which can be written in terms of the digamma function $\psi$ and the Euler constant $\gamma$ as
\be
H(x) = \gamma+ \psi(x+1) \,.
\ee

\subsection{Fermion propagator on  $S^d$}
The fermion propagator on  $S^d$ is easily computed from its known expression in flat space by performing a Weyl rescaling, see eq.~\eqref{eq:stereo}.
We have
\begin{equation}
    S^{i}_{ j}(x,0)=\langle{\psi}^i(x)\bar{\psi}_j(0)\rangle_\text{sphere}=\frac{\langle{\psi}^i(x)\bar{\psi}_j(0)\rangle_\text{flat}}{\Omega(x)^\frac{d}{2}\Omega(0)^\frac{d}{2}}=\delta^i_{ j}\frac{\Gamma\left(\frac{d}{2}\right)(R^2+x^2)^\frac{d}{2}\gamma^\mu x_\mu}{2^{(d+1)}\pi^{\frac{d}{2}}\left(x^{{2}}\right)^{\frac{d}{2}}R^d}\, ,
\end{equation}
where in the last equality we used
\begin{equation}
    \langle{\psi}^i(x)\bar{\psi}_j(0)\rangle_\text{flat}=\delta^i_{ j}\frac{\Gamma\left(\frac{d}{2}\right)\gamma^\mu x_\mu}{2\pi^{\frac{d}{2}}\left(x^{{2}}\right)^{\frac{d}{2}}}\,,
    \qquad \Omega(x)=\frac{2R^2 }{R^2+x^{2}}\,.
\end{equation}

\subsection{Vertices on the sphere}

The Feynman rules for the vertices can be  read from the interacting part of the gauge-fixed action. Namely, we have four possible interactions defined as
\begin{align}
g_0\Gamma^{\mathrm{TR}}(x)&=-\frac{1}{g_0^2}f^{abc}
\nabla_\nu A_\lambda^aA_\nu^bA_\lambda^c(x)\, ,\\
g_0^2\Gamma^{\mathrm{QU}}(x)&=-\frac{1}{4g_0^2}f^{abc}f^{ade}
g_0A_\nu^bA_\lambda^c A_\nu^dA_\lambda^e(x)\, ,   \\
g_0\Gamma^{\mathrm{GH}}(x)&=f^{abc}
\nabla_\nu\bar{c}^{\prime a}A_\nu^bc^{\prime c}(x)\, ,     \\
g_0\Gamma^{\mathrm{FE}}(x)&=T^a _{\alpha\beta}\bar{\psi^\alpha_i}\gamma^\mu\psi^\beta_i A^a_\mu(x) \, ,
\end{align}
respectively the triple gluon, the quartic gluon, the ghost-gluon and the fermion-gluon interactions. 

\section{Next to leading contribution}
In the previous section we have obtained the Feynman rules for gauge theories on the sphere. We now have all the ingredients to compute the free energy at the next-to-leading order.
For $n_f$ Dirac fermions in the fundamental representation of the gauge group $G=SU(n_c)$ we have
\begin{equation}
\begin{split}
F= & (n_c^2-1)F_{\mathrm{Max}}(d) -\frac{1}{2}(n_c^2-1)\log \left({ g_0^2 R^{4-d}} \right)+\log\left(\frac{\mathrm{vol}(SU(n_c))}{(2\pi)^{n_c^2-1}}\right)\\
& +n_f n_c F_\text{free-ferm}+F_\mathrm{curv}  -\frac{1}{2} g_0^2G_2+\dots,
\end{split}
\label{eqn:31}
\end{equation}
where $G_2$ includes all the two-loop vacuum diagrams. Note that we have kept all the couplings bare.  In section \ref{sec:diagrams} we compute the various diagrams contributing to $G_2$ in eq.~\eqref{eqn:31}: the divergent terms in a generic $\xi$-gauge
and the finite pieces in the Landau gauge $\xi =0$. Renormalization is discussed in section \ref{count1}. As a check of the validity of our results we verify in appendix \ref{app:jack} that the divergences that we obtain match with those computed with heath-kernel methods in ref.~\cite{Jack:1982sn} in the Feynman gauge $\xi=1$. 

\subsection{Computation of the diagrams}
\label{sec:diagrams}
The leading interacting part of the free energy  is given by connected vacuum diagrams up to order $g_0^2$. The corresponding contribution, which we will call $G_2$, is composed by the following two-loop diagrams:
\begin{equation}
    G_2=G_2^{\mathrm{triple}}+G_2^{\mathrm{ghost}}+G_2^{\mathrm{ferm}}+G_2^{\mathrm{quart}}+G_2^{\mathrm{CT-vec}}+G_2^{\mathrm{CT-gh}}\,.
    \label{eq:G2tot}
\end{equation}
The first four terms in \eqref{eq:G2tot} are genuine two-loop graphs: 
\begin{align}
\begin{split}
&G_2^{\mathrm{triple}}= \ \feynmandiagram[baseline=(a.base)][horizontal=a to b] {
a -- [ gluon, half left] b -- [ gluon,  half left] a--[gluon]b
}; \ =\int d^d x d^d x' \sqrt{h}\sqrt{h'}\langle \Gamma^{\mathrm{triple}}(x)  \Gamma^{\mathrm{triple}}(x') \rangle 
\, , \\
&G_2^{\mathrm{ghost}}=\feynmandiagram[baseline=(a.base)][horizontal=a to b] {
a -- [charged scalar, half left] b -- [charged scalar,  half left] a--[gluon]b
};\quad = \int d^d x d^d x'\sqrt{h}\sqrt{h'} \langle \Gamma^{\mathrm{ghost}}(x)  \Gamma^{\mathrm{ghost}}(x') \rangle 
\, ,\\
    &G_2^{\mathrm{ferm}}=\feynmandiagram[baseline=(a.base)][horizontal=a to b] {
a -- [fermion, half left] b -- [fermion,  half left] a--[gluon]b
};\, =\int d^d x d^d x'\sqrt{h}\sqrt{h'} \langle \Gamma^{\mathrm{fermion}}(x)  \Gamma^{\mathrm{fermion}}(x') \rangle 
\, ,\\
&G_2^{\mathrm{quart}}=\begin{tikzpicture}[baseline=(a.base)]
\begin{feynman}\vertex (a) ;
\vertex [left=0.2cm of a] ;
            \diagram* {a --  [gluon, out=45, in=-45, loop, min distance=1.2cm] a--  [gluon, out=-135, in=135, loop, min distance=1.2cm]a
            };
            \end{feynman}
            \end{tikzpicture}\,=2\int d^d x\sqrt{h} \langle \Gamma^{\mathrm{quart}}(x)  \rangle 
            \,.
\end{split}
\label{eq:diagram}
\end{align}
The last two ones are instead one-loop graphs with (one-loop) counterterm insertions:
\begin{eqnarray}
&&G_2^\text{CT-vec}= \feynmandiagram[baseline=(a.base)][horizontal=a to b] {
a [crossed dot]]-- [gluon, half left] b -- [ gluon,  half left] a
};\  = -2 \delta_L\ \int d^d x \langle \frac{1}{2 \xi}\left(\nabla_\mu A^{\mu a}(x)\right)^2 \rangle  
-2\delta_T\ \int d^d x \langle \frac{1}{4 } \left(\nabla_\mu A^a_\nu(x) -\nabla_\nu A^a_\mu(x) \right)^2  \rangle 
\,, \nn \\ 
&&G_2^\text{CT-gh}= \feynmandiagram[baseline=(a.base)][horizontal=a to b] {
a [crossed dot]]-- [charged scalar, half left] b -- [ charged scalar,  half left] a
};\  =-2\delta_c\int d^d x \langle \left(\bar{c}_a(x)\nabla^2 c^a(x)\right)
\rangle 
\,.
\label{eq:diagram2}
\end{eqnarray}
These counterterms are defined from the renormalized Lagrangian
\begin{equation}\label{eq:counterdef}
\frac{Z_T}{Z_{g^2} g^2} \frac{1}{4 } \left(\nabla_\mu A^a_\nu(x) -\nabla_\nu A^a_\mu(x) \right)^2 + \frac{Z_L}{2 g^2 \xi}\left(\nabla_\mu A^{\mu a}(x)\right)^2 - \frac{Z_T^{3/2}}{Z_{g^2} g^2} f^{abc} \nabla_\mu A_\nu^a A^{\mu b} A^{\nu^c} +\dots~,
\end{equation}
where $g_0^2 =Z_g^2 g^2$ is the relation between the bare and the renormalized coupling and $Z_{\bullet} = 1 + \delta_\bullet g^2$. Thanks to the vector equations of motion, we have
\be
\int d^d x \langle \frac{1}{4 } \left(\nabla_\mu A^a_\nu(x) -\nabla_\nu A^a_\mu(x) \right)^2  \rangle   
= - 
\int d^d x \langle \frac{1}{2 \xi}\left(\nabla_\mu A^\mu_a(x)\right)^2 \rangle  + \mathcal{O}(g)~,
\ee
modulo a $\delta^{(d)}(0)$ factor, which vanishes in dimensional regularization.
The counterterms $\delta_T$ and $\delta_L$ entering the vector propagator can be computed in flat space and they read
(see e.g. \cite{Schwartz:2014sze}\footnote{Comparing our Lagrangian \eqref{eq:counterdef} with the definitions in section 26.5 of \cite{Schwartz:2014sze}, we see that the relation between our counterterms and the counterterms $\delta_3$ and $\delta_{A^3}$ defined there are: $g^2 \delta_T = \delta_{A^3} -\delta_3$, and $g^2 \delta_{g^2} = \delta_{A^3} -\frac 32 \delta_3$. Moreover since there is no correction proportional to the longitudinal part of the propagator, $\delta_L = 0$. Note also that $\epsilon_{\text{there}} = -2 \epsilon_{\text{here}}$.})
\be
\delta_L = 0\,, \qquad \delta_T =  C_A\frac{3+\xi}{32\pi^2\epsilon}\ \big(1+\mathcal{O}(g_0^2)\big)\,,
\ee
with $C_A=n_c$ for the $SU(n_c)$ group. The presence of the ghost counterterm is instead a peculiarity of $S^d$, consequence of the removal of zero modes from the propagator. We refer to appendix \ref{app:ghost} for its computation. The final result is 
\be\delta_c=-C_A \frac{3-\xi }{64\pi^2\epsilon}\ \big(1+\mathcal{O}(g_0^2)\big)\,.
\label{eq:count}
\ee

Applying Wick's contraction and the previously listed Feynman rules to eq.~\eqref{eq:diagram}, we get 
\begin{align}
&G_2^\text{triple}=\ \kappa \int d^{d} x \ d^{d} x' \sqrt{h} \sqrt{h'}\ \left( \nabla^\mu\nabla^{\mu^\prime} Q^{\nu \nu'}(Q_{\mu\mu'}Q_{\nu\nu'}-Q_{\mu\nu'}Q_{\nu\mu'} ) \right.  \label{eq:g2triple}  \\
& \quad \quad+ \nabla^\nu Q^{\mu \mu '}(\nabla^{\nu '} Q_{\nu\mu'}Q_{\mu\nu'}-Q_{\nu\nu'}\nabla^{\nu'} Q_{\mu\mu'} )\left.+ \nabla_{\nu} Q^{\mu \nu '}(\nabla_{\nu '} Q^{\nu\mu'}Q_{\mu\mu'}-Q^{\nu\mu'}\nabla_{\nu'} Q_{\mu\nu'} )\right) \,, \nonumber
\\
&G_2^\text{ghost}=\  \kappa \int d^{d} x \ d^{d} x' \sqrt{h} \sqrt{h'}\ (\nabla_\mu G  \   \nabla_{\mu'}G \ Q^{\mu\mu'})\,,
\label{eq:g2ghost}
\\
&G_2^\text{ferm}=-\ n_f T_f  \left(n_c^{2}-1\right) \int d^{d} x \ d^{d} x' \sqrt{h} \sqrt{h'}\ (\mathrm{Tr}\left[\gamma_\mu S   \gamma_{\mu'} S\right] \ Q^{\mu\mu'})\,,
\\
&G_2^\mathrm{quart}=-\frac{\kappa}{2} \int d^{d} x  \sqrt{h}\ (Q^{\mu}_{\ \mu}Q^{\nu}_{\ \nu}-Q_{\mu \nu}Q^{\mu \nu})\,, \label{eq:g2quartic}
\\
&G_2^\text{CT-vec}=\kappa \frac{3+\xi}{16\pi^2\epsilon}\int d^{d} x  \sqrt{h}\ \left(\frac{1}{2\xi}\nabla^\mu\nabla^\nu Q^{\mu}_{\ \nu}\right)\,, \label{eq:g2ctvect}
\\
&G_2^\text{CT-gh}=\kappa \frac{3-\xi}{32\pi^2\epsilon} \int d^{d} x  \sqrt{h}\ (\nabla^2 G)\,.  \label{eq:g2ctghost}
\end{align}
where $T_f=1/2$ for the fundamental representation and we have defined
\be
\kappa=C_{A}\left(n_c^{2}-1\right)\,.
\ee
Note that the first term in the triple diagram \eqref{eq:g2triple} includes a double derivative of the vector propagator, which should be treated with care, because it contains a term proportional to a $\delta$-function at coincident points, which contributes to the integral.  A simple way to circumvent this problem consists in integrating by parts the first term of eq.~\eqref{eq:g2triple} getting
\begin{align}
G_2^\text{triple}&=\ \kappa \int d^{d} x \ d^{d} x' \sqrt{h} \sqrt{h'}\ \left( \nabla^{\mu^\prime} Q^{\nu \nu'}\nabla^\mu(-Q_{\mu\mu'}Q_{\nu\nu'}+Q_{\mu\nu'}Q_{\nu\mu'} ) \right. \\& + \nabla^\nu Q^{\mu \mu '}(\nabla^{\nu '} Q_{\nu\mu'}Q_{\mu\nu'}-Q_{\nu\nu'}\nabla^{\nu'} Q_{\mu\mu'} ) \left.+ \nabla_{\nu} Q^{\mu \nu '}(\nabla_{\nu '} Q^{\nu\mu'}Q_{\mu\mu'}-Q^{\nu\mu'}\nabla_{\nu'} Q_{\mu\nu'} )\right)\,.\nonumber
\end{align}
We refer to appendix \ref{app:contact} for more details on how to treat contact terms and integration by parts on $S^d$ in presence of delta function singularities.

For the first three integrals ($t=$ triple, $g=$ ghost, $f=$ fermion) we proceed as follows. As the integrals only depend on the geodesic distance, or equivalently on $z$, we can use spherical invariance to put $x'$ to zero and reduce the integration over $x'$ to a volume factor: 
\begin{equation}
    G_2^i=\int d^d x d^d x' \sqrt{h}\sqrt{h'}\ g^i\left(z\right)=\Omega_dR^d\int d^d x \sqrt{h}\ g^i\left(z\right)\,,\quad i=t,g,f\,.
\end{equation}
Then we use stereographic coordinates to convert the remaining integral in $x$ to a one-dimensional integral in the variable $z$ defined in eq.~\eqref{eq:zDef}:
\begin{equation}
\int d^d x \sqrt{h}= \Omega_{d-1}R^{2d}\int_0^\infty dx \frac{2^dx^{d-1}}{(R^2+x^2)^d} \,. 
\end{equation}
In this way we write 
\begin{equation}
    G_2^i=\int_0^1 dz\ f^i(z)\,, \quad i=t,g,f\,,
    \label{eq:fiint}
\end{equation}
for some functions $f^i(z)$. The integral \eqref{eq:fiint} cannot be computed directly as it contains UV divergences in $d=4$. We isolate them by expanding $f^i(z)$ around coincident points, i.e. $z=1$:
\begin{equation}
   f^i(z)=\sum_{k=n_i}^{N_i}\left(f_{1k}^i(d)(1-z)^{k-1}+f_{2k}^i(d)(1-z)^{k-d/2+1}+f_{3k}^i(d)(1-z)^{k-d+3}\right)+\tilde{f^i}(z)\, ,
   \label{eq:fiSum}
\end{equation}
where $f_{jk}^i(d)$ are analytic functions of $d$ and $\tilde{f}^i$ remainder terms. The lower bound $n_i$ in the sum appearing in eq.~\eqref{eq:fiSum} is the leading UV divergence of the integrand, and the upper bound $N_i$ is chosen in such a way that the integral of $\tilde{f^i}(z)$ over $z$ between 0 and 1 is finite. 
We write
\begin{equation}
G_2^i=(G_2^{i})_{N_i}+\widetilde{G_2^{i}},
\end{equation}
with
\begin{equation}
   (G_2^{i})_{N_i}=\int_0^1 dz\ \sum_{k=n_i}^{N_i}\left(f_{1k}^i(d)(1-z)^{k-1}+f_{2k}^i(d)(1-z)^{k-d/2+1}+f_{3k}^i(d)(1-z)^{k-d+3}\right)
   \label{eq:G2Ni}
 \end{equation}
 and
 \begin{equation}
\widetilde{G_2^i}=\int_0^1 dz\ \tilde{f^i}(z)\,,
\end{equation}
with $\widetilde{G_2^i}$ finite. The integral $(G_2^{i})_{N_i}$ can be computed analytically using
\begin{equation}
    \int_0^1(1-z)^{a-1}=\frac{1}{a}\, ,
    \label{eq:maintool}
\end{equation}
which is valid for $a>0$, but is extendable to any $d$-dependent $a$ by analytic continuation in $d$.\footnote{Luckily, $f^i_{1k}(d)$ is zero for $k\le 0$ in all the integrals that we have computed. Otherwise, analytic continuation of the dimension would not be sufficient to regulate the integral of eq.~\eqref{eq:maintool}. }
We then set $d=4+2\epsilon$ and extract the divergent part of eq.~\eqref{eq:G2Ni} by expanding the result in powers of $\epsilon$ and isolating the negative powers of $\epsilon$. Note that the divergence of the integral has a double source: it comes from both integration over $z$ when $k=0$ and from the expansion of the functions $f_{jk}^i(d)$ around $d=4$.\footnote{The functions $f^i_{jk}(d)$ remain separately divergent $k>0$, but for $k>N_i$ these divergences cancel when the $j = 1, 2, 3$ contributions are
summed up.} This explains the presence of double poles in the final result.

For the quartic and the counterterm diagrams the situation is simpler, as we have an integration over a single variable. Spherical invariance then means that we need to compute the integrand at coincident points and multiply it by a volume factor. 
We work out the procedure for the quartic case \eqref{eq:g2quartic} as example. We have
\begin{equation} 
    G_2^\mathrm{quart}=2\kappa R^d\Omega_d\left. \frac{\alpha(z) (d-1)(2 \beta(z) - d \alpha(z))}{
 4z^2}\right|_{z\rightarrow1}\,,
 \label{eq:G2quarticExp}
\end{equation}
where $\alpha$ and $\beta$ are the coefficients of the two components  of the vector propagator defined in eq.~\eqref{eq:gaugeprop}.
For physical values of $d$, eq.~\eqref{eq:G2quarticExp} is UV divergent. 
We expand it around coincident points for generic $d$, obtaining
\begin{equation}
   G_2^\text{quart}=\left.\ \sum_{k=0}^{N}\left(g_{1k}^\text{q}(d)(1-z)^{k}+g_{2k}^\text{q}(d)(1-z)^{-d/2+1+
   k}+g_{3k}^\text{q}(d)(1-z)^{-d+2+k}\right)\right|_{z\rightarrow1}\,,
   \end{equation}
   where $N\ge1$ and $g_{jk}^\text{q}$ are analytic functions of $d$.  For sufficiently small $d$ all terms in the expansion vanish, except  $g_{1k}^\text{q}$, with $k=0$.  We then get
  \begin{eqnarray}
    G_2^\mathrm{quart}=g_{10}^\text{q}(d)= && \hspace{-.5cm} -\frac{ \kappa R^{d-4}\Gamma (d-1)}{2^{d+2} \pi ^{\frac{d}{2}}(d-3)^2 \Gamma \left(\frac{d}{2}+1\right)} \left(\gamma  (d-3) \xi +\pi  ((d-3) \xi -d+1) \cot \Big(\frac{\pi  d}{2}\Big)\right. \nn \\
    && \left.+\big(d (\xi -1)-3 \xi +1\big) \psi(d)-\gamma  d+d+\gamma \right)^2 \,.
    \label{eq:g2QanaD}
\end{eqnarray}
The analytic continuation of eq.~\eqref{eq:g2QanaD} for any $d$ gives us the final result.
A similar computation of the integrals in eqs.~\eqref{eq:g2ctvect} and \eqref{eq:g2ctghost} gives just $-1$ and $-1/2$, respectively, for any $d$.

We finally expand eqs.~\eqref{eq:g2triple}-\eqref{eq:g2ctghost} around $\epsilon=0$ with $d=4+2\epsilon$, keeping terms up to constant order, and we get:
\begin{align}
& \!\!\! G_2^{\mathrm{triple}}\Big|_\text{div.}\!\!\! =\kappa \left( \frac{(\xi -3) (3 \xi -7)}{192 \pi ^2 \epsilon ^2}
+\frac{\xi  (31 \xi -64)-71-2 (\xi -3) (3 \xi -7) (\gamma +\log (4 \pi R^2))}{384 \pi ^2
\epsilon }\right) \,, \label{eq:triple} \\
&  \!\!\!  G_2^{\mathrm{ghost}}\Big|_\text{div.} =\kappa \left(\frac{3-\xi }{96 \pi ^2 \epsilon ^2}+\frac{-\xi -13+2 (\xi -3) (\gamma +\log (4 \pi R^2))}{192 \pi ^2 \epsilon }\right) \,,
\label{eq:ghost}\\
& G_2^{\mathrm{ferm}}\Big|_\text{div.} = (n_c^2-1)\frac{n_f T_f}{6 \pi ^2 \epsilon } \,, \\
&  \!\!\!  G_2^{\mathrm{quart}}\Big|_\text{div.} =\kappa \left(-\frac{(\xi -3)^2}{64 \pi ^2 \epsilon ^2}+\frac{(3-\xi)(3+31\xi) +6 (\xi -3)^2 (\gamma +\log (4 \pi R^2))}{384 \pi^2 \epsilon }\right) \,,
\label{eq:quartic}\\
& G_2^{\mathrm{CT-vec}}\Big|_\text{div.} =-\kappa\frac{3+ \xi}{32 \pi^2 \epsilon} \,,   \label{eq:CTvect} \\
 & \!\! G_2^{\mathrm{CT-gh}}\Big|_\text{div.} =\kappa \frac{3 - \xi}{32 \pi^2 \epsilon}\,.
\label{eq:ghostcount}
\end{align}
Summing all the contributions gives 
\begin{equation}
   \left. G_2\right|_\mathrm{div.}=-(n_c^2-1)\frac{11C_A-4n_f T_f}{48 \pi^2 \epsilon}\ .
   \label{eq:g2finalpole}
\end{equation}
Note that the results in eq.~\eqref{eq:triple}-\eqref{eq:quartic} have double poles, which cancel in the sum. Moreover, after summation the $\xi$-dependence of $G_2$ cancels, as required from gauge invariance of the total free energy.

As explained before, we compute finite terms only in the Landau gauge $\xi\rightarrow0$. These are computed numerically.
However, thanks to the integer-relation finding algorithm PSLQ \cite{Bailey:1995}, we can obtain the exact result from the approximated one:
\begin{align}
\left.G_2^{\mathrm{triple}}\right|_\text{fin.}&=\kappa \frac{-562 + 63 \pi^2 + 
 6 (\gamma + \log(4 \pi R^2)) (71 + 
    21 (\gamma + \log(4 \pi R^2)))}{2304 \pi^2}
\, ,  \label{eq:G2tFinite} \\
\left.G_2^{\mathrm{ghost}}\right|_\text{fin.}&=\kappa \frac{97 + 9 \pi^2 + 
 6 (\gamma + \log(4 \pi R^2)) (13 + 
    3 (\gamma + \log(4 \pi R^2)))}{1152 \pi^2}\, ,  \label{eq:G2gFinite} \\
  \left.  G_2^{\mathrm{ferm}}\right|_\text{fin.}&= (n_c^2-1)n_fT_f\frac{5+3 (\gamma + \log (4\pi  R^2))}{36 \pi^2},   \label{eq:G2fFinite} \\
\left.G_2^{\mathrm{quart}}\right|_\text{fin.}&=\kappa \frac{128 - 9 \pi^2 - 
 6 (\gamma + \log(4 \pi R^2)) (1 + 
    3 (\gamma + \log(4 \pi  R^2)))}{256 \pi^2}\,,
\end{align}
and zero for the counterterms, leading to 
\begin{equation}
   \left. G_2\right|_\mathrm{fin.}=(n_c^2-1)\left(C_A\frac{49 + 33 (\gamma + \log(4\pi R^2))}{144 \pi^2}-n_f T_f\frac{5+3 (\gamma + \log (4\pi  R^2))}{36 \pi^2}\right) .
   \label{eq:g2final}
\end{equation}

\subsection{Renormalization}
\label{count1}

Let us now check that the free-energy \eqref{eqn:31} is UV finite up to order $g^2$, when expressed in terms of renormalized couplings.
The bare curvature couplings in eq.~\eqref{eq:c3} renormalize as follows \cite{Jack:1990eb}:
\begin{align}
b_0&=\mu^{2\epsilon}\left(b+\frac{62(n_c^2-1)+11n_f n_c}{720(4\pi)^2 \epsilon}+\mathcal{O}(g^4)\right) \,,
 \label{eqn:4} \\
c_0&=\mu^{2\epsilon}\left(c+\mathcal{O}(g^6)\right)\,,
 \label{eqn:5}
\end{align}
while for the gauge coupling we have the well-known relation
\be
g_0^2=\mu^{-2\epsilon}\Big(g^2+\frac{11C_A-4n_f T_f}{3 \epsilon} \frac{g^4}{(4\pi)^2}+\mathcal{O}(g^6)\Big)\,,
\label{eq:g0Tog}
\ee
where $\mu$ is the RG sliding scale. Expanding in $\epsilon$ for $d=4+2\epsilon$, we get the following divergent contribution from eq.~\eqref{eqn:31} at ${\cal O}(g^0)$:
\be\begin{split}
F_\text{free-YM}|_\text{div.}& =-\frac{31 (n_c^2-1)}{90\epsilon}\,,  \\
     n_f n_c F_\text{free-ferm}|_\text{div.} & =-\frac{11 n_f n_c}{180\epsilon} \,,   \\
      F_{\text{curv}} |_\text{div.} & = \frac{31(n_c^2-1)}{90 \epsilon} + \frac{11n_f n_c}{180 \epsilon}\,,
\end{split}\ee
which cancel in the sum. At ${\cal O}(g^2)$ we have
\be\begin{split}
-\frac12    (n_c^2-1)  \log(g_0^2)|_\text{div.}  & =   - g^2(n_c^2-1)\frac{11C_A - 4 n_f T_f}{96\pi^2 \epsilon}+\mathcal{O}(g^4) \,,  \\
-\frac 12 g_0^2 G_2|_\text{div.} & = g^2 (n_c^2-1)\frac{11C_A - 4 n_f T_f}{96\pi^2 \epsilon}+\mathcal{O}(g^4) \,,
\end{split}\ee
which also cancel in the sum. 
Therefore we obtained, as expected, a finite result for the total free-energy at order ${\cal O}(g^2)$, and in any $\xi$-gauge.

\subsection{Free energy at the fixed point}

We determine here the final form of the free-energy at the fixed point obtained in the $\epsilon$ expansion up to ${\cal O}(\epsilon)$.
The fixed point is obtained by setting to zero the gauge and the curvature beta-functions $\beta_g$, $\beta_b$ and $\beta_c$.
$\beta_b$ and $\beta_c$, computed in \cite{Jack:1990eb}.
At the fixed point $g^* ,b^*,c^*$ we have
\begin{equation}
    F_{\text{conf}}(\epsilon)=F(g^*,b^*,c^*,\mu R) \,,
    \label{eq:Ffixed}
\end{equation}
of order $\epsilon$ up to two loops. Note that $F_{\text{conf}}$  has to be conformal invariant and therefore the dependence on $R$ has to cancel in the final result. 
The expressions for $\beta_g$, $\beta_b$ and $\beta_c$ --up to the order required to get $ F_{\text{conf}}(\epsilon)$ to order $\epsilon$-- are
\begin{align}
\beta_g & =\epsilon g-\Big(\frac{11}{3}C_A-\frac{4}{3}T_fn_f\Big) \frac{g^3}{(4\pi)^2}-\Big(\frac{34}{3}{C_A}^2-\frac{20}{3}C_A T_f n_f-4C_f T_f n_f\Big) \frac{g^5}{(4 \pi)^4}+\mathcal{O}(g^7) \nn \,, \\
\beta_b&=-2\epsilon b-\frac{62 (n_c^2-1)+11n_f n_c}{360(4\pi)^2 }-\frac{(n_c^2-1)}{8}\left(\frac{34}{3}{C_A}^2-\frac{20}{3}C_A T_f n_f-4C_f T_f n_f\right) \frac{g^4}{(4\pi)^6}+\mathcal{O}(g^6)\,,
\nn  \\
\beta_c&=-2\epsilon c+\mathcal{O}(g^6)\,,
\end{align}
from which we get
\begin{align}
g_*&=4\pi\sqrt{\frac{3\epsilon}{11 C_A-4n_fT_f }}\left(1-\frac{3 (17 C_A^2 - 10 C_A n_f T_f - 6 C_f n_f T_f)}{(11 C_A - 4 n_f T_f)^2}\epsilon +\mathcal{O}(\epsilon^2)\right) \,,
\nn \\
b_*&=-\left(\frac{62 (n_c^2-1)+11n_f n_c}{720(4\pi)^2 \epsilon}+\frac{(n_c^2-1)(17 C_A^2 - 10 C_A n_f T_f - 6 C_f n_f T_f)}{24\epsilon}  \frac{g_*^4}{(4\pi)^6}\right)+\mathcal{O}(\epsilon^2)\,,
\nn \\
c_*&=\mathcal{O}(\epsilon^2)\,,
 \label{eqn:11}
\end{align}
where
\be
C_f = \frac{n_c^2-1}{2n_c}\,.
\ee
Note that, since $\beta_b$ contains a constant term, $b^*$ starts at order $1/\epsilon$.

Plugging eq.~\eqref{eqn:11} in the free energy \eqref{eqn:31} and using the results for $G_2$ obtained in section \ref{sec:diagrams}, including the finite pieces computed in the $\xi=0$ gauge,  we obtain
\begin{equation} \boxed{
\begin{split}
F_\text{conf} &=(n_c^2-1)\left(F_{\mathrm{Max}} (d)-\frac{1}{2}\log \Big(\frac{48 \pi^2 \epsilon}{11 C_A - 4 n_f T_f}\Big)\right)+n_f n_c F_\text{free-ferm}(d)+\log\left(\frac{\mathrm{vol}(SU(n_c))}{(2\pi)^{n_c^2-1}}\right)\\ &
+(n_c^2-1)\left(\frac{- {n_f} {T_f} 
 (1089 {C_f}-913C_A+584 {n_f} {T_f})}{121 (11 {C_A}-4 {n_f} {T_f})^2}-\frac{386+363 \big(\gamma+\log (4 \pi ) \big)}{726}\right)\epsilon+{\cal O}(\epsilon^2)\,,
 \end{split} \label{eq:fFinal} }
\end{equation} 
where  $F_{\mathrm{Max}}$ and $F_\text{free-ferm}$ are given in eqs.~\eqref{eq:t} and \eqref{eq:r}, respectively.
The volume of the $SU(n)$ group reads (see e.g. \cite{Marino:2011nm})
 \be
 \text{vol}(SU(n))=\frac{(2\pi)^\frac{n(n+1)-2}{2}}{\prod_{k=1}^{n-1}k!} \,.
 \ee
 The cancellation of the $\log(\mu R)$ term\footnote{All the $\log R$ terms appearing in the loop computations of section \ref{sec:diagrams} arise from the expansion of an overall $R^{d-4}$ factor present in all the contributions. When moving from $g_0$ to $g$ via eq.~\eqref{eq:g0Tog} we effectively have $R\rightarrow \mu R$. \label{footnotemuR}}      
present in the two loop correction \eqref{eq:g2final}  with those arising from the replacement of the bare coupling $b_0$ in eq.~\eqref{eq:d3} and $g_0$ in the log term in eq.~\eqref{eqn:31} is a check of the result. Equation \eqref{eq:fFinal} is the main result of this work. 
 
As discussed in the introduction, the conjectured generalized $F$-theorem \eqref{eq:Ftheo} involves the modified free energy \eqref{eq:FtildeDef}. Using the expression for $F_\text{conf}$ we get
\begin{align}
\widetilde{F}_\text{conf} =&(n_c^2-1)\left(\widetilde{F}_{\mathrm{Max}} (d)+\frac{1}{2}\sin\Big(\frac{\pi d}{2}\Big) 
\log \Big(\frac{48 \pi^2 \epsilon}{11 C_A - 4 n_f T_f}\Big)\right)+n_f n_c \widetilde{F}_\text{free-ferm}(d) \nn \\
 & -\frac{1}{2}\sin\Big(\frac{\pi d}{2}\Big) \log\left(\frac{\mathrm{vol}(SU(n_c))}{(2\pi)^{n_c^2-1}}\right) \label{eq:tildef} \\
&+(n_c^2-1)\left(\frac{{n_f} {T_f} (1089 {C_f}-913C_A+584 {n_f} {T_f})}{121 (11 {C_A}-4 {n_f} {T_f})^2}+\frac{386+363 (\gamma+\log (4 \pi )}{726}\right)\pi\epsilon^2 +\mathcal{O}(\epsilon^3), \nn 
\end{align}
where
\be
\widetilde{F}_{\mathrm{Max}}  = -\sin\left(\frac{\pi d}{2}\right)F _{\mathrm{Max}} \,, \qquad 
\widetilde{F}_{\mathrm{free-ferm}}  = -\sin\left(\frac{\pi d}{2}\right)F _{\mathrm{free-ferm}} \,.
\ee

For completeness we report its expression in the Veneziano limit, where $n_c,n_f\rightarrow \infty$ with $x=n_f/n_c$ fixed. We get 
\begin{equation}
\begin{split}
F_\text{conf} &=n_c^2 \left(F_{\mathrm{Max}} (d)-\frac{1}{2}\log \Big(\frac{48 \pi^2 \epsilon}{11-2x}\Big)+x F_\text{free-ferm}(d)+\frac{3}{4}-\frac{1}{2}\log(2 \pi )\right.\\ &\left.-\left(\frac{193}{363}-\frac{737 x-584x^2}{484 (11-2x)^2}+\frac{1}{2} (\gamma+\log(4 \pi) )\right)\epsilon\right)+\mathcal{O}(n_c)\,.
\end{split}
\end{equation} 
Note that $n_c^2 \log(n_c)$ terms are induced from both log terms appearing in eq.~\eqref{eq:fFinal} and they precisely cancel.
The same cancellation occurs in the t' Hooft limit. This cancellation is expected from large $n_c$ considerations and the fact that a log term is not expected in the genus expansion. 

\section{Applications}

In this section we are going to use the conjectured monotonicity of  $\widetilde{F}$ along RG flows \cite{Giombi:2014xxa} to test some proposed RG flows in $d=3$ and $d=5$, using our result \eqref{eq:fFinal}. The perturbative expression in eq.~\eqref{eq:fFinal} is not adequate to extrapolate to physical dimensions with $|\epsilon| =1/2$. The number of available terms (three) is too limited to attempt a Borel resummation. In the same spirit of ref.~\cite{Giombi:2015haa}, we will instead look for Pad\'e approximants for $\widetilde F$. We also use the knowledge of $\widetilde F$ for special values of $d$
to effectively increase by one order the expansion in $\epsilon$. 

Note that $\widetilde F$ contains a $\log( \epsilon)$ term, which, being non-analytic, prevents the application of standard Pad\'e approximants. Moreover, the free-fermion one-loop determinant is known exactly as a function of $d$ and it is convenient to keep it not expanded in $\epsilon$. For these reasons, we split the total $\widetilde F$ in two parts, one that we keep in $d$ dimensions and contains the non-analytic term, and one that is a series in $\epsilon$. 
Following ref.~\cite{Giombi:2015haa}, we split  $\widetilde{F}_\text{conf}$ as
\be
 \widetilde{F}_\text{conf}=n_f n_c \widetilde{F}_\text{free-ferm}+\frac{1}{2}\sin\left(\frac{\pi d}{2}\right)(n_c^2-1)\log\left(\frac{2\epsilon}{11 C_A - 4 n_f T_f}\right)+\delta \widetilde{F}(\epsilon),
 \label{eq:Ftildeconf}
\ee
and we use Pad\'e approximants only on the $\delta \widetilde{F}(\epsilon)$ term. The latter includes the free photon contribution, which is evaluated numerically, and reads
\begin{align}
&  \delta \widetilde{F}(\epsilon)=(n_c^2-1) \frac{31\pi}{90} +\left((n_c^2-1)4.696- 
\pi \log\left(\frac{\text{vol}(SU(n_c))}{(2\pi)^{n_c^2-1}}\right)\right)\epsilon  \label{eq:deltatildeF}  \\
&+(n_c^2-1)\left(\frac{n_f\pi(584 {n_f} {n_c}-1089-737 n_c^2)}{484 n_c (11 n_c-2 {n_f})^2}+\frac{386\pi+363\pi (\gamma+\log (4 \pi )}{726}-10.098\right)\epsilon^2 +\mathcal{O}(\epsilon^3)\,.
 \nn
 \end{align}
 For presentation purposes we rounded to the first 4 digits the ${\cal O}(\epsilon)$ and ${\cal O}(\epsilon^2)$ contribution coming from the photon free energy, but 
 the result is available to higher precision.
Let us stress the fact that the above splitting is arbitrary and that the corresponding choice significantly affects the final results. This is a signal of the poor knowledge that we have on the series.
For the same reason we have not attempted to estimate an error bar in our results. 
 
The fixed points we get in $d=4+2\epsilon$ of QCD$_d$ with gauge group $SU(n_c)$ and $n_f$ massless Dirac fermions in the fundamental representation 
are expected to match two known CFTs:

\begin{itemize}
\item For $\epsilon=-1$ ($d=2$) the IR fixed point of QCD$_d$ with gauge group $SU(n_c)$ and $2n_f$ massless Dirac fermions in the fundamental representation
is an $SU(2n_f)_{n_c}$  Wess-Zumino-Witten model with an additional decoupled free boson \cite{Affleck:1985wa, Gepner:1984au}. This CFT has central charge 
\be c=\frac{n_c(4n_f^2-1)}{2 n_f+n_c} + 1 \,,
\ee 
and 
\be
\widetilde F_{\text{WZW}}(d=2) = \frac{\pi}{6} c \,.
\ee
Plugging $d=2$ in eq.~\eqref{eq:Ftildeconf} and identifying $\widetilde F_\text{conf}$ with $\widetilde F_{\text{WZW}}$ gives 
\begin{equation}
\delta\widetilde{F}(\epsilon=-1)=\widetilde F_{\text{WZW}} -n_c n_f \widetilde{F}_\text{free-ferm}=-\frac{\pi}{3}\frac{n_f(n_c^2-1)}{2 n_f+n_c}\,.
\label{eq:d2WZW}
\end{equation}
\item For $\epsilon=1$ ($d=6$) the theory is conjectured to have a non-unitary UV fixed point described by a Lagrangian with a higher-derivative kinetic term $F^a_{\mu\nu}\nabla^2F_a^{\mu\nu}$ \cite{Gracey:2015xmw,Casarin:2019aqw}, whose anomaly coefficient is $a=-(n_c^2-1)\frac{55}{84}$ \cite{Giombi:2015haa}. This leads to 
\be\label{eq:d6bc}
\delta\widetilde{F}_{d=6}=\frac{\pi}{2}a=-\frac{55\pi}{168}(n_c^2-1)\,.
\ee
\end{itemize}

To improve the numerical estimate of our result we constrain the Pad\'e approximants of $\delta\widetilde{F}$ to these known points. In order to avoid misleading results, we exclude approximants with poles in the range between the constraint and $d=4$.

\subsection{${F}$-Theorem in $d=3$}

Non-abelian $3d$ gauge theories have received particular attention in the last years due to their possible emergence in quantum phase transitions with deconfined criticality \cite{Wang:2017txt}
and as theories governing domain walls among different vacua in non-abelian $4d$ gauge theories \cite{Gaiotto:2017tne}. Theoretically, they are of course also interesting theories by themselves.
 
 \begin{figure}[t!]
  \centering 
 \includegraphics[width=0.45
 \columnwidth]{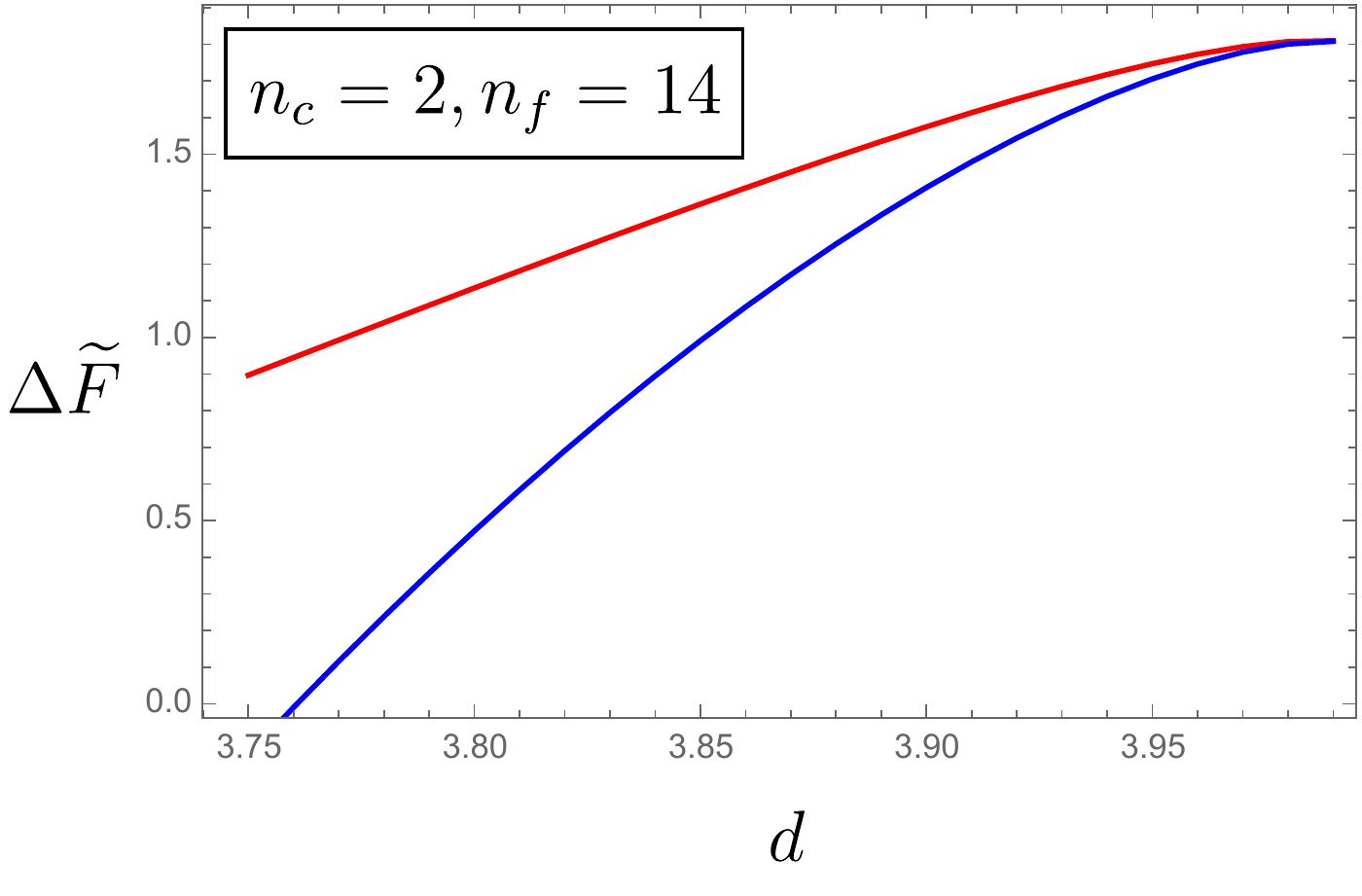}  
  \includegraphics[width=0.535
 \columnwidth]{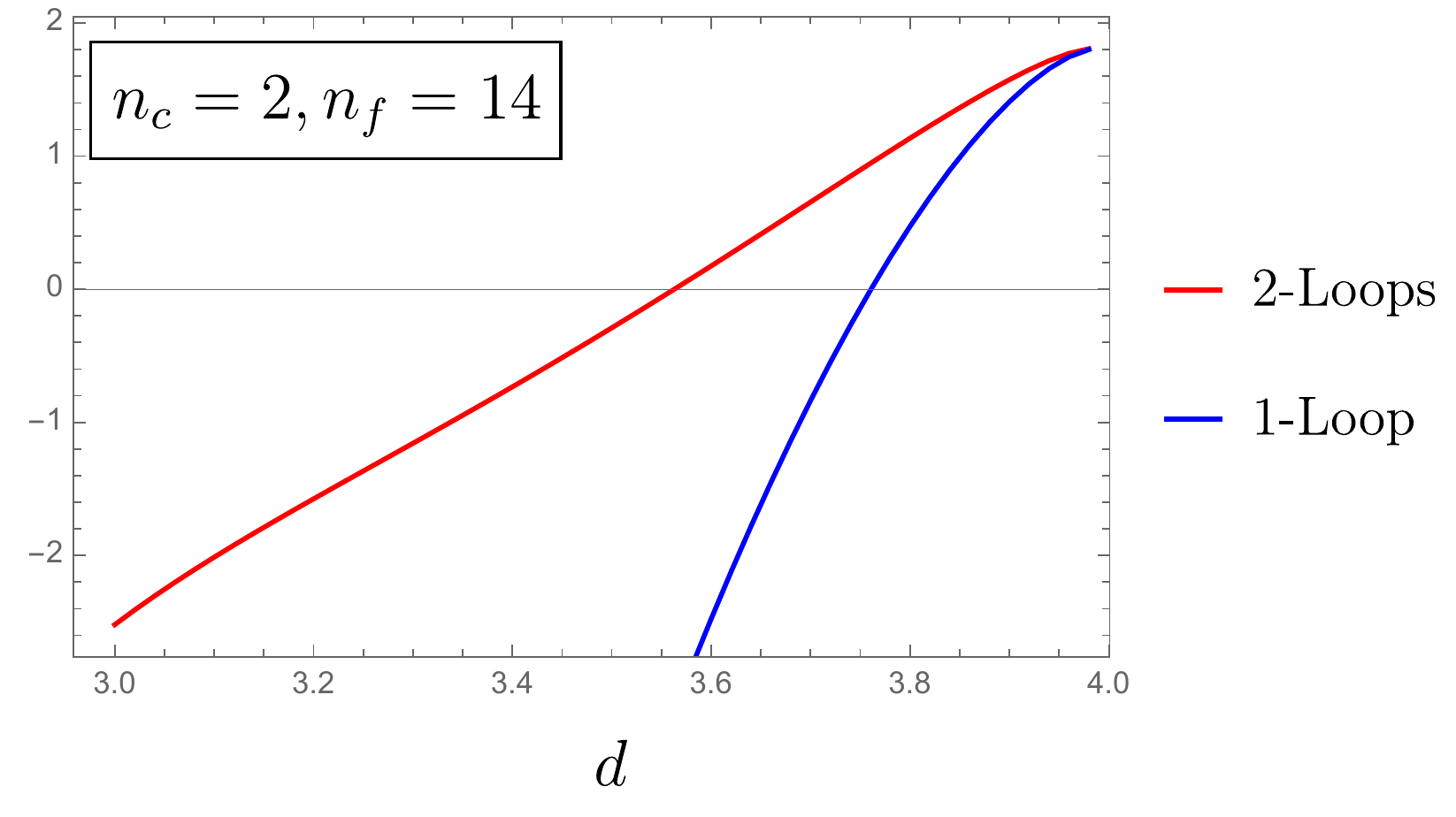}  
\caption{Left panel: Comparison between $\Delta \widetilde{F}$ as a function of the dimension $d$ for small $\epsilon$  computed by using 
the result for $\widetilde F$ in eq.~\eqref{eq:tildef} (red) or only its free part given by the first two rows of eq.~\eqref{eq:tildef} (blue). Right panel: Same comparison extended up to $d=3$.}
\label{fig:nf14dfrac}
\end{figure}

It is known since the early work  \cite{Appelquist:1989tc} that at large $n_f$  QCD$_3$ flows in the IR to a CFT. For $n_f\leq n_f^*$, with $n_f^*$ an unknown parameter, a phase with spontaneous symmetry breaking (SB) of the $U(2n_f)$ global symmetry is expected. The only pattern of spontaneous breaking of the global symmetry $U(2n_f)$ compatible with the results of \cite{Vafa:1983tf,Vafa:1984xh} is
\be
U(2n_f)\rightarrow U(n_f) \times U(n_f)\,.
\label{eq:SSBphase}
\ee
More recently, a qualitative phase diagram of the theory as a function of the number of flavors $n_f$, a fermion mass term, and the level $k$ of a possible Chern-Simons term has been suggested  \cite{Komargodski:2017keh}. We will focus on $k=0$ in the following and use the $F$-theorem to put an upper bound on $n_f^*$.
A naive way to check if the spontaneous symmetry breaking phase \eqref{eq:SSBphase} can be realized would be to compare $F_{\text{IR}}= F_{\text{SB}}$ as given by $2n_f^2$ Goldstone bosons 
(free in the deep IR), with  $F_{\text{UV}}$ given in the deep UV by $n_c^2-1$ free photons and $n_f n_c$ free fermions. Unfortunately, due to the log term in \eqref{eq:s}, 
$F_{\text{UV}}$ diverges and no useful information can be extracted. We overcome this problem by assuming that conformality is lost at $n_f=n_f^*$  by annihilation between the critical QCD$_3$ fixed point with another one, known as QCD$_3^*$ \cite{Kaplan:2009kr}. A similar analysis for QED$_3$ has been performed in \cite{Giombi:2015haa}.
Treating $n_f$ as a continuous parameter, for $n_f= n_f^*+\eta$ and $0<\eta \ll 1$, the theory flows to the IR fixed point QCD$_3$. On the other hand, for $n_f= n_f^*-\eta$ the theory is expected to undergo a weak first-order phase transition \cite{Gorbenko:2018ncu} (i.e. a walking regime, see \cite{Benini:2019dfy} for an explicit realization in $4d$ gauge theories) with a slow RG passing close to the (now complex) fixed points, reaching eventually the spontaneously broken phase \eqref{eq:SSBphase}. By continuity and the generalized $F$-theorem, we then expect that 
\be
\Delta\widetilde F(n_f^*)=  \widetilde F_{\text{conf}}(n_f^*) - \widetilde F_{\text{SB}}(n_f^*)>0 \,.
\label{eq:DeltaFDef}
\ee
Note that values of $n_f$ such that  $\Delta\widetilde F(n_f)<0$ are incompatible with a symmetry breaking phase. On the other hand, values of $n_f$ with $\Delta\widetilde F(n_f)>0$ 
are compatible with either a CFT or a symmetry breaking phase. For this reason we can only determine an upper bound $n_f^*\leq n_f^0$, where $\Delta\widetilde F(n_f^0) = 0$.

An early previous estimate of $n_f^*$ was based on Schwinger-Dyson gap equations \cite{Appelquist:1989tc} and resulted in $n_f^* \approx 128(n_c^2-1)/(3\pi^2 n_c)$.
More recently, a lattice analysis \cite{Karthik:2018nzf} found $n_f^*\leq 4$ for $n_c=2$. 
An estimate based on the $F$-theorem already appeared in \cite{Sharon:2018apk}, where as UV theory it was used a SUSY version of QCD$_3$, a genuine CFT with finite $F$ which can flow to QCD$_3$ by appropriate deformations. By comparing $F_{\text{SUSY}}$ computed by means of supersymmetric localization with $F_{\text{SB}}$ (and assuming that we can flow from the IR SCQD$_3$ fixed point to the IR QCD$_3$ fixed point), it was found $n_f^*< 13/2$ for $n_c=2$.

 \begin{figure}[t!]
  \centering 
 \includegraphics[width=0.7
 \columnwidth]{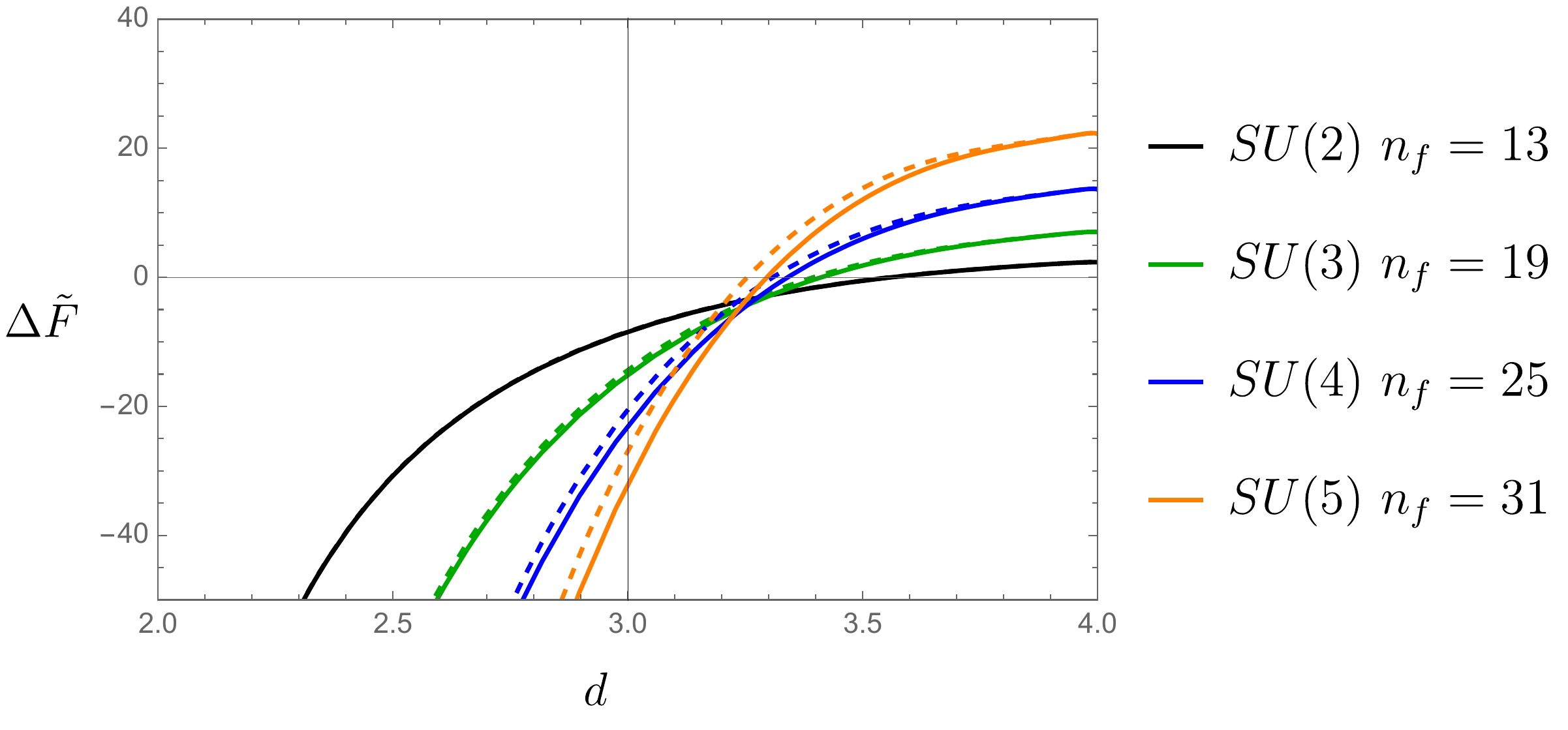}  
\caption{Values of $\Delta \widetilde{F}$ for $SU(n_c)$ as a function of the dimension $d$ computed with Pad\'e-approximants [2/1] (continuous line) and [1/2] (dashed line) at $n_c=2,3,4,5$. The value of $n_f$ is set to the smallest integer without poles in both approximants in $2<d<4$ satisfying $g^{*2}>0$.} 
\label{fig:su3}
\end{figure}

The value of $\widetilde F_{\text{SB}}(n_f)$ is easily computed by noting that the $2n_f^2$ Goldstone bosons associated to the breaking pattern \eqref{eq:SSBphase}
become free in the deep IR.
The contribution to the free energy for a single real scalar  reads \cite{Giombi:2014xxa}
\be
\begin{split}
F_\text{free-sc}& =-\frac{1}{\sin(\frac{\pi d }{2})\Gamma(1+d)}\int_0^1 du\ u \sin(\pi u)\Gamma\left(\frac{d}{2}+u\right) \Gamma\left(\frac{d}{2}-u\right)\,, \\
\widetilde{F}_{\mathrm{free-sc}} &  = -\sin\left(\frac{\pi d}{2}\right)F _{\mathrm{free-sc}} \,.
\end{split}
\ee
We then have
\be
\widetilde F_{\text{SB}}(n_f) = 2n_f^2 \widetilde{F}_{\mathrm{free-sc}}\,.
\ee
For $d=3$ it reads
\begin{equation}
\widetilde{F}_\text{SB}=2n_f^2 \left(\frac{\log 2}{8}-\frac{3\zeta(3)}{16\pi^2}\right)\,.
\end{equation}
\begin{table}
\centering
\begin{tabu}{|c|c|c|c|c|c|}
\hline$n_f$ & 12 & 13 & 14 & 15 & 16  \\
\hline \rowfont{\color{red}}SB & $18.38$&$21.57$&$ 25.01$&$ 28.71$&$ 32.67$ \\
\hline ${[2/1]}$ & $12.1$ & $13.1$ & $13.6$ & $13.9$ &$14.16$ \\
\hline ${[1/2]}$ & $-$ & $13.2$ & $13.9$ & $15.01$&$16.10$ \\
\hline
\end{tabu}
\caption{Comparison between the $3d$ values of $\widetilde F$ in the broken phase $\widetilde{F}_\text{SB}$ (red) with those obtained from  Pad\'e-approximants [2/1] and [1/2] 
of  $\widetilde{F}_\text{conf}$ for QCD$_3$ with $n_c=2$. In all cases $\Delta \widetilde{F}< 0$.}
\label{tab:1}
\end{table}
Before presenting the results of our extrapolations to $d=3$, it is useful to see the effect of the 2-loop correction to the free energy with respect to the one-loop
free theory contribution in the controlled regime with $|\epsilon| \ll 1$. This is shown in fig. \ref{fig:nf14dfrac} where we plot $\Delta \widetilde{F}$ (for $n_c=2$ and $n_f=14$)
defined as in eq.~\eqref{eq:DeltaFDef} as a function of the dimension $d$. We compare the result for $\widetilde F_{\text{conf}}$ obtained using eq.~\eqref{eq:tildef} (red line) 
with the one obtained using only the first two rows of the same equation (blue line), i.e. only its free part.  
We note that the effect of the interactions is to favor the SB phase with respect to the conformal one and that the latter is more favored as we lower the space-time dimensions. More importantly, we see from the left panel in the figure that when $|\epsilon| \approx 0.1$ the one and two-loop results differ significantly and that there is no hope to get 
reliable results from perturbation theory in $d=3$ (for illustration purposes we report in the right panel of fig. \ref{fig:nf14dfrac} the same plot extended up to $d=3$).
As anticipated at the beginning of the section, we then consider Pad\'e approximants of \eqref{eq:deltatildeF}. For $d<4$ we augment the approximant by one more term by imposing the constraint \eqref{eq:d2WZW}.

\begin{figure}[t!]
  \centering
 \includegraphics[width=0.46
 \columnwidth]{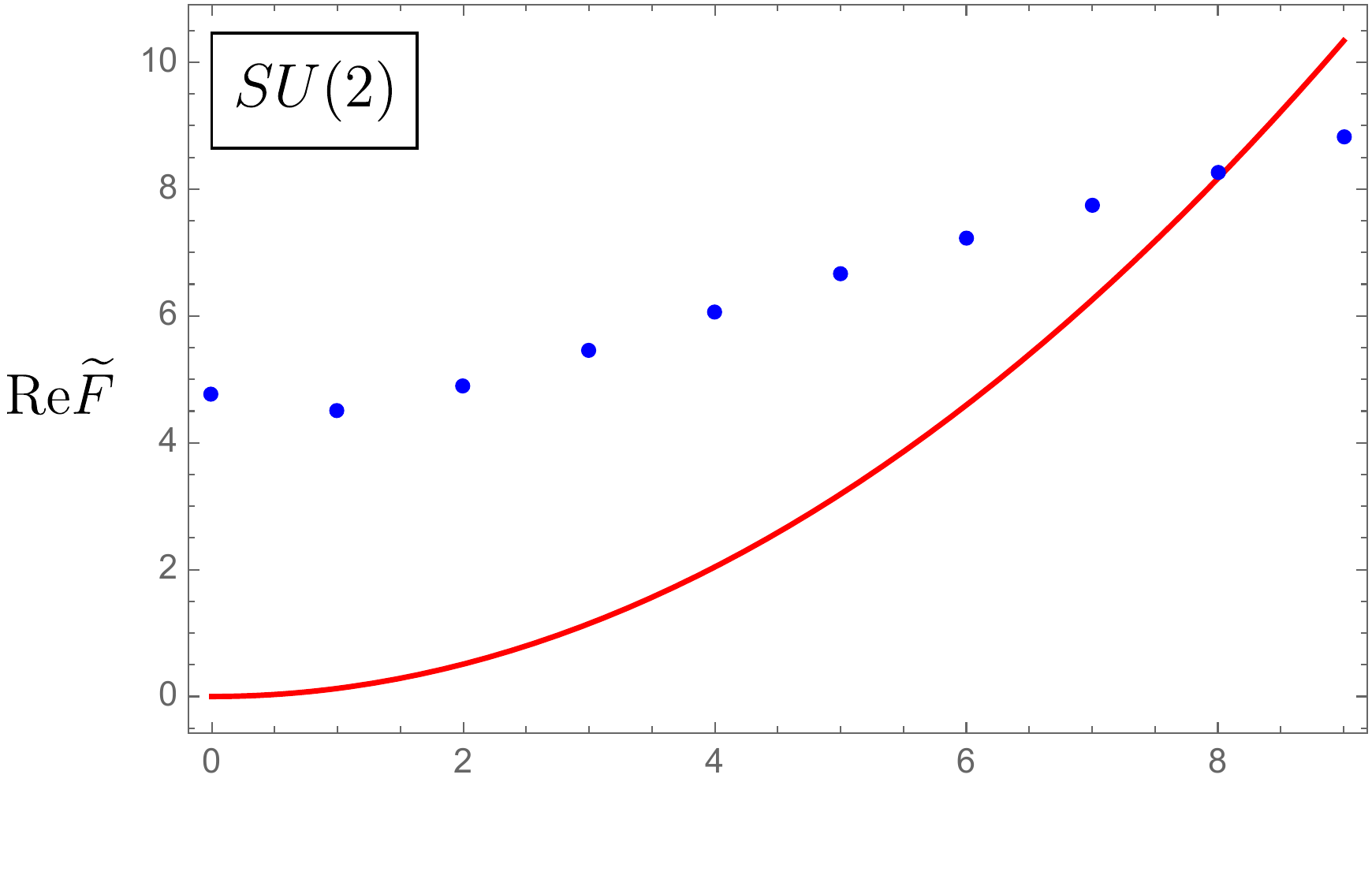}   \includegraphics[width=0.5
 \columnwidth]{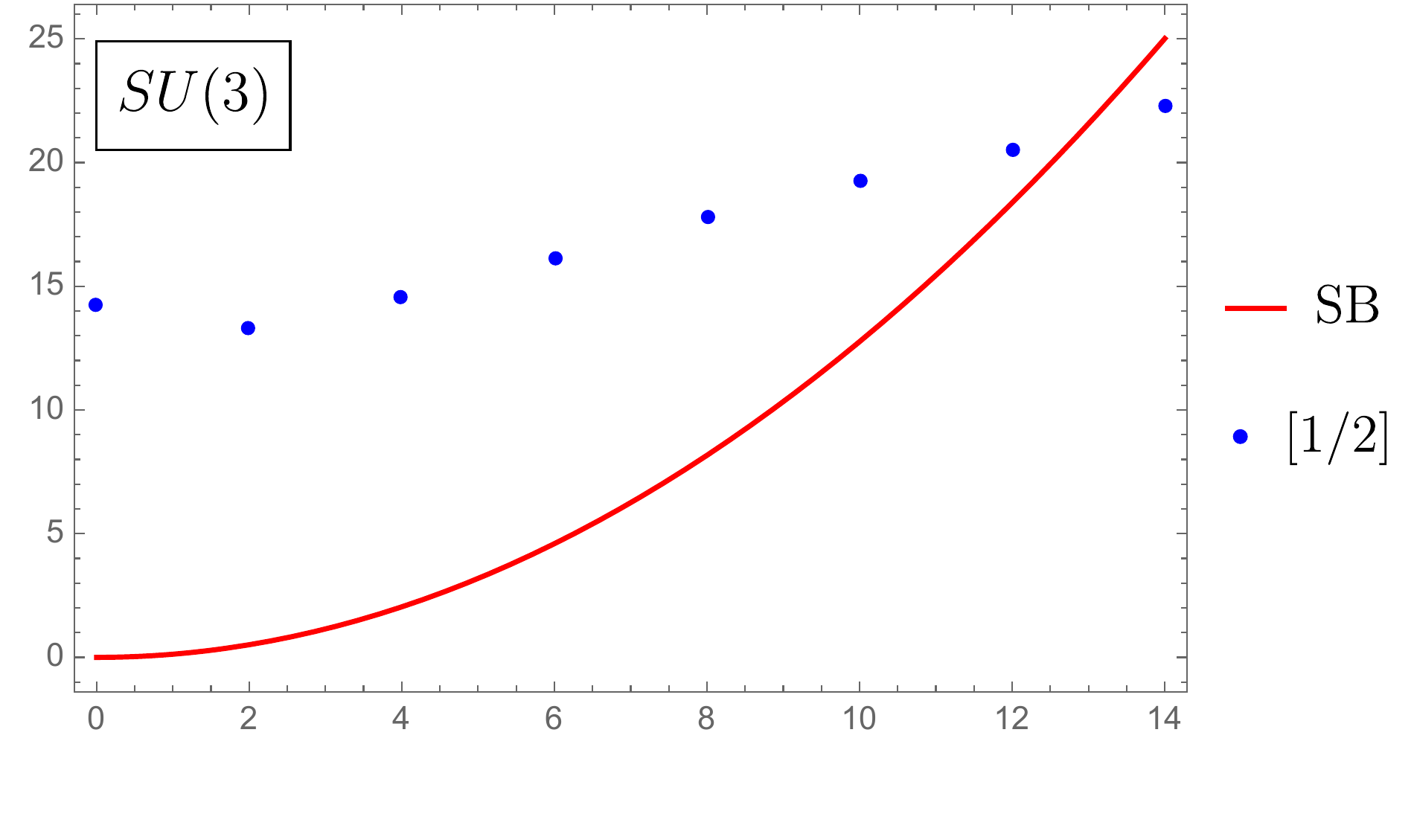} \,  \includegraphics[width=0.46
 \columnwidth]{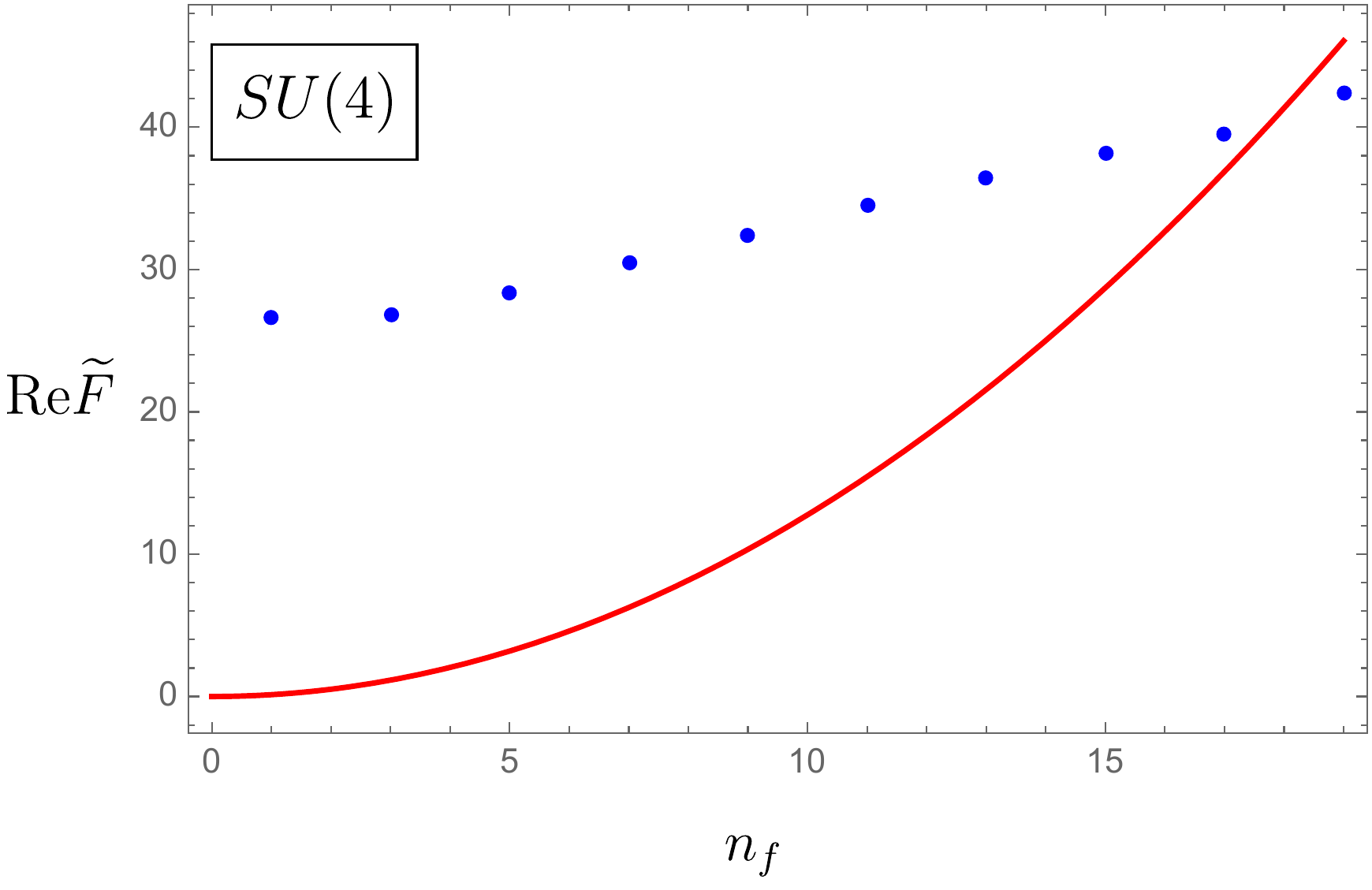}  \includegraphics[width=0.424
 \columnwidth]{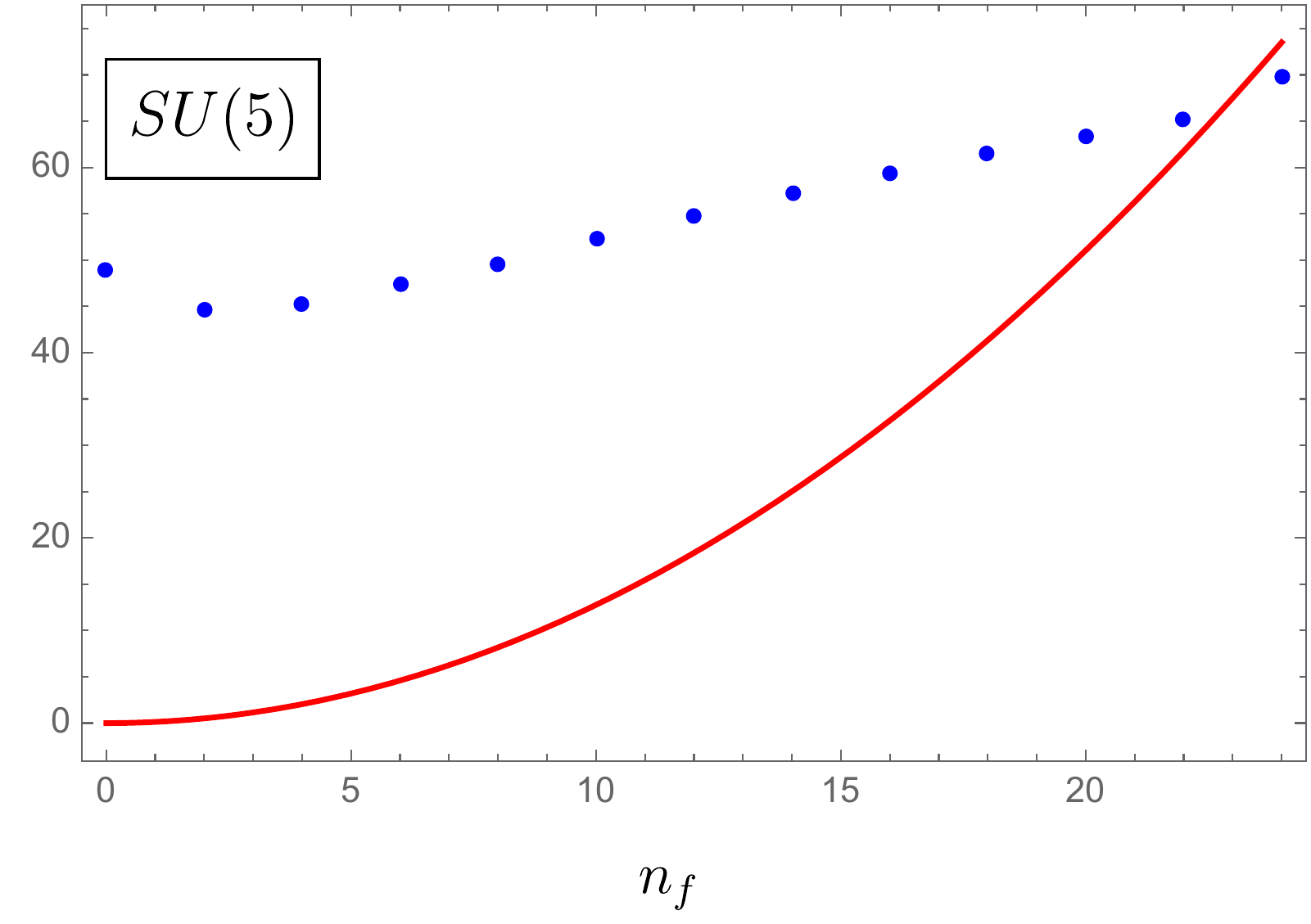}  \qquad \quad\,\caption{Comparison between the $3d$ value of $\widetilde{F}_\text{SB}$ (red line) and of the real part of $\widetilde{F}_\text{conf}$  (blue points)  as a function of $n_f$ for $n_c=2,3,4,5$
.  The [1/2] approximants provide $\text{Re}[\widetilde{F}_\text{conf}]>\widetilde{F}_\text{SB}$ for $n_f\le8,12,17,22$ suggesting that a chiral symmetry breaking may occur in these ranges of values.}
\label{fig:su2nflow}
\end{figure}

In fig.~\ref{fig:su3} we show the value of $\Delta\widetilde{F}$ as a function of the dimension $d$ for $n_c=2,3,4,5$ and $n_f$ equal to the smallest integer without poles in approximants [1/2] and [2/1] satisfying $g^{*2}>0$, i.e. $n_f=13,19,25,31$ respectively.\footnote{Note that regions in $n_f$ close to $11n_c/2$ are more subject to instabilities as $g^{*2}$ blows up there,  producing a pole of order two in the free energy. This is another reason to avoid smaller values of $n_f$ which still satisfy $g^{*2}>0$ (i.e. $n_f=12$ for $n_c=2$). }
We see that at $d=3$ $\Delta\widetilde{F}<0$ in all these cases, indicating the presence of the conformal phase. As expected, this behavior persists for higher values of $n_f$: 
we report in tab.~\ref{tab:1} the comparison between the free energy $\widetilde F_{\text{conf}}$  and that of the broken phase for $n_c=2$, $12\le n_f\le 16$. Not only the value of $\widetilde F_\text{SB}$ remains above $\widetilde{F}_\text{conf}$, but also the gap between the two values gets larger and larger. 

\subsubsection{Small $n_f$}

The one-loop beta-function of the gauge coupling  vanishes  at $n_f=11 n_c/2$ and changes sign below that, making $g^{*2}_{\text{one-loop}}<0$. 
Of course, a unitary fixed point in $d=3$ does not necessarily appear as a real {\it one-loop} fixed point when $\epsilon \ll 1$.\footnote{A notable example of this sort is provided by the abelian Higgs model of $n$ complex scalar fields. It is known that in this theory a real one-loop Wilson-Fisher fixed-point appears for $n> 183$ \cite{Halperin:1973jh} and this number greatly varies with the order, see e.g. \cite{Ihrig:2019kfv}. It is in fact likely that the $3d$ abelian Higgs theory has an IR conformal phase for values of $n$ well below 183.} 
As mentioned, lattice results for $SU(2)$ find that $n_f^*\leq 4$, suggesting that even if $g^{*2}_{\text{one-loop}}<0$, there exists a range in $n_f$ where the $3d$ theory is conformal in the IR.
For $n_f<11n_c/2 $ we could still use the free energy to extract information on the RG flow. For $g^{*2}<0$, the free energy becomes complex, due to the $\log$ term in eq.~\eqref{eq:Ftildeconf},
with an opposite phase depending on which of the two imaginary fixed points is chosen: 
\begin{equation}
\log(g^{*2})=\log(|g^{*2}|)\pm i \log(\pi)\,.
\end{equation}
We propose to estimate the value of $F$ at the strongly coupled real fixed point by an extrapolation of the half-sum of the two complex values obtained with the $\epsilon$-expansion, i.e. of their real part. The stability of the conformal phase then requires this value to be smaller than $\widetilde{F}_\text{SB}$. As a result, our more speculative criterion in the range $n_f<11n_c/2$ is
\be
{\rm Re}\, \Delta\widetilde F(n_f^*)=  \widetilde F_{\text{conf}}(n_f^*) - {\rm Re}\, \widetilde F_{\text{SB}}(n_f^*)>0 \,.
\label{deltaftilde}
\ee
We report in fig.~\ref{fig:su2nflow} the real part of $\widetilde{F}_\text{conf}$ compared to $\widetilde{F}_\text{SB}$ for $n_c=2,3,4,5$ computed with the Pad\'e approximant [1/2]. We see that in all cases there is a wide range of $n_f$ for which the conformal phase appears to be unstable. We have 
\begin{align}
\begin{split}
n_f^* & \lesssim 8 \,,\qquad \qquad \;\;SU(2) \,, \\
n_f^* & \lesssim12  \,,\qquad \qquad SU(3)\,,  \\
n_f^*& \lesssim17 \,,\qquad \qquad SU(4)\,,   \\
n_f^* & \lesssim 22 \,,\qquad \qquad SU(5) \,.
\label{eq:nfbounds}
\end{split}
\end{align}
The upper bound for $SU(2)$ is consistent with the bound $n_f^*< 13/2$ of  \cite{Sharon:2018apk}, and $n_f^*\leq 4$ of \cite{Karthik:2018nzf}. A similar analysis can be done in the Veneziano limit, by taking the large $n_c,n_f$ limit of eq.~\eqref{deltaftilde}. The resulting bound is 
\be
x^*\lesssim4.5\,.
\label{eq:Veneziano}
\ee

\begin{figure}[t!]
  \centering 
 \includegraphics[width=0.7
 \columnwidth]{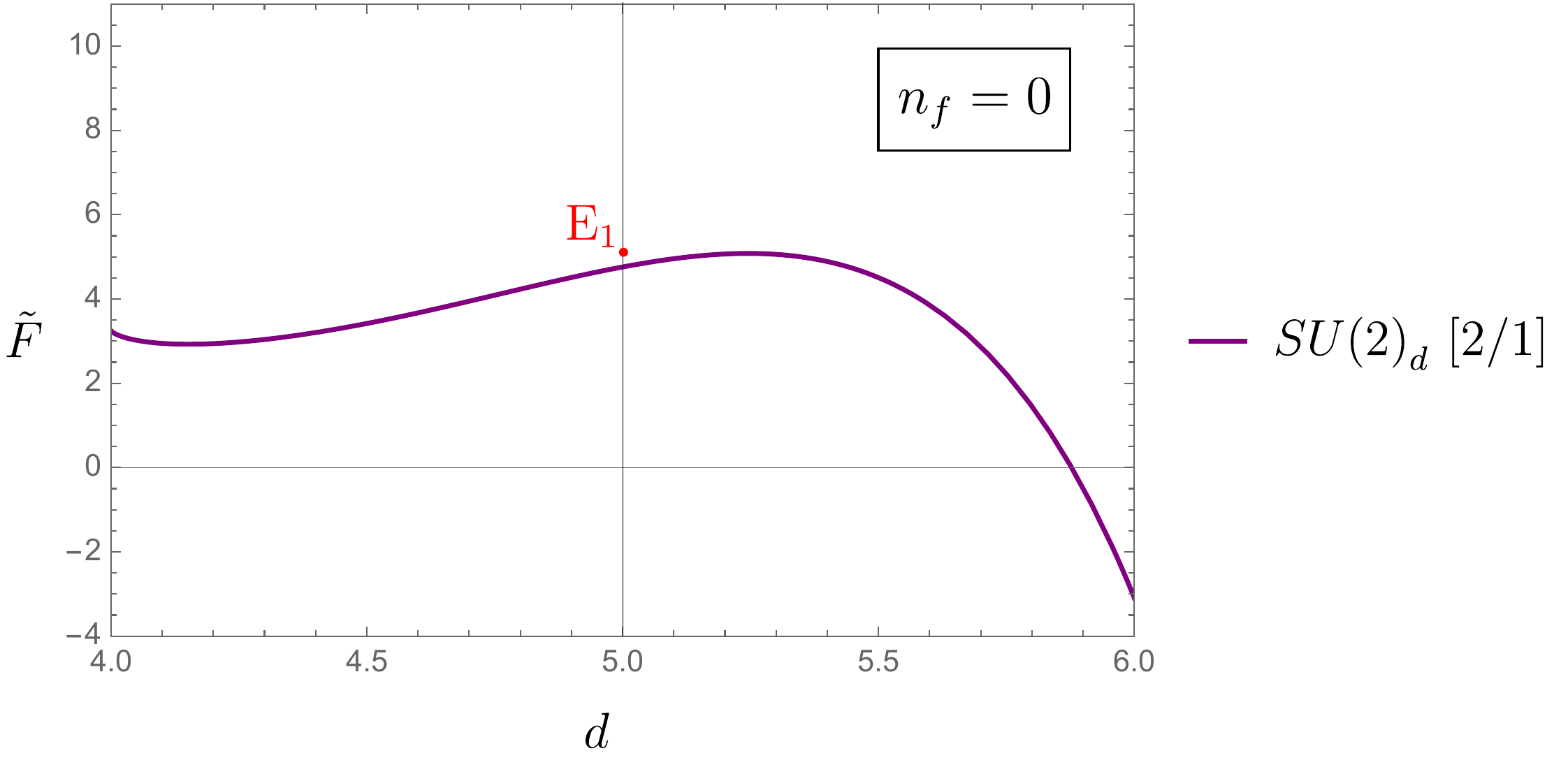}  
\caption{ Values of $\widetilde{F}_\text{conf}$ for pure $SU(2)$ YM as a function of the dimension $d$ computed with the Pad\'e-approximant [2/1] (purple line), compared to the value of the $5d$ supersymmetric fixed point ${E_1}$, the UV completion of $SU(2)$ SYM gauge theory (red point). }
\label{fig:su2}
\end{figure}

\subsection{${F}$-Theorem in $d=5$}
In this section we extrapolate $\widetilde{F}$ to $5d$ to test a proposed construction of an interacting CFT that provides a UV completion of $5d$ $SU(2)$ YM theory. Ref. \cite{BenettiGenolini:2019zth} proposed to construct this CFT as the IR fixed point of a supersymmetry-breaking deformation of the interacting superconformal field theory known as $E_1$ theory \cite{Seiberg:1996bd}. The latter is known to provide the UV completion of $SU(2)$ supersymmetric YM theory (SYM). Ref. \cite{BenettiGenolini:2019zth} studied the various phases in the two-dimensional space of relevant deformations of the $E_1$ theory, which includes both the  supersymmetric deformation to SYM and the non-supersymmetric one, and suggested the existence of a second-order transition between two phases that are described by $SU(2)$ YM theory and a different symmetry-protected topological order. The CFT capturing this phase transition would therefore be a UV completion of YM, and provide an example of a non-supersymmetric interacting CFT in $d>4$. This scenario was further explored in \cite{Bertolini:2021cew}, that showed that actually the phase transition should be viewed as separating the YM phase from a phase with spontaneous breaking of the instantonic $U(1)$, and in \cite{Bertolini:2022osy} where a certain generalization of the theory admitting a large $N$ limit was argued to have a second order transition in that limit.

A possible test for the proposal of ref. \cite{BenettiGenolini:2019zth, Bertolini:2021cew} relies on the $F$-theorem: the sphere free energy $\widetilde{F}_{E_1}$ of the SCFT and that of the non-supersymmetric CFT $\widetilde{F}_{\text{CFT}}$ should satisfy $\widetilde{F}_{E_1} > \widetilde{F}_{\text{CFT}}$. The quantity $\widetilde{F}_{E_1}$ has been computed using localization in \cite{Chang:2017cdx}. It is natural to conjecture that the non-supersymmetric fixed point is the continuation to $d=5$ of the UV fixed point visible in the $\epsilon$ expansion in $d=4+2\epsilon$, and therefore to estimate $\widetilde{F}_{\text{CFT}}$ by an extrapolation of our result $\eqref{eq:fFinal}$. An evidence for the persistence of the $d=4+2\epsilon$ fixed point up to $d=5$ was obtained in ref.~\cite{DeCesare:2021pfb} using the five-loop $\overline{{\rm MS}}$ $\beta$-function and Pad\'e-Borel resummation techniques, both for the pure $SU(2)$ YM theory and for the theory with $n_f$ fundamental Dirac fermions, with $n_f\leq 4$. Note that the continuation from $d=4+2\epsilon$ suggests that the critical point should separate a free YM phase from a confined phase (the only phase realized in $d=4$) rather than a second YM phase, similarly to the refined proposal of \cite{Bertolini:2021cew} and in agreement with a recent lattice study that sees hints of a second order confinement/deconfinement transition \cite{Florio:2021uoz}.
\begin{table}
\centering
\begin{tabu}{|c|c|c|c|c|c|c}
\hline$n_f$ & 0 & 1 & 2& 3 &4  \\
\hline \rowfont{\color{red}}$E_{n_f+1}$ & $5.097$ & $6.140$ & $7.395$ & $8.959$ & $11.007$ \\
\hline [2,1] & $4.8$ & $5.1$ & $5.4$ & $5.7$ &$6.2$ \\
\hline
\end{tabu}
\caption{Comparison between the value of $\widetilde{F}_{E_{n_f+1}}$ (red) and the $[2,1]$ Pad\'e  approximant of $\widetilde{F}_\text{conf}$ in $d=5$ (black) as a function of $n_f$ for $0\le n_f\le 4$.}
\label{tab:3}
\end{table}
We therefore proceed to extrapolate $\widetilde{F}_\text{conf}$ using the only available Pad\'e approximant that is constrained also by the $d=6$ boundary condition \eqref{eq:d6bc} and without poles in the interval $4\leq d \leq 6$. In fig.~\ref{fig:su2} we plot the resulting extrapolation of $\widetilde{F}_\text{conf}$ as a function of the dimension. The value ranges between a local minimum of $\sim 2.9$ and a maximum of $\sim5.0$, before turning negative in the vicinity of $d=6$. The value in $d=5$ is $\sim4.8$, remarkably close to the known value $\sim5.1$ of $\widetilde{F}$ in the $E_1$ theory, and below it consistently with the proposals of \cite{BenettiGenolini:2019zth, Bertolini:2021cew}. 

The UV completion of the supersymmetric theory is also known in the case with $0<n_f \leq 7$ flavors and is given by the $E_{n_f+1}$ SCFT \cite{Seiberg:1996bd}.  The value of $\widetilde{F}$ can be obtained from localization similarly to the $E_1$ case \cite{Chang:2017cdx}. It is possible that also these theories flow to a non-supersymmetric fixed point when perturbed by a susy-breaking deformation. This fixed point would then provide a UV completion of the non-supersymmetric $SU(2)$ gauge theory with $n_f$ flavors. We test this possibility by comparing our extrapolation of $\widetilde{F}_\text{conf}$ to $\widetilde{F}_{E_{n_f+1}}$. We limit ourselves to the range $n_f\leq 4$ in which the fixed point in $d=4+2\epsilon$ was seen to persist up to $d=5$ in \cite{DeCesare:2021pfb}. We collect the values of the two $\widetilde{F}$'s in tab.~\ref{tab:3}. We always find $\widetilde{F}_{E_{n_f+1}}>\widetilde{F}_\text{conf}$,  consistently with the existence of the RG flow. 

\section{Conclusion}

In this paper we obtained the NLO result for the free energy on $S^d$ in non-abelian gauge theories in Euclidean $d$ dimensions evaluated at their perturbative fixed point. We extrapolated the result to compute the quantity $F$ for the corresponding CFTs in $d=3$ or $d=5$ and used our best estimates together with the monotonicity property of $F$ to test the existence and/or proposed constructions of these CFTs.

While successful in many contexts, the $\epsilon$ expansion is not a rigorous method. Going forward, it would be interesting to assess its reliability in the context of gauge theories. A possible verification could come from comparison with lattice and/or conformal bootstrap results. To that end it would be useful to compute scaling dimensions of operators in addition to those obtained in \cite{DeCesare:2021pfb}, or to improve the precision of the predictions by computing at higher loop order. 

Another possibility is to apply the $\epsilon$ expansion to cases in which the existence of a fixed point, and the associated data, are known from other methods such as supersymmetry or holography. For instance, one could apply it to the $4d$ theory with the same matter content as $5d$ $\mathcal{N}=1$ SU(2) SYM with $n_f$ fundamental flavors, and check if $\epsilon$ expansion finds a UV fixed point that extrapolates to the $E_{n_f}$ SCFT in $5d$. Note that when continuing the fields to $4d$ one does not land on a supersymmetric theory:  the $5d$ vector multiplet contains a real scalar, a $5d$ vector, and a symplectic Majorana fermion, all in the adjoint representation, and their continuation to $4d$ gives rise to a real scalar, a $4d$ vector, and a Dirac fermion, which is not the content of a supersymmetric theory in $4d$.\footnote{At least, this is the case with the method we are currently using to continue vector fields. One could imagine a different continuation, more in the spirit of DRED \cite{Siegel:1979wq}, in which the number of components of the vector is kept fixed. In this putative approach, $3d$ gauge theories with matter would not be obtainable with $\epsilon$ expansion, because the vector in $4d$ would give rise to additional scalars coupled to matter fields (and gauge fields as well, in the non-abelian case).} 
As a result, supersymmetry is expected to emerge only in the limit $d\to5$. To check the existence of fixed points in $d=4+2\epsilon$ one then needs the coupled system of $\beta$ functions for the gauge coupling in the presence of both fermionic and bosonic adjoint matter, and of the Yukawa coupling, see e.g. the Lagrangian (15) in \cite{Mirabelli:1997aj}. Note that these $\beta$ functions are known at lower loop order compared to the case with only fermionic matter that was used in \cite{DeCesare:2021pfb}, see \cite{Davies:2021mnc, Bednyakov:2021qxa}. 
We leave this as direction for future studies.

The perturbative expansion of the free energy is insensitive to the global structure of the gauge group, except the log term in eq.~\eqref{eq:fFinal} where the volume of the gauge group appears. 
It would be interesting to compare our results for $F$ with those computed using localization (or some other method) in SCFTs based on gauge theories such as $PSU(n_c) = SU(n_c)/{\bf Z}_{n_c}$. 
This analysis might be useful to shed some light on the nature of the transition delimited by our fixed points for $d>4$, since a confinement/deconfinement transition has the one-form symmetry ${\bf Z}_{n_c}^{(1)}$  as order parameter, while the latter is gauged in $PSU(n_c)$ theories and replaced by an emergent magnetic symmetry.

\section*{Acknowledgments}

We thank Francesco Benini, Matteo Bertolini, Simone Giombi, Francesco Mignosa, Jesse van Muiden and Yifan Wang for useful discussions. Work partially supported by INFN Iniziativa Specifica ST\&FI. LD also acknowledges support by the program ``Rita Levi Montalcini'' for young researchers.

\appendix
\section{Computation of the vector propagator}
\label{app:prop}
In this appendix we follow ref. \cite{Allen_1986} for the computation of the vector propagator. We report the main steps, generalizing the computation to an arbitrary choice of the gauge. 

We have seen in the main text that the vector propagator 
$
Q_{\nu \lambda}^{ab}(x,x')=\delta^{ab}g_0^2Q_{\nu \lambda}(x,x')
$
satisfies eq.~\eqref{eqn:44} and can be written as in eq.~\eqref{eq:gaugeprop} where $\alpha$ and $\beta$ are generic functions of the geodesic distance.
Using the relations in eq.~\eqref{prop} we can decompose eq.~\eqref{eq:gaugeprop} in two parts, respectively proportional to $g_{\nu \lambda'}$ and $n_{\nu}n_{\lambda'}$:
\begin{eqnarray}
 \alpha''+&& \hspace{-.5cm}(d-1)A\alpha' +\left((A+C)^2+(d-1)\right)\alpha+2AC\beta \nn \\
& &\hspace{-.5cm}-\left(1-\frac{1}{\xi}\right)C\left(\beta'-\alpha'-(d-1)(A+C)\alpha+(d-1)A\beta\right)=-\delta(x,x')  \,, \nn  \\
 \beta''+&&\hspace{-.5cm} (d-1)A\beta' +\left((A+C)^2-d(A^2+C^2)-(d-1)\right)\beta+(d-2)(A+C)^2\alpha   \label{eq:alpha}  \\
 &&\hspace{-.5cm}+\left(1-\frac{1}{\xi}\right)\Big(\beta''-\alpha''+((d-1)A+C)\beta'-(d(A+C)-A)\alpha' \nn \\
& &\hspace{-.5cm}+\big((d-1)AC+(d-1)A'\big)\beta+A' \big(-(d-1)C(A+C)+(1-d)A'+(1-d)C'\big)\alpha \Big)=0\ .  \nn
\end{eqnarray}
This is a system of two coupled second order differential equations, which is in general hard to solve. To make the computation easier it is convenient to introduce a new maximally symmetric gauge invariant bitensor defined as
\be
\langle F^a_{\mu \nu} F_b^{\mu^{\prime} \nu^{\prime}}\rangle=4\delta^a_{\ b}\ \nabla_{[\mu} \nabla^{\left[\mu^{\prime}\right.} Q_{\nu]}^{\left.\nu^{\prime}\right]}=\delta^a_{\ b}\ \left(\sigma(\mu) h_{[\mu}^{\left[\mu^{\prime}\right.} h_{\nu]}^{\left.\nu^{\prime}\right]}+\tau(\mu) n_{[\mu} h_{\nu]}^{\left[\nu^{\prime}\right.} n^{\left.\mu^{\prime}\right]}\right),
\label{eq:FF}
\ee
with square brackets meaning antisymmetrized indices and $\tau$ and $\sigma$ being generic functions of the geodesic distance. From the definition of $Q_{\nu \lambda^{\prime}}$ in terms of $\alpha$ and $\beta$ and  eq.~\eqref{prop}, we get
\be
\sigma=4 C\left[\alpha^{\prime}+(A+C) \alpha-C \beta\right], \label{eq:sigma}\ee
\be
\tau=C^{-1}\left[\sigma^{\prime}+2(A+C) \sigma\right] .
\label{eq:tau}
\ee
Now, taking the covariant derivative of eq.~\eqref{eq:FF} and using eq.~\eqref{eqn:44} properly antisymmetrized, one can find the equation of motion for $\sigma$ and $\tau$:
\begin{equation}
    \nabla^\mu\nabla_{[\mu} \nabla^{\left[\mu^{\prime}\right.} Q_{\nu]}^{\left.\nu^{\prime}\right]}=\frac{1}{\xi}\nabla_\nu \left(\nabla^{\mu} \nabla^{\left[\mu^{\prime}\right.} Q_{\mu}^{\left.\nu^{\prime}\right]}\right) = 0 \,.
    \label{eq:sigmatau}
\end{equation}
The last equality in \eqref{eq:sigmatau} derives from the fact that the bitensor in parenthesis has two primed antisymmetrized indices, while the only (0,2) bitensor structures are symmetric. In terms of $\sigma$ and $\tau$ defined in eq.~\eqref{eq:FF}, eq.~\eqref{eq:sigmatau} reads
\be
\sigma^{\prime}-\frac{1}{2} \tau^{\prime}+(d-2)(A+C) \sigma-\frac{1}{2}(d-2) A \tau=0.
\ee
Plugging the expression for $\tau$ in eq.~\eqref{eq:tau},  we get a second order differential equation for $\sigma$, which will be useful in the following to solve the system for $\alpha$ and $\beta$:
\be
\sigma^{\prime \prime}+(d+1)A\sigma'-2(d-1)\sigma=0\ .
\ee
This equation can be rewritten as a function of the variable $z$ defined in eq.~\eqref{eq:zed}:
\be
z(1-z)\frac{d^2\sigma}{dz^2}+\frac{1}{2}(d+2)(1-2z)\frac{d\sigma}{dz}-2(d-1)\sigma=0\ ,
\ee
which is solved by two linearly independent hypergeometric functions. The correct solution is chosen by imposing regularity at antipodals point ($z=0$) and the correct limit of coincident points ($z=1$). The last condition can be computed by starting from the expression in coordinate space of the gauge propagator in flat space
\be
\langle A^\mu_a(x) A_{\nu'}^b(x')\rangle_\mathrm{flat}=\delta_{\ a}^{b}\left(\frac{\Gamma\left(\frac{d}{2} - 1\right)(1+\xi)}{2(4 \pi)^\frac{d}{2} |x-x'|^{d-2}}\delta^\mu_{\ \nu'}+\frac{\Gamma\left(\frac{d}{2} \right)(1-\xi)}{(4 \pi)^\frac{d}{2} |x-x'|^{d}}x^\mu x_{\nu'} \right)
\label{eq:AAflatspace}
\ee
and the flat space expression for $\sigma$:  
\be
\sigma(z)_\mathrm{flat}= \frac{2\Gamma\left(\frac{d}{2}\right)}{(4 \pi)^{\frac{d}{2}}(1-z)^{\frac{d}{2}}}\,.
\ee
We find then
\be
\sigma(z)=p \ _2F_1\Big(d-1,2,\frac{d}{2}+1,z\Big)\ ,
\ee
with 
\be
p=\frac{\Gamma(d-1)}{\Gamma (\frac{d}{2}+1)2^{d-1}\pi^{\frac{d}{2}}}\ .
\label{eq:pfactor}
\ee
We can use this result to compute $\alpha$, proceeding as follows: we compute $\beta$ as a function of $\alpha$ and $\sigma$ from eq.~\eqref{eq:sigma} and we replace the result in eq.~\eqref{eq:alpha}. This leads to the following inhomogenous equation for $\alpha$:
\begin{equation}
\begin{split}
 \alpha''+&(d+1)A\alpha'-d\alpha-\frac{A}{2C}\sigma-\left(1-\frac{1}{\xi}\right)\left(\alpha''+\left(\frac{C'}{C}-Ad\right)\alpha'\right.\\&\left.-\left(d-1-A'+\frac{AC}{C'}\right)\alpha-\frac{1}{4C}\sigma+\left(\frac{C'}{2C}-\frac{A}{4C}(d-1)\right)\sigma\right)=-\delta(x,x').
 \end{split}
\end{equation}
The solution is given by the sum of the solution of the corresponding homogenous equation (again we should impose the flat space limit and regularity at $z=0$) and a particular solution to reproduce the correct source term. We have 
\begin{equation}
\alpha(z)=q \ _2F_1\Big(d,1,\frac{d}{2}+1,1-z\Big)+\tilde{\alpha}(z),
\label{eq:alphasol}
\end{equation}
where the first term is the solution of the homogeneous equation, with a normalization $q$ to be fixed, and $\tilde{\alpha}$ is a particular solution of the full equation, that plugging the expression for $\sigma$, $C$ and $A$ takes the form
\be
\begin{split}
& z(1-z)\frac{d^2\tilde{\alpha}}{dz^2}+\Big(\frac{d}{2}+1-(d+2) z\Big)\frac{d\tilde{\alpha}}{dz}-d\tilde{\alpha}=
\frac{\pi ^{-\frac{d}{2}} \Gamma (d-1)}{2^{d+1}}\\ & \left((d+2-\xi) (1-2z)  {}_2{F}_1\Big(2,d-1;\frac{d}{2}+1;z\Big)-4 (d-\xi)
(z-1) z \, {}_2{F}_1\Big(3,d;\frac{d}{2}+2;z\Big)\right)\, .
\end{split}
\label{eq:alphatilde}
\ee
A solution for $\tilde \alpha$ can be found as follows \cite{Allen_1986}.  We introduce the hypergeometric operator
\begin{equation}
    H(a,b,c)=z(1-z)\frac{d^2}{dz^2}+\left( \big(c-(a+b+1)z \frac{d}{dz} \big) -ab\right) \,,
\end{equation}
in order to rewrite the left-hand side of eq.~\eqref{eq:alphatilde} as 
\be 
H\left(a_{1}+1, b_{1}-1, c_{1}\right)\tilde{\alpha}\, ,
\ee 
with $a_1=d-1$, $b_1=2$, $c_1=\frac{d}{2}+1$. Then, we rewrite the right-hand-side
of eq.~\eqref{eq:alphatilde} as $H(a_{1}+1, b_{1}-1, c_{1})f$, with $f$ a function to be determined, 
using identities among hypergeometric functions (see e.g. chapter 15 of ref.~\cite{abramowitz+stegun}).
A particular solution would then be $\tilde \alpha = f$.

The right-hand-side of eq.~\eqref{eq:alphatilde} is first rewritten as a function of ${}_2F_1\left(a_{1}, b_{1} , c_{1} , z\right)$, ${}_2F_1\left(a_{1}-1, b_{1} , c_{1} , z\right)$ and ${}_2F_1\left(a_{1}+1, b_{1}-1 , c_{1} , z\right)$ only. Then, the following identities are used:
\begin{eqnarray}
 _2F_1\left(a_{1}, b_{1} , c_{1} , z\right) \!\!\!& = &\!\!\! \frac{1}{d-2}H\left(a_{1}+1, b_{1}-1, c_{1}\right)  {}_2F_1\left(a_{1}, b_{1} , c_{1} , z\right) , \\
{}_2F_1\left(a_{1}-1, b_{1} , c_{1} , z\right)\!\!\!& = &\!\!\!  \frac{1}{2(d-3)}H\left(a_{1}+1, b_{1}-1, c_{1}\right)\left( {}_2F_1\left(a_{1}-1, b_{1} , c_{1} , z\right)+ {}_2F_1(a_1, b_{1} , c_{1} , z)\right) , \nn \\
{}_2F_1\left(a_{1}+1, b_{1}-1,c_1,z\right)\!\!\!& = &\!\!\!  \frac{1}{b_1-a_1-2}
\left.H\left(a_{1}+1, b_{1}-1, c_1\right)\left(\frac{\partial}{\partial a}-\frac{\partial}{\partial b}\right)  {}_2F_1(a,b,c_1,z)\right|_{\substack{a=a_{1}+1 \\ b=b_{1}-1}} . \nn
\end{eqnarray}
Matching with the left hand side of eq.~\eqref{eq:alphatilde} gives the particular solution $\tilde{\alpha}=f$:
\be
\begin{split}
\tilde{\alpha}=\frac{p}{4(d-3)^2} \bigg(-2\ _2F_1\left(a_1-1,b_1+1,c_1,z\right)+(d-4) \ _2F_1\left(a_{1}-1,b_1,c_1,z\right)\\
+\big(2+(d-3)(1-\xi)\big)(3-d)\left(\frac{\partial}{\partial a}-\frac{\partial}{\partial b}\right)\  _2F_1(a,b,c_1,z)\Big|_{\substack{a=a_{1}+1 \\ b=b_{1}-1}}\bigg)\,.
\label{eq:alphatilda}
\end{split}
\ee
The value of the coefficient $q$ appearing in eq.~\eqref{eq:alphasol} is determined by imposing the correct flat space limit of $\alpha$. which is the term proportional 
to $\delta^\mu_{\nu'}$ in eq.~\eqref{eq:AAflatspace}. We get
\be
q=p\frac{(d-1)(d-2)-(2+(d-3)(1-\xi))(d-3)(\psi(d)-\psi(1))}{4(d-3)^2}\, .
\label{eq:qfactor}
\ee
Finally, we obtain the expression for $\beta$ by replacing $\alpha$ and $\sigma$ in eq.~\eqref{eq:sigma}:
\begin{eqnarray}
\beta=&&\hspace{-.5cm}-\frac{ (z-1) \Gamma (d-1) }{2^{d} \pi ^{\frac{d}{2}}(d-3) \Gamma \left(\frac{d}{2}+1\right)} \nn \\
&&\hspace{-.5cm}\left(\frac{d}{dz}\Big(-z(2+(d-3)(1-\xi))(3-d)\left(\frac{\partial}{\partial a}-\frac{\partial}{\partial b}\right)\  \left._2F_1(a,b,c_1,z)\right|_{\substack{a=a_{1}+1 \\ b=b_{1}-1}}\Big)\right.  \label{eq:betaExp}\\
&&\hspace{-.5cm}+\Gamma \left(\frac{d}{2}+1\right) \Big( {}_2{F}_1\Big(2,d;\frac{d}{2}+1;z\Big) \Big(-\big((d-3) (1-\xi) +2\big) (\psi ^{(0)}(d)+\gamma )-2 d z+d+2 z-4\Big) \nn \\
&&\hspace{-.5cm} \left.-4 (z-1) \, _2{F}_1\Big(3,d;\frac{d}{2}+1;z\Big)\Big)\right) \nn \,.
\end{eqnarray}
Summarizing, the gauge propagator on $S^d$ is obtained by replacing in eq.~\eqref{eq:gaugeprop}  
the expression for $\alpha$ in eqs.(\ref{eq:alphasol},\ref{eq:alphatilda}), the one for $\beta$ just reported, together with the expressions for the coefficients $p$ and $q$ in eqs.(\ref{eq:pfactor},\ref{eq:qfactor}).

Let us now explain how to expand $\alpha$ and $\beta$ around coincident points ($z=1$). First, note that the hypergeometrics have branch points in $z=1$. In order to expand in powers of $(z-1)$ it is then convenient to use an identity to obtain only hypergeometric functions with argument $1-z$. In this way the non-analytic dependence on $z-1$ will be captured completely by the power-law prefactors. The identity that we will use for this purpose is
    \begin{equation}
\begin{aligned}
{ }_2 F_1(a, b , c, z)=& \frac{\Gamma(c) \Gamma(c-a-b)}{\Gamma(c-a) \Gamma(c-b)}{ }_2 F_1(a, b , a+b+1-c , 1-z) \\
&+\frac{\Gamma(c) \Gamma(a+b-c)}{\Gamma(a) \Gamma(b)}(1-z)^{c-a-b}{ }_2 F_1(c-a, c-b , 1+c-a-b , 1-z)\,.
\label{eq:hypidentity}
\end{aligned}
\end{equation}
Derivatives of the hypergeometrics with respect to the parameter $a$ and $b$ appear in both $\alpha$ and $\beta$. In order to obtain the expansion in $(1-z)$ for these derivatives, we first apply the identity in eq.~\eqref{eq:hypidentity} and then we expand the hypergeometric as
\be
{ }_2 F_1\left(a, b, c, 1-z\right)=\sum_{n=0}^{\infty} \frac{\left(a\right)_n\left(b\right)_n}{\left(c\right)_n} \frac{(1-z)^n}{n !}\,,
\ee
where  $(x)_n$ are the Pochammer symbols.
We truncate the series at a sufficiently high order and then we apply the derivatives with respect to $a$ and $b$ to this truuncated series.  
In order to improve the efficiency of the numerical integration of hypergeometrics needed to get the finite terms \eqref{eq:G2tFinite}-\eqref{eq:G2fFinite}, 
it is useful to split the interval of integration $0\leq z \leq 1$ in two parts (i.e. [0,1/2] and [1/2,1]) and expand respectively around 0 and around 1 the hypergeometrics.  

\section{Computation of the ghost counterterm}
\label{app:ghost}
The ghost wave function renormalization can be computed by imposing finiteness of the ghost propagator at one loop. We compute the divergence in configuration space:
\begin{equation}\begin{split}
\left.\feynmandiagram [horizontal=a to b] {a -- [charged scalar] b -- [charged scalar] c-- [charged scalar] d,b--[gluon,half left] c};\right|_{\text{div.}}=\delta_{ab} g_{0}^{2}\ C_A \int d^{d} x_1 \ d^{d} x_2 \sqrt{h} \sqrt{h'}\  G(x,x_1) \   \nabla_{\mu_1}G(x_1,x_2)  \ \\Q^{\mu_1\mu_2}(x_1,x_2)\nabla_{\mu_2}G(x_2,0) \,.
\end{split}
\label{eq:AppB1}
\end{equation}
Since we are dealing with UV divergences, we can take the limit of coincident points $x_1\sim x_2$. Taylor expanding the propagator $G(x_2,0)$ around $x_1$, we get
\begin{equation}
G(x_2,0)=G(x_1,0)+(x_2^{\mu}-x_1^{\mu})\nabla_{\mu}G(x_1,0)+\frac{1}{2}(x_2^{\mu}-x_1^{\mu})(x_2^{\nu}-x_1^{\nu})\nabla_\nu\nabla_{\mu}G(x_1,0)+\dots
\end{equation}
Replacing this expression in eq.~\eqref{eq:AppB1} we find that the only non-vanishing  contribution comes from the third term: the first vanishes when derived with respect to $x_2$, while the second is zero because of Lorentz invariance. All other terms in the expansion provide convergent result and are therefore not relevant for the computation of counterterms. We have then
\begin{equation}
 \delta_{ab}\ \frac{g_{0}^{2}}{2}\ C_A \int d^{d} x_1  \sqrt{h}\  G(x,x_1)I^{\mu\nu}\,\nabla_\nu\nabla_{\mu}G(x_1,0),
\end{equation}
with
\begin{equation}
I^{\mu\nu}=  \int d^{d} x_2  \sqrt{h'} \left( \nabla_{\mu_1}G(x_1,x_2)  \ Q^{\mu_1\mu_2}(x_1,x_2)\nabla_{\mu_2}(x_2^{\mu}-x_1^{\mu})(x_2^{\nu}-x_1^{\nu})\right) \,.
\end{equation}
By spherical invariance this integral does not depend on the position of $x_1$, which can be set to zero. We can use Lorentz invariance to rewrite the integral as
\begin{equation}
    I^{\mu\nu}=\frac{g^{\mu\nu}}{d}(I^{\lambda\sigma}g_{\lambda\sigma})\,.
\end{equation}
The divergence can be computed by using stereographic coordinates and expanding around coincident points, as done in sec.\ref{sec:diagrams}:
we get 
\begin{equation}
    \left.\feynmandiagram [horizontal=a to b] {a -- [charged scalar] b -- [charged scalar] c-- [charged scalar] d,b--[gluon,half left] c};\right|_{\text{div.}}=\delta_{ab}\frac{3-\xi}{64\pi^2\epsilon}\ g_{0}^{2}  C_A \int d^{d} x_1  \sqrt{h}\  G(x,x_1)  \,\nabla^2G(x_1,0)\,.
\end{equation}
This divergence can be removed by taking the following wave function renormalization:
\begin{equation}
c=Z_c^{\frac{1}{2}}c_R\ ,\qquad \bar{c}=Z_c^{\frac{1}{2}}\bar{c}_R \,,
\end{equation}
with
\begin{equation}
    Z_c=1+\delta_c=1-g_0^2\frac{3-\xi}{64\pi^2\epsilon}C_A +\mathcal{O}(g^4)\,,
\end{equation}
which reproduces eq.~\eqref{eq:count}.
Note that since there is no divergence proportional to 
\begin{equation}
 \int d^{d} x_1  \sqrt{h}\  G(x,x_1)G(x_1,0),
\end{equation}
there is no mass renormalization, as expected.

\section{Subtleties on contact terms and integration by parts}
\label{app:contact}

In this appendix we show the subtleties that can arise when integrating propagators derived multiple times on $S^d$. This analysis is relevant for our purposes in presence of two derivatives acting on the same propagator.
For simplicity we will consider a scalar propagator satisfying the equation
\begin{equation}
(-\nabla^2+m^2)G(x,x')=\delta(x,x')\,,
\label{eq:scalprop}
\end{equation}
but the same remarks hold for the vector propagator and can be applied to eq.~\eqref{eq:g2triple}. Let us consider the integral
\begin{equation}
    \int_{S^d} d^dx\sqrt{h}\ f(\mu)\nabla^2 G(x,0) \,,
    \label{eq:basicInt}
\end{equation}
where $f$ is a function of the geodesic distance $\mu=\mu(x,0)$, which is taken to be smooth and bounded on $S^d$. 
If one tries to compute this integral by specifying some coordinate system and writing explicitly the action of the laplacian on the resulting function in the chosen coordinates, one gets a wrong answer. This is because the resulting expression for $\nabla^2 G(x,0)$ misses the contact term, and the answer one gets would correspond to substituting simply $\nabla^2 G(x,0) = m^2 G(x,0)$ inside the integral.

A strategy to obtain the correct answer is to integrate by parts 
\begin{equation}\label{eq:intcorrect}
    \int_{S^d} d^dx\sqrt{h}\ f(\mu(x,0))\nabla^2 G(x,0) \,=-\int_{S^d } d^dx\sqrt{h}\ \nabla^\nu f(\mu(x,0))\nabla_\nu G(x,0)~.
\end{equation}
To check that this works, let us start by separating two regions in the integral
\begin{equation}
       \int_{S^d \backslash B_\delta} d^dx\sqrt{h}\ f(\mu)\nabla^2 G(x,0) \,+   \int_{B_\delta} d^dx\sqrt{h}\ f(\mu)\nabla^2 G(x,0)~,
       \label{eq:intparts}
\end{equation}
where $B_\delta$ is defined as a small $d$-dimensional ball of radius $\delta$ centered at the origin. In the second integral, for $\delta\rightarrow 0$, we get the contact term $-f(0)$.
In the first term,  we integrate by parts 
\begin{equation}
\int_{S^d \backslash B_\delta} \!\!\!\!\! \!d^dx\sqrt{h} f(\mu)\nabla^2 G(x,0) = 
-\int_{S^d \backslash B_\delta} \!\!\! \!\!\! d^dx\sqrt{h} \nabla^\nu f(\mu)\nabla_\nu G(x,0)
+\int_{S^d \backslash B_\delta} \!\!\! \!\! \! d^dx\sqrt{h} \nabla^\nu \big(f(\mu)\nabla_\nu G(x,0)\big) \,.
    \label{eq:bound1}
\end{equation}
In the first integral the limit $\delta\rightarrow0$ is straightforward, while the second integral requires more care. It is a boundary term that we can rewrite using
the first relations in eqs.~\eqref{prop},~\eqref{eq:A&C} and the chain rule as
\begin{equation}
\begin{aligned}
    &\int_{S^d \backslash B_\delta} d^dx\sqrt{h}\ \nabla^\nu (f(\mu)\nabla_\nu G(x,0))=\int_{S^d \backslash B_\delta} d^dx\sqrt{h}\ \nabla^\nu \Big( f(\mu(z))G'(z)\frac{\partial z}{\partial \mu}n_\nu\Big)\\
    &=\int_{S^d \backslash B_\delta} d^d x \sqrt{h} \left(A(d-1)f(\mu(z))G'(z)\frac{\partial z}{\partial \mu}+\frac{\partial }{\partial z}\Big(f(\mu(z))G'(z)\frac{\partial z}{\partial \mu}\Big)\frac{\partial z}{\partial \mu}\right)\,.
    \end{aligned}
\end{equation}
Here we used the variable $z$ defined in \eqref{eq:zdef}. By changing the integration variable to $z$ we get
\begin{equation}
   \lim_{\delta'\rightarrow0} \frac{2^d\pi^{\frac{d}{2}}} {\Gamma\left(\frac{d}{2}\right)}\int^{1-\delta'}_0 \!\! dz \frac{\partial}{\partial z}\left((z(1-z))^{\frac{d-1}{2}}f(\mu(z))G'(z)\frac{\partial z}{\partial \mu}\right)\,.
   \label{eq:bound}
\end{equation}
The above integral would vanish for well-defined functions on $S^d$, as expected from Stokes theorem, but the propagator is actually a distribution
which is singular at coincident points $z\rightarrow 1$, so care is required. 
In the limit $z\rightarrow 1$ the scalar propagator can be approximated to
\begin{equation}
    G(z)\simeq \frac{ \pi^{1 - \frac{d}{2}} }{2^d\Gamma\left(
 2 - \frac{d}{2}\right)\sin\left(\frac{d \pi}{2}\right)} (1 - z)^{1-\frac{d}{2}}+\dots\,
 \label{eq:PropAppc}
\end{equation}
Replacing eq.~\eqref{eq:PropAppc} in eq.~\eqref{eq:bound} gives a non-vanishing result:
\begin{equation}
    \lim_{\delta'\rightarrow0}\left.f(\mu(z))z^{\frac{d}{2}}\right|_{z=1-\delta'}=f(\mu=0)\,.
\end{equation}
This boundary term exactly cancels the contribution  coming from the second term
in eq.~\eqref{eq:intparts}, proving eq. \eqref{eq:intcorrect}. Summarizing, the evaluation of eq.~\eqref{eq:basicInt} without integrating by parts would require
to pay attention to contact terms by introducing a regulator, while upon integrating by parts the contact term contribution is compensated
by another contact term arising from a total derivative contribution.

\section{Check with Jack \cite{Jack:1982sn}}
\label{app:jack}

The poles of the diagrams \eqref{eq:triple}, \eqref{eq:ghost}, \eqref{eq:quartic}, and \eqref{eq:CTvect} can also be computed with a different procedure. This procedure, which is based on the heat-kernel expansion of the propagators, is more general because it works on any curved background. We show in this section how the divergences which were obtained in this way for pure Yang-Mills theory in ref.~\cite{Jack:1982sn} in the Feynman gauge $\xi=1$ agree with our previous results. 
Matching with the results of \cite{Jack:1982sn} requires a bit of manipulations. It is then useful to briefly recall the key results found in \cite{Jack:1983sk,Jack:1982sn} using heat kernel methods. 

Let us consider an elliptic differential operator of the form
\begin{equation}
M(x)=-\nabla^2+Y(x)
    \label{eq:elliptic}
\end{equation}
and the corresponding propagator satisfying
\begin{equation}
    M(x) G_M(x,x')=\delta(x-x')\,.
\end{equation}
Around coincident points $x\sim x'$  the following expansion holds \cite{Jack:1983sk}:
\begin{equation}
    G_M\sim -\frac{1}{16\pi^2\epsilon}a^M_{1\ \text{diag}}+H_{\text{diag}}^M\, , 
    \label{eq:defH}
\end{equation}
where $H_{\text{diag}}^M$ is in general a complicated non-local expression satisfying
\begin{equation}
    M(x)H_\text{diag}^M=\frac{1}{16\pi^2}a^M_{2\ \text{diag}}  \,  .
    \label{eq:Mh}
\end{equation}
The coefficients $a_{1\ \text{diag}} ^M$ and $a_{2 \ \text{diag}}^M$ admit instead a local expression in terms of the curvature tensors and they can be computed for any elliptic operator $M$. From the propagator equations in sec.~\ref{rules}, we see that the ghost differential operator is indeed of the form \eqref{eq:elliptic}, while the vector one is not for a generic choice of the gauge. This is why ref.~\cite{Jack:1983sk} provides results only in the Feynman gauge $\xi=1$, for which also the vector operator is of the form \eqref{eq:elliptic}. The coefficients then read  
\begin{equation}\begin{split}
    a_{1\text{diag}}^\text{gh} & =\frac{1}{6}\mathcal{R} \,, \\
    a_{2\text{diag}}^\text{gh}& =\frac{1}{180}(\mathcal{R}_{\mu\nu\rho\sigma}\mathcal{R}^{\mu\nu\rho\sigma}-\mathcal{R}_{\mu\nu}\mathcal{R}^{\mu\nu})+\frac{1}{72}\mathcal{R}^2 \,,
\end{split} \end{equation}
for the ghost and
\begin{equation} \begin{split}
    a_{1\text{diag}\ \mu\nu}^\text{vec}& =\frac{1}{6}\mathcal{R}g_{\mu\nu}-\mathcal{R}_{\mu\nu} \,, \\
    a_{2\text{diag}}^\text{vec}& =\frac{1}{360}(2(d-15)\mathcal{R}_{\mu\nu\rho\sigma}\mathcal{R}^{\mu\nu\rho\sigma}
    -2(d-90)\mathcal{R}_{\mu\nu}\mathcal{R}^{\mu\nu}+5(d-12)\mathcal{R}^2) \,,
    \end{split}
\end{equation}
for the vector. Ref.~\cite{Jack:1982sn} provides the expressions for the poles of diagrams as a function of the curvature tensors and of the derivatives of $H_{\text{diag}}$,  
more specifically, $\nabla^2 H^\text{gh}_\text{diag}$, $H^{\text{vec}\ \mu\nu}_\text{diag}$, $g_{\mu\nu} \nabla^2 H^{\text{vec}\ \mu\nu}_\text{diag}$ and $\nabla_{\mu}\nabla_{\nu}H^{\text{vec}\ \mu\nu}_\text{diag}$.\footnote{In ref.~\cite{Jack:1982sn} $H^\text{gh}_\text{diag}$ is denoted $H^\text{0}_\text{diag}$, while $H^{\text{vec}\ \mu\nu}_\text{diag}$ is denoted
$H^{\text{1}\ \mu\nu}_\text{diag}$.} Now, from eq.~\eqref{eq:Mh} we have
\begin{align}
    -\nabla^2&H^\text{gh}_\text{diag}=\frac{1}{16\pi^2}a^\text{gh}_{2\ \text{diag}}\,,  \\
    -\nabla^2&H^{\text{vec}\ \mu\nu}_\text{diag}g_{\mu\nu}=\frac{1}{16\pi^2}a^\text{vec}_{2\ \text{diag}} -H^{\text{vec}\ \mu\nu}_\text{diag}{\cal R}_{\mu\nu}\,.
    \label{eq:Hghdiag}
    \end{align}
The first equation allows us to find a simple expression for $\nabla^2 H^\text{gh}_\text{diag}$ in terms of the curvature tensors. However we cannot solve the second equation to obtain a similar simple expression for $H^{\text{vec}\ \mu\nu}_\text{diag}$.
A way to compute $\nabla_{\mu}\nabla_{\nu}H^{\text{vec}\ \mu\nu}_\text{diag}$ is by imposing the cancellation of poles in the total free-energy inserting the expression for the diagrams obtained in ref.~\cite{Jack:1982sn}  (detailed in footnote \ref{footnoteJack} below) in eq.~\eqref{eqn:31}. Note that only $\nabla_{\mu}\nabla_{\nu}H^{\text{vec}\ \mu\nu}_\text{diag}$ and $\nabla^2 H^\text{gh}_\text{diag}$ enter the expression for these diagrams, not $H^{\text{vec}\ \mu\nu}_\text{diag}$. In order to obtain a result valid on a generic manifold, we use the one-loop free energies computed in refs.~\cite{Jack:1983sk,Jack:1982sn} and the renormalization of the curvature coefficient $a$. We  get 
 \begin{align}
    \nabla_{\mu}\nabla_{\nu}H^{\text{vec}\ \mu\nu}_\text{diag}=\frac{1}{8\pi^2}\left(\frac{109}{3960}\mathcal{R}_{ \mu  \nu  \rho  \sigma} \mathcal{R}^{ \mu  \nu  \rho  \sigma}- \frac{229}{3960 } \mathcal{R}_{ \mu  \nu} \mathcal{R}^{ \mu  \nu}+ \frac{5}{1584} \mathcal{R}^2+3 \right)\,.
    \label{eq:HdiagNabla}
\end{align}
The above relations apply on any manifold. 

We can now focus on $S^d$ to get explicit results.  
The term $H^{\text{vec}\ \mu\nu}_\text{diag}$ can be computed by expanding the propagator around coincident points and using eq.~\eqref{eq:defH}. Taking the expression for the gauge propagator of eq.~\eqref{eq:gaugeprop} for $\xi=1$ we get
\begin{equation}
    \begin{split}
    Q^\mu_{\ \nu}(z)=R^{2-d}\left(\frac{\Gamma \left(\frac{d}{2}-1\right)}{2^{d} \pi ^{\frac{d}{2}}} (1-z) ^{1-\frac{d}{2}} +\frac{\Gamma \left(\frac{d}{2}-2\right)(d^2-6d+4) }{2^{d+2} \pi ^{\frac{d}{2}}} (1-z) ^{2-\frac{d}{2}}\right.\\\left.
+\frac{ \Gamma \left(\frac{d-3}{2}\right) \left(-d+2 \pi  \cot \left(\frac{\pi  d}{2}\right)+2
\left(\psi(d)+\gamma \right)\right)}{8\pi ^{\frac{d+1}{2}} d}\right)\delta^\mu_{\ \nu}+\dots \,.
\end{split}
\end{equation}
Using analytic continuation in $d$, we can set to zero the powers $(1-z) ^{1-\frac{d}{2}}$ and $(1-z) ^{2-\frac{d}{2}}$ of the expansion. The remaining part can be computed at $d=4+2\epsilon$ and expanded around $\epsilon=0$.  Plugging the result in eq.~\eqref{eq:defH} gives
\begin{equation}
   H^{\text{vec}\ \mu\nu}_\text{diag} (S^4)=-\frac{1 + 3\gamma + 3\log(4\pi \mu^2 R^2)}{48\pi^2  R^2} \delta^{\mu\nu}\,.
  \end{equation}
From eq.~\eqref{eq:HdiagNabla} we have
\begin{equation}
    \nabla_{\mu}\nabla_{\nu}H^{\text{vec}\ \mu\nu}_\text{diag}(S^4)=\frac{61}{240\pi^2 R^4}\,.
\end{equation}
Using eq.~\eqref{eq:Hghdiag} we can similarly  get the expressions for  $\nabla^2 H^\text{gh}_\text{diag}$ and $\nabla_{\mu}\nabla_{\nu}H^{\text{vec}\ \mu\nu}_\text{diag}$:
 \begin{align}
g_{\mu \nu}     \nabla^2H^{\text{vec}\ \mu\nu}_\text{diag}(S^4) & =-\frac{232 + 120 \bigl(-1 + 3 \gamma+\log(4\pi  \mu^2 R^2)\bigr)}{ 480 \pi^2 R^4}\,, \\
   \nabla^2H^{\text{gh}\ }_\text{diag}(S^4)& =-\frac{29}{240 \pi^2 R^4} \,.
\end{align}
Substituting in the results of ref.~\cite{Jack:1982sn}, we find\footnote{See eq.~(2.55) of ref.~\cite{Jack:1982sn} for $G_{2J}^{\mathrm{triple}}$, eq.~(2.52) for $G_{2J}^{\mathrm{ghost}}$,  eq.~(2.33) for $G_{2J}^{\mathrm{quart}}$ and eqs.~(2.31),(2.59) for $G_{2J}^{\mathrm{CT-vect}}$. Note that in our convention $d=4+2\epsilon$, while in ref.~\cite{Jack:1982sn} the authors used $d=4-\epsilon$. Moreover, all diagrams in ref.~\cite{Jack:1982sn} are multiplied by a factor 1/2, which we factorized instead outside $G_2$. \label{footnoteJack}}
\begin{align}
\left. G_{2J}^{\mathrm{triple}}\right|_\text{div.}& =\kappa \left(\frac{1}{24 \pi ^2 \epsilon ^2}-\frac{13+2(\gamma +\log (4 \pi  \mu^2 R^2))}{48 \pi ^2 \epsilon }\right) \,,\label{eq:tripleJ}\\
\left.G_{2J}^{\mathrm{ghost}}\right|_\text{div.}& =\kappa \left(\frac{1}{48 \pi ^2 \epsilon ^2}-\frac{ 1+2 (\gamma+\log (4 \pi  \mu^2R^2))}{96 \pi ^2 \epsilon }\right) \,, \label{eq:ghostJ} \\
\left.G_{2J}^{\mathrm{quart}}\right|_\text{div.}&=\kappa \left(-\frac{1}{16 \pi ^2 \epsilon ^2}+\frac{ 17+6(\gamma+\log (4 \pi  \mu^2R^2))}{96 \pi ^2 \epsilon }\right) \,,
\label{eq:quarticJ}\\
\left.G_{2J}^{\mathrm{CT-vect}}\right|_\text{div.}&=\kappa \left(-\frac{1}{8 \pi ^2 \epsilon }\right)\,.\label{eq:CTvectJ}
\end{align}
Eqs.~\eqref{eq:tripleJ}, \eqref{eq:quarticJ} and \eqref{eq:CTvectJ} match respectively eqs.~\eqref{eq:triple}, \eqref{eq:quartic} and \eqref{eq:CTvect} evaluated at $\xi=1$.
As explained, the ghost counterterm \eqref{eq:ghost} arises because of ghost zero modes, specific for $S^d$. Heat kernel methods apply to generic manifolds and therefore there is no ghost counterterm in ref.~\cite{Jack:1982sn}.  The ghost contribution \eqref{eq:ghostJ} should then match the sum of eq.~\eqref{eq:ghost} and the counterterm \eqref{eq:ghostcount} for $\xi=1$,
and this is indeed the case. We then have a check diagram by diagram of our computation.

\section{Alternative gauge-fixing procedure}

In section \ref{sec:Landau} we have seen that the quantization of non-abelian gauge theories on $S^d$ using an ordinary Faddeev-Popov formalism
leads to ghost zero modes. In this appendix we would like to show that our heuristic treatment of the zero modes is confirmed by a more rigorous
treatment using a Batalin-Vilkovisky formalism and ghosts for ghosts, see e.g. ref.~\cite{weinberg_1996}  for a nice introduction or ref.~\cite{Gomis:1994he} for a more detailed treatment.  
We start by briefly recalling the method in Yang-Mills theories on flat space and then apply it on $S^d$, where we 
reproduce the action presented in ref.~\cite{Pestun:2007rz}. We then compute the ghost contribution in eq.~\eqref{eq:G2tot} using the new action and show that it matches with eq.~\eqref{eq:g2ghost} obtained with the more heuristic treatment discussed in the main text.

\subsection{Gauge theories on $S^d$}

Yang-Mills theories on flat space do not require ghosts for ghosts and can be treated with the Faddeev-Popov method. 
Let us briefly review how the same gauge-fixing can be obtained with the Batalin-Vilkovisky formalism. 
Recall that in this formalism for each field $\phi_A$ we introduce an antifield $\phi^*_A$ and we require the master equation
\be
(S, S)=0\,,
\label{eqn:29}
\ee
where
\be
(F, G)\equiv \frac{\delta_{R} F}{\delta \phi^{A}} \frac{\delta_{L} G}{\delta \phi_{A}^{*}}-\frac{\delta_{R} F}{\delta \phi_{A}^{*}} \frac{\delta_{L} G}{\delta \phi^{A}}\,.
\ee
In the Yang-Mills theory case, $\phi^A = \{A, c\}$, where $c$ are the ghost fields needed to take into account of the gauge redundancy of the classical action.
The action satisfying the master equation \eqref{eqn:29} reads
\be S_\text{flat}=S_\text{YM}[A]+\int d^dx\, \Big(A^*D{c}-{i}{c}^{*}{c}^2 +\bar{{c}}^{*} B \Big)\,,
\label{eqn:34b}
\ee
where $D=\partial-i[A,\cdot\ ]$,~$[\phi_1,\phi_2]=i f_{abc}T^a\phi_1^b\phi_2^c$, with $T^a$ and $f_{abc}$  the generators in the fundamental representation and the structure constants of the Lie algebra, respectively.  In eq.~\eqref{eqn:34b} trace over group indices and Lorentz indices are implicit and 
we have added an auxiliary pair of fields $\bar{{c}} /B$ and their corresponding antifields, which do not affect the 
master equation. Note that only $\bar{c}^*$ and $B$ enter the action but also $\bar{c}$ and $B^*$ are integrated over in the path integral. A gauge-fixing  is introduced through a fermionic functional $\Psi[\phi]$  which fixes the value of the antifields:
\be
\phi_{A}^{*}=\frac{\delta \Psi[\phi]}{\delta \phi^{A}}\,,
\label{eqn:13}
\ee
where now  $\phi^A = \{A, c, \bar c, B\}$.
An appropriate choice for the gauge-fixing functional is 
\be
\Psi=\int d^dx \ \bar{{c}}\left(-\frac{\xi}{2}B-\partial A\right),
\label{eq:fermfun}
\ee
which leads to 
\be
S^\text{g.f.}_\text{flat}=S_\text{YM}[A]+\int \! d^dx\left( \bar{c}\ \partial D{c}-B\Big(\frac{\xi}{2}B+\partial A\Big) \right)\,,
\label{eq:flat}
\ee
which is the usual $R_\xi$ gauge fixing of the Yang-Mills action. 

On $S^d$ an important difference arises. Covariantly constant modes leave the gauge field invariant, so the transformation $c\rightarrow c +\theta \tilde{a}_{0}$, with $\theta$ a Grassmann constant parameter, leaves the gauge field invariant, provided that 
\begin{equation}
    D {\tilde{a}_0}=0\,.
    \label{eq:eta}
\end{equation}The mode $\tilde{a}_0$ is a (bosonic) ghost for ghost. We should then add $\tilde{a}_0$ to the set of fields in the action, together with its antifield. $\tilde{a}_0$ is actually not a field, but a single mode of a field, the covariantly constant one. For simplicity we keep this implicit.
The solution to the master equation reads now
\begin{equation}
    S=S_\text{YM}[A]+\int \! d^dx\,\sqrt{h}\Big( A^{*}D{c}+{c}^{*}{a}_0-i{c}^{*}{c}^2+i{a_0}^*[c,{a_0}]+\bar{{c}}^{*} B\,+{\bar{a}_0}^{*} \bar{c}_0+{b_0}^{*} {c_0} \Big)\,,
    \label{eq:red}
\end{equation}
where we have added two pairs $\bar{a}_0/\bar{c}_0$ and $ {b_0}/ {c_0}$ of fields (and their antifields), composed only of a covariantly constant mode, like $\tilde{a}_0$.
In a perturbative treatment, where we expand in modes the quadratic action, the covariantly constant mode $\tilde{a}_0$ should be replaced by a constant mode ${a_0}$
satisfying 
\begin{equation}
    \nabla{a_0}=0\,,
\end{equation}
which corresponds to the ghost zero modes found in the ordinary Faddeev-Popov procedure followed in section \ref{sec:Landau}.
The action \eqref{eq:red} no longer solves the classical master equation if $\tilde{a}_0 \rightarrow {a_0}$. We now have  
\be
(S,S)=2A^{*} D{a_0}=2 iA^{*} [{a_0},A] \neq 0\,,
\ee
However, adding appropriate terms to the action we can introduce a new action $\tilde{S}=S+\delta S$  such that
\be
(\tilde{S},\tilde{S})=2i\phi_A^*[{a_0},\phi^A]\,.
\label{eq:stilda}
\ee
In this way, after gauge-fixing we have 
\be
(\tilde{S},\tilde{S})=2i\frac{\delta \Psi}{\delta \phi^A}[{a_0},\phi^A]\,=2i[{a_0},\Psi[\phi]] \,.
\label{eq:tildeSgf}
\ee
For appropriate choices of the fermionic functional $\Psi$ (gauge-fixing), the last term in eq.~\eqref{eq:tildeSgf} vanishes and the master equations are satisfied, 
together with gauge-fixing independence of correlations function of gauge-invariant operators. 
In order to satisfy eq.~\eqref{eq:stilda}, we add to $S$ in eq.~\eqref{eq:red} a term
\be
\delta S=i\int d^d x \sqrt{h}\Big( B^*[{a_0},\bar{c}]+ \bar{c}_0^*[{a_0},{\bar{a}_0}]+c_0^*[{a_0}, {b_0}]+{a_0}^*[{a_0},c]\Big)\,.
\ee
The BRST transformation of fields is given by  $\delta_{\theta} \phi^{A}=\theta(-1)^{\epsilon_A} (S, \phi^{A})$,
$\delta_{\theta} \phi_{A}^{*}=-\theta(-1)^{\epsilon_A}(S, \phi_{A}^{*})$, with $\epsilon_A=0,1$ depending on the statistics of $\phi_A$ . Explicitly we get
\be
\begin{split}
\delta_{\theta} A& =\theta D c\,, \qquad  \;\;
\delta_{\theta} c =\theta\big(-{a_0}+ic^2 \big), \quad \delta_{\theta} \bar{c} =-\theta B, \quad \delta_{\theta} B =i\theta [{a_0},\bar{c}]\,, \\  \qquad \delta_{\theta}{a_0}=0 \,,
\quad \delta_{\theta}  \bar{c}_0  & =-i\theta [{a_0},{\bar{a}_0}]\,,\quad \delta_{\theta}  {c_0}=-i\theta [{a_0}, {b_0}],
\qquad \delta_{\theta} {\bar{a}_0}  =\theta  \bar{c}_0\,, \quad \delta_{\theta} {b_0} =\theta  {c_0}\,, \qquad\qquad
\end{split} \nn
\ee
so that $\delta_{\theta_1}\delta_{\theta_2} \phi=-i\theta_1\theta_2[{a_0},\phi]$ for any field $\phi$.
The gauge-fixing fermionic functional is taken as 
\be
\Psi=\int d^dx \ \sqrt{h}\left( \bar{{c}}\Big(-\frac{\xi}{2}B-\nabla A- {b_0}\Big)+{\bar{a}_0} c \right),
\ee
providing
\be
\begin{split}
\quad 
A^{*}& =-\nabla \bar{c}\,,\quad  {c}^{*}={\bar{a}_0}\,,\quad \bar{c}^{*}=-\frac{\xi}{2}B-\nabla A- {b_0}\,, \\
B^{*}& =-\frac{\xi}{2}\bar{c},\quad \; b_0^*= \bar{c}\,,\quad  {{\bar{a}_0}}^{*}=c\,.
\end{split}
\ee
We then get
\be
\begin{split}
S^\text{g.f.}=S_\text{YM}[A]+\int d^dx\,\sqrt{h}\bigg( \bar{c} \nabla  D {c}+B\big(-\frac{\xi}{2}B+\nabla A+ {b_0}\big)+{\bar{a}_0}{a_0} +c \bar{c}_0+\bar{c} {c_0}
+i{\xi}\ \bar{c}^2 {a_0}-{i}{c}^2 {\bar{a}_0} \bigg)\,,
\end{split}
\label{eq:pest}
\ee
which is the same action of eq.(4.2) in ref.~\cite{Pestun:2007rz}.\footnote{The precise matching, in the notation of ref.~\cite{Pestun:2007rz}, is $\bar{c}\rightarrow i\tilde{c}$, $B\rightarrow -i b$, $\bar a_0\rightarrow \tilde a_0-\xi_2a_0/2$, where $\xi_2$  is another gauge fixing parameter which does not affect the total path integral. Note that ref.~\cite{Pestun:2007rz} takes the fields to be antihermitian and not hermitian as in our case.}
We have then the following path integral:
\begin{equation}
\begin{split}
    Z_\text{YM}&=\frac{1}{\text{vol}(\mathcal{G},g)}\int \mathcal{D}A\exp(-S_\text{YM})\\&=\frac{1}{\mathrm{vol}(G)\sqrt{ \mathrm{vol}(S^d)^{\text{dim}(G)}}}\int\mathcal{D}A\ \mathcal{D}{c}\ \mathcal{D}\bar{c} \ \mathcal{D}B \ \mathcal{D}{a_0} \ \mathcal{D}{\bar{a}_0}\ \mathcal{D} {b_0}\ \mathcal{D} \bar{c}_0\ \mathcal{D} {c_0}  \exp(-S^\text{g.f.})\,.
    \end{split}
    \label{eq:pathPestun}
\end{equation}
The volume factor obtained after gauge fixing is the same found with the procedure used in the rest of this work: we can indeed verify that integrating out all fields and proceeding in reverse order to what we did in  sec. \ref{sec:Landau}, we reproduce the path integral in the first line of eq.~\eqref{eq:pathPestun}.
The integration of $ \bar{c}_0, {c_0}$ and $ {b_0}$  removes the zero modes respectively of $c,\bar{c}$ and $B$. Integrating out $a_0$ (along an imaginary contour to have a convergent path integral) sets also ${\bar{a}_0} $ to zero, while the gaussian integration in $B$ reproduces the usual gauge fixing term $(\nabla A)^2/(2\xi)$. 
\subsection{Computation of the ghost propagator}
In the previous section we explained how to perform the gauge-fixing of Yang-Mills theories on $S^d$ with the field-antifield formalism. The action that we obtained contains many fields that were not present in our main computation. As mentioned, one possibility is to integrate them out: in such a way we recover our original action and we can proceed as we already did. The other possibility is to keep the action \eqref{eq:pest} as it is and compute Feynman rules directly from it. We will focus in particular on the ghost action, which is
\begin{equation}
    S_\text{ghost}=\int d^dx\,\sqrt{h}\ \bar{c}\nabla D{c}+ c\bar{c}_0+\bar{c} {c_0}
+i{\xi}\ \bar{c}^2{a_0}-{i}c^2{\bar{a}_0}\,.
\end{equation}
In order to compute the propagator, we should rewrite the quadratic part of this action as
\begin{equation}
  S_\text{ghost}= \int d^d \sqrt{h}\ \frac{1}{2} \begin{pmatrix}
\bar{c} & c
\end{pmatrix} M \begin{pmatrix}
{c}\\ \bar{c}
\end{pmatrix} +c \bar{c}_0+ \bar{c}{c_0}\,,
\label{eq:Sghostp}
\end{equation}
with 
\begin{equation}
    M^{ab}=\begin{pmatrix}
\delta^{ab}\nabla^2  & -{\xi}f^{abc} {a_0} _{c} \\  f^{abc}{\bar{a}_{0c}}& -\delta^{ab}\nabla^2  
\end{pmatrix} .
\end{equation}
As explained before, the terms linear in $ \bar{c}_0$ and $ {c_0}$ set to zero the constant modes. As opposed to the standard ghost action, we do not have only the ghost-antighost term, but also terms quadratic in ghosts and antighosts.
The propagator will then be a matrix 
\begin{equation}
    G_{ab}=\begin{pmatrix}
\feynmandiagram [horizontal=a to b] {a -- [charged scalar] b -- [charged scalar] c}; &\feynmandiagram [horizontal=a to b] {a -- [anti charged scalar] b -- [ charged scalar] c};\\  \feynmandiagram [horizontal=a to b] {a -- [charged scalar] b -- [anti charged scalar] c}; & \feynmandiagram [horizontal=a to b] {a -- [anti charged scalar] b -- [anti charged scalar] c};
\end{pmatrix}=\begin{pmatrix}
\langle\bar{c}'_a c'_b \rangle  &\langle{c'}_a c'_b \rangle \\  \langle\bar{c}'_a \bar{c}'_b \rangle & \langle{c}'_a \bar{c}'_b \rangle
\end{pmatrix} 
\label{eq:ghostprop}
\end{equation}
with all entries different from zero, satisfying
\begin{equation}
    M^{ij}_{ab} G_{jk}^{bc}=-\delta^i_k\ \delta^c_a\  \delta(x-x')\,.
    \label{eq:propghost}
\end{equation}
Let us consider the following ansatz and verify if there exists such a solution:
\begin{equation}
    G_{ab}=\begin{pmatrix}
\delta_{ab} f_1(z)+ {a_0} _{b}{\bar{a}_{0a}} f_2(z)& f_{abc}{{a_0}} ^cg_1(z)  \\   f_{abc}{\bar{a}_0}^c h_1(z)  &  -\delta_{ab}f_1(z)-{a_0} _{a}{\bar{a}_{0b}} f_2(z) \,
\end{pmatrix} \,,
\end{equation}
with $f_1(z),f_2(z),g(z),h(z)$ generic functions of the stereographic coordinates. 
By replacing in eq.~\eqref{eq:propghost} and using the identities
\begin{equation}
f_{a b c} f_{a'b'c}=\frac{n_c}{n_c^2 - 2}\left( \delta_{a a'}\delta_{b b'} - 
\delta_{a b'} \delta_{ba'}\right),
\end{equation}
we get
\begin{equation}
\begin{split}
    \nabla^2 f_1&=-\xi\frac{n_c}{n_c^2 - 2} ({a_0}{\bar{a}_0}) h_1(z)+\delta(x-x') \,,\quad   \nabla^2 h_1= -f_1 \,,\\
        \nabla^2 f_2&=\xi\frac{n_c}{n_c^2 - 2} h_1(z) 
        \,, \qquad \qquad \qquad \qquad\qquad\quad  g_1 = \xi h_1\,.
\end{split}
\end{equation}
This is a system of coupled ordinary differential equations.
The solution is easily found by decomposing in spherical harmonics each function of $z$:
\begin{equation}
    f(z)=\sum_{\ell>0}f_\ell\  Y_\ell(x)Y_\ell(x')\,,
\end{equation}
where we exclude the constant mode $\ell=0$ because of the linear terms in eq.~\eqref{eq:Sghostp}.
We get
\begin{equation}
\begin{split}
    f_{1\ell}& =\frac{1}{2}\left(\frac{1}{-\ell(\ell+d-1)+m^2}+\frac{1}{-\ell(\ell+d-1)-m^2}\right) \,, \\
    h_{1\ell}& =\frac{1}{2m^2}\left(\frac{1}{-\ell(\ell+d-1)+m^2}-\frac{1}{-\ell(\ell+d-1)-m^2}\right) \,, \\
    f_{2\ell}& =\frac{\xi}{2 m^4}\left(\frac{1}{-\ell(\ell+d-1)+m^2}+\frac{1}{-\ell(\ell+d-1)-m^2}-\frac{2}{-\ell(\ell+d-1)}\right) \,, \\
    g_{1\ell} & = \xi f_{1\ell}\,,
    \end{split}
    \end{equation}
   where 
\be
m^2\equiv \sqrt{\frac{\xi n_c }{n_c^2 - 2} } (a_0\bar{a}_0)^{\frac 12}\,.
\ee    
Following the notation of sec.\ref{sec:sphere}, we denote by $G_\text{reg}(x,m^2)$ the solution of the scalar propagator equation on $S^d$  with zero modes removed:
\begin{equation}
    G_\text{reg}(z,m^2)=\sum_{\ell>0}\frac{1}{-\ell(\ell+d-1)+m^2}Y_\ell(x)Y_\ell(x').
\end{equation}
Summing over the non-constant modes we then find
\begin{equation}
\begin{split}
    &f_{1}=\frac{1}{2}\left(G_\text{reg}(z,m^2)+G_\text{reg}(z,-m^2)\right) \,, \\
     &f_2=\frac{\xi}{2m^4}\left(G_\text{reg}(z,m^2)+G_\text{reg}(z,-m^2)-2G_\text{reg}(z,0)\right) \,, \\
    &h_1=\frac{1}{2m^2}\left(G_\text{reg}(z,m^2)-G_\text{reg}(z,-m^2)\right) \,,\\
    &g_1=\xi h_1\,.
    \end{split}
    \end{equation}

\subsection{Match with eq.~\eqref{eq:g2ghost}}
Let us now consider how the ghost contribution $G_2^\text{ghost}$ is modified when the propagator in eq.~\eqref{eq:ghostprop} is used. As in this case also Wick contractions of two ghosts or two antighosts are allowed, the number of ghost diagrams increases. We have 
\begin{equation}
 \widetilde    G_2^\text{ghost}=\begin{tikzpicture}[baseline=(a.base)]
\begin{feynman}\vertex (a2);
\vertex[right=1cm of a2] (a3);
\vertex at ($(a2)!0.5!(a3)!0.5cm!90:(a3)$) (d);
\vertex at ($(a2)!0.5!(a3)!-0.5cm!90:(a3)$) (e);
            \diagram* {(a2) --  [charged scalar, quarter left] (d)-- [charged scalar, quarter left] (a3)--  [charged scalar, quarter left] (e)--  [charged scalar, quarter left] (a2)--  [gluon] (a3)
            };
            \end{feynman}
            \end{tikzpicture}\,+    \begin{tikzpicture}[baseline=(a.base)]
\begin{feynman}\vertex (a2);
\vertex[right=1cm of a2] (a3);
\vertex at ($(a2)!0.5!(a3)!0.5cm!90:(a3)$) (d);
\vertex at ($(a2)!0.5!(a3)!-0.5cm!90:(a3)$) (e);
            \diagram* {(a2) --  [anti charged scalar, quarter left] (d)-- [charged scalar, quarter left] (a3)--  [anti charged scalar, quarter left] (e)--  [charged scalar, quarter left] (a2)--  [gluon] (a3)
            };
            \end{feynman}
            \end{tikzpicture}\,=  \widetilde  G_2^\text{ghost1}+ \widetilde  G_2^\text{ghost2},
\end{equation}
with
\begin{align}
 \widetilde  G_2^\text{ghost1} & =g_{0}^{2}\!\! \int \!\! d^{d} x \ d^{d} x' \sqrt{h} \sqrt{h'} \Big(n_c\left(n_c^{2}-1\right)\nabla_\mu f_1    \nabla_{\mu'}f_1 +\frac{n_c}{n_c^{2}-2}({a_0}^2{\bar{a}_0}^2-({a_0}{\bar{a}_0})^2)\nabla_\mu f_2    \nabla_{\mu'}f_2\Big)  Q^{\mu\mu'}, \nn \\
 \widetilde  G_2^\text{ghost2}  & =g_{0}^{2}  \frac{n_c^2}{n_c^{2}-2}\xi  ({a_0}{\bar{a}_0})\int d^{d} x \ d^{d} x' \sqrt{h} \sqrt{h'}\  (\nabla_{\mu'}\nabla_\mu  h_1)\,h_1   Q^{\mu\mu'}\,.
\label{eq:noname}
\end{align}
The evaluation of eq.~\eqref{eq:noname} for generic $\xi$, which includes integrating over $a_0$ and $\bar a_0$, is a non-trivial task.   The computation remarkably simplifies in the Landau gauge $\xi\rightarrow 0$. In this limit $m^2\rightarrow 0$, the functions $f_1$ and $h_1$ are of order 1, while $f_2$ and $g_1$ are subleading in $\xi$. 
The only contribution  left in the limit is given by the first term in $G_2^\text{ghost1}$, the one involving the product of two $f_1$. For $\xi\rightarrow 0$,
 $f_1\rightarrow G_\text{reg}(z,0)$, which coincides with the ghost propagator in eq.~\eqref{eqn:13b}, and we
 reproduce exactly the result in eq.~\eqref{eq:g2ghost}:
\be
\lim_{\xi\rightarrow 0} \widetilde G_2^\text{ghost} =  G_2^\text{ghost} \,.
\ee
This is a sanity check of the validity of the heuristic Faddeev-Popov approach followed in the main text.

\bibliographystyle{JHEP}
\bibliography{Refs}

\providecommand{\href}[2]{#2}\begingroup\raggedright\begin{thebibliography}{10}

\bibitem{Jack:1982sn}
I.~Jack, {\it {Two Loop Background Field Calculations for Gauge Theories With
  Scalar Fields}},  {\em J. Phys. A} {\bf 16} (1983) 1083.

\bibitem{Zamolodchikov:1986gt}
A.~B. Zamolodchikov, {\it {Irreversibility of the Flux of the Renormalization
  Group in a 2D Field Theory}},  {\em JETP Lett.} {\bf 43} (1986) 730--732.

\bibitem{Cardy:1988cwa}
J.~L. Cardy, {\it {Is There a c Theorem in Four-Dimensions?}},  {\em Phys.
  Lett. B} {\bf 215} (1988) 749--752.

\bibitem{Komargodski:2011vj}
Z.~Komargodski and A.~Schwimmer, {\it {On Renormalization Group Flows in Four
  Dimensions}},  {\em JHEP} {\bf 12} (2011) 099,
  [\href{http://arxiv.org/abs/1107.3987}{{\tt arXiv:1107.3987}}].

\bibitem{Komargodski:2011xv}
Z.~Komargodski, {\it {The Constraints of Conformal Symmetry on RG Flows}},
  {\em JHEP} {\bf 07} (2012) 069, [\href{http://arxiv.org/abs/1112.4538}{{\tt
  arXiv:1112.4538}}].

\bibitem{Elvang:2012st}
H.~Elvang, D.~Z. Freedman, L.-Y. Hung, M.~Kiermaier, R.~C. Myers, and
  S.~Theisen, {\it {On renormalization group flows and the a-theorem in 6d}},
  {\em JHEP} {\bf 10} (2012) 011, [\href{http://arxiv.org/abs/1205.3994}{{\tt
  arXiv:1205.3994}}].

\bibitem{Cordova:2015fha}
C.~Cordova, T.~T. Dumitrescu, and K.~Intriligator, {\it {Anomalies,
  renormalization group flows, and the a-theorem in six-dimensional (1, 0)
  theories}},  {\em JHEP} {\bf 10} (2016) 080,
  [\href{http://arxiv.org/abs/1506.03807}{{\tt arXiv:1506.03807}}].

\bibitem{Jafferis:2011zi}
D.~L. Jafferis, I.~R. Klebanov, S.~S. Pufu, and B.~R. Safdi, {\it {Towards the
  F-Theorem: N=2 Field Theories on the Three-Sphere}},  {\em JHEP} {\bf 06}
  (2011) 102, [\href{http://arxiv.org/abs/1103.1181}{{\tt arXiv:1103.1181}}].

\bibitem{Klebanov:2011gs}
I.~R. Klebanov, S.~S. Pufu, and B.~R. Safdi, {\it {F-Theorem without
  Supersymmetry}},  {\em JHEP} {\bf 10} (2011) 038,
  [\href{http://arxiv.org/abs/1105.4598}{{\tt arXiv:1105.4598}}].

\bibitem{Casini:2011kv}
H.~Casini, M.~Huerta, and R.~C. Myers, {\it {Towards a derivation of
  holographic entanglement entropy}},  {\em JHEP} {\bf 05} (2011) 036,
  [\href{http://arxiv.org/abs/1102.0440}{{\tt arXiv:1102.0440}}].

\bibitem{Casini:2012ei}
H.~Casini and M.~Huerta, {\it {On the RG running of the entanglement entropy of
  a circle}},  {\em Phys. Rev. D} {\bf 85} (2012) 125016,
  [\href{http://arxiv.org/abs/1202.5650}{{\tt arXiv:1202.5650}}].

\bibitem{Casini:2006es}
H.~Casini and M.~Huerta, {\it {A c-theorem for the entanglement entropy}},
  {\em J. Phys. A} {\bf 40} (2007) 7031--7036,
  [\href{http://arxiv.org/abs/cond-mat/0610375}{{\tt cond-mat/0610375}}].

\bibitem{Casini:2017vbe}
H.~Casini, E.~Test\'e, and G.~Torroba, {\it {Markov Property of the Conformal
  Field Theory Vacuum and the a Theorem}},  {\em Phys. Rev. Lett.} {\bf 118}
  (2017), no.~26 261602, [\href{http://arxiv.org/abs/1704.01870}{{\tt
  arXiv:1704.01870}}].

\bibitem{Wilson:1971dc}
K.~G. Wilson and M.~E. Fisher, {\it {Critical exponents in 3.99 dimensions}},
  {\em Phys. Rev. Lett.} {\bf 28} (1972) 240--243.

\bibitem{Wilson:1973jj}
K.~G. Wilson and J.~B. Kogut, {\it {The Renormalization group and the epsilon
  expansion}},  {\em Phys. Rept.} {\bf 12} (1974) 75--199.

\bibitem{Giombi:2014xxa}
S.~Giombi and I.~R. Klebanov, {\it {Interpolating between $a$ and $F$}},  {\em
  JHEP} {\bf 03} (2015) 117, [\href{http://arxiv.org/abs/1409.1937}{{\tt
  arXiv:1409.1937}}].

\bibitem{Giombi:2015haa}
S.~Giombi, I.~R. Klebanov, and G.~Tarnopolsky, {\it {Conformal QED$_d$,
  $F$-Theorem and the $\epsilon$ Expansion}},  {\em J. Phys. A} {\bf 49}
  (2016), no.~13 135403, [\href{http://arxiv.org/abs/1508.06354}{{\tt
  arXiv:1508.06354}}].

\bibitem{Faddeev:1967fc}
L.~D. Faddeev and V.~N. Popov, {\it {Feynman Diagrams for the Yang-Mills
  Field}},  {\em Phys. Lett. B} {\bf 25} (1967) 29--30.

\bibitem{Appelquist:1989tc}
T.~Appelquist and D.~Nash, {\it {Critical Behavior in (2+1)-dimensional
  {QCD}}},  {\em Phys. Rev. Lett.} {\bf 64} (1990) 721.

\bibitem{Vafa:1983tf}
C.~Vafa and E.~Witten, {\it {Restrictions on Symmetry Breaking in Vector-Like
  Gauge Theories}},  {\em Nucl. Phys. B} {\bf 234} (1984) 173--188.

\bibitem{Vafa:1984xh}
C.~Vafa and E.~Witten, {\it {Eigenvalue Inequalities for Fermions in Gauge
  Theories}},  {\em Commun. Math. Phys.} {\bf 95} (1984) 257.

\bibitem{Komargodski:2017keh}
Z.~Komargodski and N.~Seiberg, {\it {A symmetry breaking scenario for
  QCD$_{3}$}},  {\em JHEP} {\bf 01} (2018) 109,
  [\href{http://arxiv.org/abs/1706.08755}{{\tt arXiv:1706.08755}}].

\bibitem{Sharon:2018apk}
A.~Sharon, {\it {QCD$_{3}$ dualities and the F-theorem}},  {\em JHEP} {\bf 08}
  (2018) 078, [\href{http://arxiv.org/abs/1803.06983}{{\tt arXiv:1803.06983}}].

\bibitem{Karthik:2018nzf}
N.~Karthik and R.~Narayanan, {\it {Scale-invariance and scale-breaking in
  parity-invariant three-dimensional QCD}},  {\em Phys. Rev. D} {\bf 97}
  (2018), no.~5 054510, [\href{http://arxiv.org/abs/1801.02637}{{\tt
  arXiv:1801.02637}}].

\bibitem{BenettiGenolini:2019zth}
P.~Benetti~Genolini, M.~Honda, H.-C. Kim, D.~Tong, and C.~Vafa, {\it {Evidence
  for a Non-Supersymmetric 5d CFT from Deformations of 5d $SU(2)$ SYM}},  {\em
  JHEP} {\bf 05} (2020) 058, [\href{http://arxiv.org/abs/2001.00023}{{\tt
  arXiv:2001.00023}}].

\bibitem{Bertolini:2021cew}
M.~Bertolini and F.~Mignosa, {\it {Supersymmetry breaking deformations and
  phase transitions in five dimensions}},  {\em JHEP} {\bf 10} (2021) 244,
  [\href{http://arxiv.org/abs/2109.02662}{{\tt arXiv:2109.02662}}].

\bibitem{Chang:2017cdx}
C.-M. Chang, M.~Fluder, Y.-H. Lin, and Y.~Wang, {\it {Spheres, Charges,
  Instantons, and Bootstrap: A Five-Dimensional Odyssey}},  {\em JHEP} {\bf 03}
  (2018) 123, [\href{http://arxiv.org/abs/1710.08418}{{\tt arXiv:1710.08418}}].

\bibitem{DeCesare:2021pfb}
F.~De~Cesare, L.~Di~Pietro, and M.~Serone, {\it {Five-dimensional CFTs from the
  \ensuremath{\varepsilon}-expansion}},  {\em Phys. Rev. D} {\bf 104} (2021),
  no.~10 105015, [\href{http://arxiv.org/abs/2107.00342}{{\tt
  arXiv:2107.00342}}].

\bibitem{Pestun:2007rz}
V.~Pestun, {\it {Localization of gauge theory on a four-sphere and
  supersymmetric Wilson loops}},  {\em Commun. Math. Phys.} {\bf 313} (2012)
  71--129, [\href{http://arxiv.org/abs/0712.2824}{{\tt arXiv:0712.2824}}].

\bibitem{Rubin:1984tc}
M.~A. Rubin and C.~R. Ordonez, {\it {Symmetric Tensor Eigen Spectrum of the
  Laplacian on $n$ Spheres}},  {\em J. Math. Phys.} {\bf 26} (1985) 65.

\bibitem{Allen_1986}
B.~Allen and T.~Jacobson, {\it Vector two-point functions in maximally
  symmetric spaces},  {\em Communications in Mathematical Physics} {\bf 103}
  (1986), no.~4 669--692.

\bibitem{Schwartz:2014sze}
M.~D. Schwartz, {\em {Quantum Field Theory and the Standard Model}}.
\newblock Cambridge University Press, 3, 2014.

\bibitem{Bailey:1995}
D.~H. Bailey, {\it A fortran 90-based multiprecision system},  {\em ACM Trans.
  Math. Softw.} {\bf 21} (dec, 1995) 379–387.

\bibitem{Jack:1990eb}
I.~Jack and H.~Osborn, {\it {Analogs for the $c$-Theorem for Four-dimensional
  Renormalizable Field Theories}},  {\em Nucl. Phys. B} {\bf 343} (1990)
  647--688.

\bibitem{Marino:2011nm}
M.~Marino, {\it {Lectures on localization and matrix models in supersymmetric
  Chern-Simons-matter theories}},  {\em J. Phys. A} {\bf 44} (2011) 463001,
  [\href{http://arxiv.org/abs/1104.0783}{{\tt arXiv:1104.0783}}].

\bibitem{Affleck:1985wa}
I.~Affleck, {\it {On the Realization of Chiral Symmetry in (1+1)-dimensions}},
  {\em Nucl. Phys. B} {\bf 265} (1986) 448--468.

\bibitem{Gepner:1984au}
D.~Gepner, {\it {Nonabelian Bosonization and Multiflavor {QED} and {QCD} in
  Two-dimensions}},  {\em Nucl. Phys. B} {\bf 252} (1985) 481--507.

\bibitem{Gracey:2015xmw}
J.~A. Gracey, {\it {Six dimensional QCD at two loops}},  {\em Phys. Rev. D}
  {\bf 93} (2016), no.~2 025025, [\href{http://arxiv.org/abs/1512.04443}{{\tt
  arXiv:1512.04443}}].

\bibitem{Casarin:2019aqw}
L.~Casarin and A.~A. Tseytlin, {\it {One-loop $\beta$-functions in 4-derivative
  gauge theory in 6 dimensions}},  {\em JHEP} {\bf 08} (2019) 159,
  [\href{http://arxiv.org/abs/1907.02501}{{\tt arXiv:1907.02501}}].

\bibitem{Wang:2017txt}
C.~Wang, A.~Nahum, M.~A. Metlitski, C.~Xu, and T.~Senthil, {\it {Deconfined
  quantum critical points: symmetries and dualities}},  {\em Phys. Rev. X} {\bf
  7} (2017), no.~3 031051, [\href{http://arxiv.org/abs/1703.02426}{{\tt
  arXiv:1703.02426}}].

\bibitem{Gaiotto:2017tne}
D.~Gaiotto, Z.~Komargodski, and N.~Seiberg, {\it {Time-reversal breaking in
  QCD$_{4}$, walls, and dualities in 2 + 1 dimensions}},  {\em JHEP} {\bf 01}
  (2018) 110, [\href{http://arxiv.org/abs/1708.06806}{{\tt arXiv:1708.06806}}].

\bibitem{Kaplan:2009kr}
D.~B. Kaplan, J.-W. Lee, D.~T. Son, and M.~A. Stephanov, {\it {Conformality
  Lost}},  {\em Phys. Rev.} {\bf D80} (2009) 125005,
  [\href{http://arxiv.org/abs/0905.4752}{{\tt arXiv:0905.4752}}].

\bibitem{Gorbenko:2018ncu}
V.~Gorbenko, S.~Rychkov, and B.~Zan, {\it {Walking, Weak first-order
  transitions, and Complex CFTs}},  {\em JHEP} {\bf 10} (2018) 108,
  [\href{http://arxiv.org/abs/1807.11512}{{\tt arXiv:1807.11512}}].

\bibitem{Benini:2019dfy}
F.~Benini, C.~Iossa, and M.~Serone, {\it {Conformality Loss, Walking, and 4D
  Complex Conformal Field Theories at Weak Coupling}},  {\em Phys. Rev. Lett.}
  {\bf 124} (2020), no.~5 051602, [\href{http://arxiv.org/abs/1908.04325}{{\tt
  arXiv:1908.04325}}].

\bibitem{Halperin:1973jh}
B.~i. Halperin, T.~C. Lubensky, and S.-k. Ma, {\it {First order phase
  transitions in superconductors and smectic A liquid crystals}},  {\em Phys.
  Rev. Lett.} {\bf 32} (1974) 292--295.

\bibitem{Ihrig:2019kfv}
B.~Ihrig, N.~Zerf, P.~Marquard, I.~F. Herbut, and M.~M. Scherer, {\it {Abelian
  Higgs model at four loops, fixed-point collision and deconfined
  criticality}},  {\em Phys. Rev. B} {\bf 100} (2019), no.~13 134507,
  [\href{http://arxiv.org/abs/1907.08140}{{\tt arXiv:1907.08140}}].

\bibitem{Seiberg:1996bd}
N.~Seiberg, {\it {Five-dimensional SUSY field theories, nontrivial fixed points
  and string dynamics}},  {\em Phys. Lett. B} {\bf 388} (1996) 753--760,
  [\href{http://arxiv.org/abs/hep-th/9608111}{{\tt hep-th/9608111}}].

\bibitem{Bertolini:2022osy}
M.~Bertolini, F.~Mignosa, and J.~van Muiden, {\it {On non-supersymmetric fixed
  points in five dimensions}},  {\em JHEP} {\bf 10} (2022) 064,
  [\href{http://arxiv.org/abs/2207.11162}{{\tt arXiv:2207.11162}}].

\bibitem{Florio:2021uoz}
A.~Florio, J.~a. M. V.~P. Lopes, J.~Matos, and J.~a. Penedones, {\it {Searching
  for continuous phase transitions in 5D SU(2) lattice gauge theory}},  {\em
  JHEP} {\bf 12} (2021) 076, [\href{http://arxiv.org/abs/2103.15242}{{\tt
  arXiv:2103.15242}}].

\bibitem{Siegel:1979wq}
W.~Siegel, {\it {Supersymmetric Dimensional Regularization via Dimensional
  Reduction}},  {\em Phys. Lett. B} {\bf 84} (1979) 193--196.

\bibitem{Mirabelli:1997aj}
E.~A. Mirabelli and M.~E. Peskin, {\it {Transmission of supersymmetry breaking
  from a four-dimensional boundary}},  {\em Phys. Rev. D} {\bf 58} (1998)
  065002, [\href{http://arxiv.org/abs/hep-th/9712214}{{\tt hep-th/9712214}}].

\bibitem{Davies:2021mnc}
J.~Davies, F.~Herren, and A.~E. Thomsen, {\it {General gauge-Yukawa-quartic
  $\beta$-functions at 4-3-2-loop order}},  {\em JHEP} {\bf 01} (2022) 051,
  [\href{http://arxiv.org/abs/2110.05496}{{\tt arXiv:2110.05496}}].

\bibitem{Bednyakov:2021qxa}
A.~Bednyakov and A.~Pikelner, {\it {Four-Loop Gauge and Three-Loop Yukawa Beta
  Functions in a General Renormalizable Theory}},  {\em Phys. Rev. Lett.} {\bf
  127} (2021), no.~4 041801, [\href{http://arxiv.org/abs/2105.09918}{{\tt
  arXiv:2105.09918}}].

\bibitem{abramowitz+stegun}
M.~Abramowitz and I.~A. Stegun, {\em Handbook of Mathematical Functions with
  Formulas, Graphs, and Mathematical Tables}.
\newblock Dover, New York, ninth dover printing, tenth gpo printing~ed., 1964.

\bibitem{Jack:1983sk}
I.~Jack and H.~Osborn, {\it {Background Field Calculations in Curved
  Space-time. 1. General Formalism and Application to Scalar Fields}},  {\em
  Nucl. Phys. B} {\bf 234} (1984) 331--364.

\bibitem{weinberg_1996}
S.~Weinberg, {\em The Quantum Theory of Fields}, vol.~2.
\newblock Cambridge University Press, 1996.

\bibitem{Gomis:1994he}
J.~Gomis, J.~Paris, and S.~Samuel, {\it {Antibracket, antifields and gauge
  theory quantization}},  {\em Phys. Rept.} {\bf 259} (1995) 1--145,
  [\href{http://arxiv.org/abs/hep-th/9412228}{{\tt hep-th/9412228}}].

\end{thebibliography}\endgroup

\end{document}